\documentclass[a4paper]{article}

\usepackage{amsmath,amssymb,amsfonts,amsbsy,amsthm,epsfig,graphicx}
\usepackage[left=7.5pc,right=7.5pc,top=1in,bottom=1in]{geometry}

\usepackage[numbers,sort&compress]{natbib}

\usepackage{color}
\usepackage[usenames,dvipsnames,table]{xcolor}
\usepackage{arydshln}
\usepackage{tabularx, url}

\theoremstyle{plain}
\newtheorem{theorem}{Theorem}[section]

\theoremstyle{definition}
\newtheorem{definition}[theorem]{Definition}

\theoremstyle{remark}
\newtheorem{remark}[theorem]{Remark}

\theoremstyle{result}
\newtheorem{result}[theorem]{Result}

\providecommand{\bo}{\mathbf}
\providecommand{\bs}{\boldsymbol}
\providecommand{\cov}{\mathrm{COV}}
\providecommand{\E}{\mathrm{E}}

\newcommand{\indep}{\bot\!\!\,\!\!\bot}

\begin{document}

\title{Sliced Average Variance Estimation for Multivariate Time Series}

\author{
M. Matilainen\textsuperscript{a}$^{\ast}$\footnote{$^\ast$Corresponding author. Email: markus.matilainen@utu.fi},
C. Croux\textsuperscript{b},
K. Nordhausen\textsuperscript{c}
and H. Oja\textsuperscript{d} \\
$ $ \\
{\small \textsuperscript{a} Department of Mathematics and Statistics, University of Turku, Turku, Finland} \\
{\small and Turku PET Centre, Turku, Finland;}\\
{\small \textsuperscript{b} Faculty of Economics and Business, K. U. Leuven, Belgium;}\\
{\small \textsuperscript{c}
 Institute of Statistics \& Mathematical Methods in Economics,} \\
 {\small Vienna University of Technology, Austria;}\\
{\small\textsuperscript{d} Department of Mathematics and Statistics, University of Turku, Turku, Finland}
}
\date{}
\maketitle

\begin{abstract}
Supervised dimension reduction for time series is challenging as there may be temporal dependence between the response $y$ and the predictors $\bo x$.
Recently a time series version of sliced inverse regression, TSIR, was suggested, which applies approximate joint diagonalization of several supervised lagged covariance matrices to consider the temporal nature of the data. In this paper we develop this concept further and propose a time series version of sliced average variance estimation, TSAVE. As both TSIR and TSAVE have their own advantages and disadvantages, we consider furthermore a hybrid version of TSIR and TSAVE. Based on examples and simulations we demonstrate and evaluate the differences between the three methods and show also that they are superior to apply their iid counterparts to when also using lagged values of the explaining variables as predictors.
\end{abstract}

\section{Introduction}

Linear supervised dimension reduction has a long tradition for independent and identically distributed (iid) observations with a rich literature reviewed for example in \cite{MaZhu2013}. The idea is to find all linear combinations of a predictor vector $\bo x$ which are needed to model a response $y$ even when the true functional relationship between the response and explaining variables is not known. In multivariate time series context with temporal dependence, the goal is similarly to model a response time series value $y_t$ at $t$ as a function of the previous history
of a stationary multivariate predictor time series $(\bo x_{t-j})_{j=1, 2, \ldots}$.
Popular time series dimension reduction methods such as those based on (dynamic) factor models, reviewed for example in \cite{Ensor2013}, are not supervised, and supervised dimension reduction methods are still rare in the literature. Both Xia et al. \cite{XiaTongLiZhu2002} and Becker and Fried \cite{BeckerFried2003} propose the use of standard supervised iid dimension reduction methods
simply  by  explaining a response series value $y_t$ with a vector of lagged predictor time series values in $\bo x_{t-1}, \ldots, \bo x_{t-j}$.  This may then increase the dimension of the problem dramatically and at the same time reduces the sample size. Barbarino and Bura \cite{BarbarinoBura2015, BarbarinoBura2017} combine ideas from factor models and standard iid supervised dimension reduction methods.

Recently, Matilainen et al. \cite{MatilainenCrouxNordhausenOja2017} proposed a procedure that finds most relevant linear combinations of the predictor series with their most relevant lags when modelling the response series.
The approach, called TSIR, is an extension of the sliced inverse regression (SIR), introduced by Li \cite{Li1991},
and is based on the approximate joint diagonalization of the covariance matrices of conditional expected values $\E(\bo x_{t}|y_{t+j})$, $j=1, 2, \ldots $ which naturally consider the temporal nature of the data.
Considering for example for a $p$-variate explaining time series $\bo x_t$, $k$ lags means to jointly approximately diagonalize $k+1$ $p \times p$ matrices while in approaches like \cite{XiaTongLiZhu2002} and \cite{BeckerFried2003} two $p(k+1) \times p(k+1)$ matrices need to be simultaneously diagonalized.

In sliced average variance estimation (SAVE; Cook and Weisberg \cite{CookWeisberg1991,Cook2000}) for iid observations,
one considers the variation of conditional covariance $\cov(\bo x|y)$ rather than the variation of the conditional expectation $\E(\bo x|y)$ to detect better the cases of nonlinear dependence.
In this paper we suggest the similar use of $\cov(\bo x_{t}|y_{t+j})$, $j=1,2, \ldots$, in a time series context. This is a time series extension of SAVE, and is called here TSAVE.
As TSIR and TSAVE have their own specific drawbacks, a hybrid of TSIR and TSAVE, denoted as TSSH,  is also introduced. It can be seen as a weighted combination of TSIR and TSAVE generalizing the hybrid in Zhu et al. \citep{ZhuOhtakiLi2007} to the time series context.

One aim here is also to see whether in a time series context the number of slices and the weight coefficient of the hybrid have similar preferred values as their iid counterparts. This was also not investigated in \citep{MatilainenCrouxNordhausenOja2017} for TSIR and the number of slices was just assumed to be the same as in the iid case. We will also investigate if these tuning parameters depend much on the underlying stochastic processes.  

The structure of the paper is as follows. We first recall SIR and SAVE for iid data. Then we move to the time series context, where first TSIR is reviewed and then in Section \ref{sec::TSAVE} TSAVE and in Section \ref{sec::TSSH} also the hybrid of TSIR and TSAVE are introduced.
Section \ref{sec::ExSim} then includes examples and simulation studies.
In Section \ref{sec::H} we conduct a simulation study to find some guidelines to how many slices we need in practice to estimate the matrices that we approximately jointly diagonalize. Then in Section \ref{sec::a} we have another simulation study to find the appropriate weights of TSIR and TSAVE parts for method TSSH and show the hybrid can sometimes be more efficient than these methods separately.
Finally in \ref{sec::comparison} we show that also TSAVE is often better than TSIR and that that both methods beat their iid counterparts applied to time series, such as the method in \cite{BeckerFried2003}.

\section{Supervised dimension reduction for iid data}\label{sec::iid}

In this section we review iid supervised dimension reduction methods SIR, SAVE and their hybrid version.
We formulate the supervised dimension reduction problem as an estimation problem in a blind source separation (BSS) model for the joint distribution
of the response variable $y$ and  the $p$-variate vector of observable explaining variables $\bo x$.
The BSS model then assumes that
\begin{equation}\label{IIDmodel}
  \bo x = \bo \Omega \bo z + \bs \mu,
\end{equation}
where the full rank $p \times p$ matrix  $\bo\Omega$ is called the mixing matrix and $\bs \mu$ is the location $p$-vector. The latent $p$-vector $\bo z$ can be partitioned as $\bo z = \left(\bo {z^{(1)}}^\top, \bo {z^{(2)}}^\top\right)^\top$ with the respective dimensions $k$ and $p-k$, and
\begin{enumerate}
\item[(I1)] $\E(\bo z)= \bo 0$ and $\cov(\bo z)=\bo I_p$ and
\item[(I2)] $(y,\bo z^{(1)}{}^\top)^\top \indep \bo z^{(2)}$.
\end{enumerate}

Hence in this model $\bo z^{(1)}$ carries all the information needed to model the response $y$ and $\bo z^{(2)}$ can be considered  as the noise part.
Note that assumption (I2) made in this paper for both SIR and SAVE is slightly stronger than those made in the original papers, i.e.
\begin{align*}
&\bo z^{(2)} \indep y \vert \bo z^{(1)}, \\
&\E(\bo z_t^{(2)}|\bo z_t^{(1)})= \bo 0 \ \mbox{(a.s.)  (for SIR) and} \\
&\cov(\bo z_t^{(2)}|\bo z_t^{(1)}) = \bo I_{p-k} \ \mbox{(a.s.) (for SAVE).}
\end{align*}
The assumption (I2) implies all these three assumptions and is needed  in \cite{NordhausenOjaTyler2016} to build  asymptotic and bootstrap tests for the true subspace dimension $k$ in the SIR methodology.

As there may be several partitions of $\bo z$  fulfilling (I1) and (I2), we choose the one with the smallest $k$.
 The aim is to find a $k \times p$ unmixing matrix $\bs \Gamma$ such that $\bs \Gamma \bo x = \bo z^{(1)}$ up to a pre-multiplication by a $k\times k$ orthogonal matrix. Note that the latent $\bo z^{(1)}$ as stated in our model has the same indeterminacy.

A direct consequence of assuming this model is the following.

\begin{result} \label{RES1}
Let $y$ denote the response and $\bo z$ have the properties as stated in (I1) and (I2). Then
   \[
\cov[\E(\bo z| y)]
      = \left(
        \begin{array}{cc}
         \cov[\E(\bo z^{(1)}| y)] & 0  \\
         0  & 0 \\
        \end{array}
   \right)
\]
and
   \[
\E[(\bo I_p - \cov(\bo z| y))^2]
      = \left(
        \begin{array}{cc}
         \E[(\bo I_k - \cov(\bo z^{(1)}| y))^2] & 0  \\
         0  & 0 \\
        \end{array}
   \right).
\]
\end{result}

As in \cite{MiettinenTaskinenNordhausenOja2015}, $\bo x = \bo \Omega \bo z + \bs \mu$ implies that  there exists an orthogonal matrix $\bo U$ such that
\begin{align}\label{4thMOMpaper}
\bo z = \bo U\ {\cov(\bo x)}^{-1/2}({\bo x} - \E(\bo x)).
\end{align}
 Then it is shown in \cite{MatilainenCrouxNordhausenOja2017} that, based upon  Result~\ref{RES1} (the first equation) and \eqref{4thMOMpaper}, one can define
 an unmixing matrix in the  sliced inverse regression (SIR) \cite{Li1991} using the following steps.
\begin{definition}
  The SIR functional ${\bs \Gamma}_{SIR}(\bo x;y)$ is defined as follows.
\begin{enumerate}
\item Consider the standardized variable ${\bo x}^{st} := {\cov(\bo x)}^{-1/2}({\bo x} - \E(\bo x))$.
\item Find the $k\times p$  matrix $\bo W_{SIR}=(\bo w_1,\ldots,\bo w_k)^\top$ with orthonormal rows $\bo w_1,\ldots,\bo w_k$ which maximizes
\begin{align*}
\sum_{i=1}^k \left[\bo w_i^\top  \cov[\E(\bo {\bo x}^{st}| y)] \bo w_i  \right]^2.
\end{align*}
\item ${\bo \Gamma_{SIR}}(\bo x;y) := {\bo W_{SIR} \cov(\bo x)}^{-1/2}$.
\end{enumerate}
\end{definition}

Assume now that $\bo V_1 \bo \Lambda_1 \bo V_1^\top$ is the eigenvector-eigenvalue decomposition of $\E[(\bo I_k - \cov(\bo z^{(1)}| y))^2]$. Let $\bo U_1$ be the $p\times k$  matrix
consisting of the first $k$ columns of $\bo U$. Based upon the second equation of Result~\ref{RES1} and \eqref{4thMOMpaper}
\begin{eqnarray*}
\E[(\bo I_p - \cov(\bo x^{st}| y))^2] &=& \E[(\bo I_p - \bo U \cov(\bo z| y)\bo U^\top)^2] \\
&=& \bo U\E[(\bo I_p - \cov(\bo z| y))^2]\bo U^\top \\
&=& \bo U_1\E[(\bo I_k - \cov(\bo z^{(1)}| y))^2]\bo U_1^\top \\
&=& \bo U_1 \bo V_1 \bo \Lambda_1 \bo V_1^\top\bo U_1^\top.
\end{eqnarray*}
Write now $\bo W_{SAVE}=(\bo U_1 \bo V_1)^\top$. Then  $\bo W_{SAVE} \E[(\bo I_p - \cov(\bo x^{st}| y))^2] \bo W_{SAVE}^\top=\bo \Lambda_1 $ is diagonal.
The sliced average variance estimation (SAVE) \cite{CookWeisberg1991,Cook2000} then has the following steps.

\begin{definition}
The SAVE functional ${\bs \Gamma_{SAVE}}(\bo x;y)$ is defined as follows.
\begin{enumerate}
\item Consider the standardized variable  ${\bo x}^{st} := {\cov(\bo x)}^{-1/2}({\bo x} - \E(\bo x))$.
\item Find the $k\times p$  matrix $\bo W_{SAVE}=(\bo w_1,\ldots,\bo w_k)^\top$ with orthonormal rows $\bo w_1,\ldots,\bo w_k$ that maximizes
\begin{align*}
\sum_{i=1}^k \left[\bo w_i^\top \E[(\bo I_p - \cov(\bo x^{st}| y))^2] \bo w_i  \right]^2.
\end{align*}
\item ${\bs \Gamma_{SAVE}}(\bo x;y) = {\bo W_{SAVE} \cov(\bo x)}^{-1/2}$.
\end{enumerate}
\end{definition}

For a random sample  $(y_1, \bo x_1), \ldots, (y_n, \bo x_n)$,  the population values of the two functionals can be consistently estimated if $y$ is discrete with a finite number of values.  In practice, the continuous $y$ is then often replaced by its discretized version utilizing  $H$ disjoint slices $S_1,...,S_H$, $S_1+ \ldots + S_H=\mathbb{R}$.
One can for example define a discrete variable $y^{sl}\in \{1,...,H\}$ by the condition $y^{sl}=h\Leftrightarrow y\in S_h$, $h=1,...,H$. Note that the condition (I2)
and the model still holds true for $(y^{sl},\bo x)$, but the dimension $k$ and the functionals may change.

\begin{remark}
Both SIR and SAVE can be seen as an approach which jointly diagonalizes two matrices \cite{LiskiNordhausenOja2014}, the regular covariance matrix and the supervised matrices $\cov[\E(\bo x| y)]$
or $\E[(\bo I_p - \cov(\bo x)^{-1/2}\cov(\bo x| y) \cov(\bo x)^{-1/2})^2]$, respectively. Hence, both methods can be solved using
a generalized eigenvector-eigenvalue decomposition, where under model~\eqref{IIDmodel} there are $\hat{k} \le k$ non-zero eigenvalues, and the functional is given by the corresponding generalized eigenvectors. Hence the functionals are unique if all non-zero eigenvalues are distinct.
\end{remark}

\begin{remark}
The estimation of the dimension of the subspace $k$ has some issues.
For example the number of slices $H$ can change the estimated value for the subspace, see e.g. \cite{BuraCook2001b}. However, the block diagonal structures in Result~\ref{RES1} still exist for different values of $H$, but the block sizes may be different.
Also the method used may not find the whole subspace. In case of SIR for example when $y$ is a quadratic function of a component, SIR fails to capture it (see e.g. \cite{CookWeisberg1991}). In general SAVE is able to capture a larger portion of the subspace than SIR, as Cook and Critchley have shown in  \citep{CookCritchley2000}.
Due to these issues, it is possible that $\hat{k} < k$.
\end{remark}

In the practical data analysis with unknown $k$, the estimated eigenvalues and the variation of the eigenvectors have been used to estimate $k$, see for example especially in the context of SIR, see \cite{Portier2016,LuoLi2016,NordhausenOjaTyler2016} and the references therein. The magnitude of the eigenvalues indicates the relevance of the corresponding source to model the response.

As \cite{CookCritchley2000} show, SAVE is in general considered more comprehensive when estimating the subspace of interest and SIR can be seen in certain situations as a special case. This increased flexibility of SAVE is however considered costly and it is usually said that SAVE needs more data than SIR \cite{CookCritchley2000}. This is also reflected when considering the numbers of slices used.

For SIR the slices are often chosen so that $\mathbb{P}(y\in \mathbb{S}_h)=1/H$, $h=1,\ldots,H$, with $H=10$. In simulations in \cite{Li1991} it was shown that  SIR is not very sensitive to the choice of $H$.
The rank of $\cov(\E(\bo x|y^{sl}))$ and the maximum number of non-zero eigenvalues is $H-1$, which however gives the restriction $H>\hat{k}+1$.

SAVE, unlike SIR, is more sensitive to the choice of $H$ as it uses higher moments and therefore needs more observations per slice than SIR, see for example \cite{Cook2000,CookCritchley2000, LiZhu2007}. Zhu et al. have conducted some simulations for SAVE \cite{ZhuOhtakiLi2007}. With a data length of $n = 480$, SAVE with $H = 6$ still produces proper results in all of their settings, but with $H = 24$ not anymore. However, we should note that these results are based only on some specific simulation settings.

Asymptotic properties of SAVE estimator $\bo W_{SAVE}$ have been investigated e.g. in \cite{CookNi2005} in order to find a way an estimate of the subspace dimension $k$. Li and Zhu \cite{LiZhu2007} have examined the consistency of the SAVE estimator. SAVE can achieve consistency and in the case where the response is discrete and takes finite values, SAVE can also achieve $\sqrt{n}$ consistency. However, generally SAVE cannot achieve this, unlike SIR. For asymptotic properties of the SIR estimator, including $\sqrt{n}$ consistency and asymptotic normality of the estimator, see \cite{ZhuNg1995}.

In \cite{CookCritchley2000} it is argued that SAVE with sufficient data is superior to SIR but in practice it would be better to use both and complement them to uncover the structures of interest.
A hybrid method based on SIR and SAVE, using a convex combination $(1-a)\cov(\E(\bo x^{st}|y^{sl})) + a\E[(\bo I_p - \cov(\bo x^{st}| y^{sl}))^2]$, with $a \in [0,1]$, has been proposed. It is discussed first briefly in \cite{YeWeiss2003} and then more closely with the discussion of the choice of the coefficient $a$ in \cite{ZhuOhtakiLi2007}. In Section \ref{sec::TSSH} we introduce a time series version of this hybrid method.

Also \cite{ShakerPrendergast2011} have combined the strengths of SIR and SAVE by suggesting the $SAVE|SIR$ method. As SIR is efficient in finding the linear relationships, it can be used to find a partial dimension reduction subspace and SAVE is then used to find the rest.

Note also that SIR$\alpha$, mentioned already in the rejoinder of Li's SIR paper \citep{Li1991} and developed further in \citep{Saracco2001} and \citep{GannounSaracco2003}, is the first kind of hybrid method of the first and second moments in supervised dimension reduction. However, this is not a combination of SIR and SAVE.

As recently  \cite{MatilainenCrouxNordhausenOja2017} extended SIR to the time series framework it is therefore natural also to extend SAVE, which will be done in the following sections.

\section{Linear supervised dimension reduction for time series}

\subsection{The blind source separation model for linear supervised dimension reduction for time series}

Assume that $y=(y_t)_{t\in \mathbb{Z}}$ and $\bo x =(\bo x_t)_{t\in \mathbb{Z}}  $ are (weakly and jointly) stationary univariate and $p$-variate time series, respectively.
In this paper the term time series is used for both the observed realizations and the stochastic process producing them.

In the time series prediction problem, it is usually assumed that the response series $y$ at time $t$ is an unspecified function of $\bo x_{t},\bo x_{t-1},...$ and
$\epsilon_{t},\epsilon_{t-1},...$ where $\epsilon=(\epsilon_t)_{t\in \mathbb{Z}}$ is
an unspecified stationary noise process independent from $\bo x$, i.e.,
\begin{align*}
y_{t} = f(\bo x_t, \bo x_{t-1}, \ldots; \epsilon_t, \epsilon_{t-1}, \ldots).
\end{align*}

As in the iid case we assume the blind source separation (BSS) model which states that only $k \ll p$ linear combinations of $\bo x$ are needed in the prediction model.
In the following, if $\bo A$ is a $k\times p$ matrix and $\bo b$ a $k$-vector, $\bo A\bo x+\bo b$ is a $k$-variate time series with the value $\bo A\bo x_t+\bo b$ at $t$.
In the time series BSS model,
we assume that
\begin{equation*}
  \bo x = \bo \Omega \bo z + \bs \mu,
\end{equation*}
where $\bo\Omega$ is a full rank $p \times p$  mixing matrix and $\bs \mu$ is a location vector. Furthermore, just like in the iid case,
the stationary $p$-variate source time series $\bo z$ can be partitioned as $\bo z = \left(\bo {z^{(1)}}^\top, \bo {z^{(2)}}^\top\right)^\top$
with the dimensions $k$ and $p-k$ of the subseries, respectively.
Dimension  $k$ is the smallest one to fulfil the conditions
\begin{enumerate}
\item[(T1)] $\E(\bo z_t)= \bo 0$ and $\cov(\bo z_t)=\bo I_p$ and
\item[(T2)] $(y,\bo {z^{(1)}}^\top)^\top \indep \bo z^{(2)}$.
\end{enumerate}

As in Section \ref{sec::iid}, from (T2) it follows therefore that
\begin{align*}
&\bo z^{(2)} \indep y \vert \bo z^{(1)}, \\
&\E(\bo z_{t+s}^{(2)}|\bo z_t^{(1)})= \bo 0 \ \mbox{(a.s.) for all $s\in \mathbb{Z}$ and} \\
&\cov(\bo z_{t+s}^{(2)}|\bo z_t^{(1)}) = \bo I_{p-k} \ \mbox{(a.s.) for all $s\in \mathbb{Z}$.}
\end{align*}

All the information needed to model $y$ is therefore contained in the process $\bo z^{(1)}$ and one can write
\begin{align}\label{eq::predmodel}
y_{t} = f(\bo x_t, \bo x_{t-1}, \ldots; \epsilon_t, \epsilon_{t-1}, \ldots)= f_0(\bo z_t^{(1)}, \bo z_{t-1}^{(1)}, \ldots; \epsilon_t, \epsilon_{t-1}, \ldots)
\end{align}
with another unspecified function $f_0$, possibly depending on $\bo \Omega$ and  $\bs \mu$.

Also in this time series case the model is ill-defined in the sense that both $\bo z^{(1)}$ and $\bo z^{(2)}$ can be multiplied by respective orthogonal matrices
and they still fulfil (T1) and (T2).
The goal is therefore the estimation of the unmixing matrix $\bs \Gamma$ such that $\bs \Gamma \bo x = \bo z^{(1)}$ up to orthogonal transformations. The function $f_0$ should  then be parametrized to allow
all linear combinations of the elements of $\bo z_t^{(1)}$, for example.
One should also identify which  lagged values $\bo z_t^{(1)},\bo z_{t-1}^{(1)},... $ contribute in the model.

In this section, the TSIR method from \cite{MatilainenCrouxNordhausenOja2017} is first reviewed and then the methods TSAVE and TSSH, a combination of TSIR and TSAVE are introduced. Finally we recall the choosing of the number of important sources and lags mentioned in \cite{MatilainenCrouxNordhausenOja2017}.

\subsection{SIR for time series}

In \cite{MatilainenCrouxNordhausenOja2017} the sliced inverse regression for time series uses the matrices
\[
G_{0,j}(\bo z, y) = \cov(\E(\bo z_t \vert y_{t+j})), \ \ j \in \mathbf{Z}_+
\]
with the following important property.
\begin{result} \label{RES2}
Under (T1) and (T2)
\[
\cov[\E(\bo z_t| y_{t+j})]
      = \left(
        \begin{array}{cc}
         \cov[\E(\bo z_t^{(1)}| y_{t+j})] & 0  \\
         0  & 0 \\
        \end{array}
   \right),
\]
for all lags $j \in \mathbf{Z}_+$.
\end{result}

Based on Result~\ref{RES2}, \cite{MatilainenCrouxNordhausenOja2017} then finds  the  time series sliced inverse regression (TSIR) estimate
of the unmixing matrix with the following three steps.
\begin{definition}\label{def::TSIR}
  The TSIR functional ${\bs \Gamma}_{TSIR}(\bo x;y)$ is defined as follows.
\begin{enumerate}
\item[1.] Find ${\bo x}^{st} := {\cov(\bo x)}^{-1/2}({\bo x} - \E(\bo x))$.
\item[2.]
Find the $k\times p$  matrix $\bo W_{TSIR}=(\bo w_1,\ldots,\bo w_k)^\top$ with orthonormal rows $\bo w_1,\ldots,\bo w_k$ which maximizes
\begin{align}\label{eq::TSIRmax}
\sum_{j \in S} \sum_{i=1}^k \left[\bo w_i^\top G_{0,j}(\bo x^{st}, y) \bo w_i  \right]^2,
\end{align}
for a chosen set of lags $S=\left\{S_1,\ldots, S_s \right\}$ with $S_j \geq 1$.

\item[3.] ${\bo \Gamma_{TSIR}}(\bo x;y) = {\bo W_{TSIR} \cov(\bo x)}^{-1/2}$.
\end{enumerate}
\end{definition}

The matrix $\bo W_{TSIR}$ is a $k \times p$ matrix. If the approach for time series used in \cite{XiaTongLiZhu2002} and \cite{BeckerFried2003} were applied here, this matrix would be of size $k \times (\vert S \vert + 1)p$. This would make the method less stable when the number of time series and the number of lags used increases.

\subsection{SAVE for time series}\label{sec::TSAVE}

To make a time series version of SAVE a natural extension for the matrix of interest is
\[
G_{1,j}(\bo z, y) = \E((\bo I_p - \cov(\bo z_t\vert y_{t+j}))^2),\ \  j \in \mathbf{Z}_+
\]
that depends a joint distribution of $y$ and $\bo x$. We then have the following.

\begin{result}\label{RES3}
Under (T1) and (T2)
\[
\E((\bo I_p - \cov(\bo z_t\vert y_{t+j}))^2)
      = \left(
        \begin{array}{cc}
         E((\bo I_p - \cov(\bo z^{(1)}_t\vert y_{t+j}))^2) & 0  \\
         0  & 0 \\
        \end{array}
   \right),
\]
for all lags $j \in \mathbf{Z}_+$.
\end{result}

The following unmixing matrix estimate is then called the time series sliced average variance estimator (TSAVE) functional:

\begin{definition}\label{def::TSAVE}
The TSAVE functional ${\bs \Gamma_{TSAVE}}(\bo x;y)$ is defined as follows.
\begin{enumerate}
\item[1.] Find ${\bo x}^{st} := {\cov(\bo x)}^{-1/2}({\bo x} - \E(\bo x))$.
\item[2.] Find the $k\times p$  matrix $\bo W_{TSAVE}=(\bo w_1,\ldots,\bo w_k)^\top$ with orthonormal rows $\bo w_1,\ldots,\bo w_k$ that maximizes
\begin{align}\label{eq::TSAVEmax}
&\sum_{j \in S}\sum_{i=1}^k \left[\bo w_i^\top G_{1,j}(\bo x^{st}, y) \bo w_i  \right]^2,
\end{align}
for a chosen set of lags $S=\left\{S_1,\ldots, S_s \right\}$ with $S_j \geq 1$.
\item[3.] ${\bo \Gamma_{TSAVE}}(\bo x;y) = {\bo W_{TSAVE} \cov(\bo x)}^{-1/2}$.
\end{enumerate}

\end{definition}

TSIR and TSAVE can be seen as procedures which jointly diagonalize $|S|+1$ matrices (in terms of the Frobenius norm), that is, the covariance matrix of $\bo x_t$ and $|S|$ matrices
depending on the joint distributions of $\bo x$ and $y$ with lags in $S$.
Under the blind source separation model assumed in this paper all the matrices of interest can be jointly diagonalized. However, for finite data this can be done only approximately anymore and it has to be solved using algorithms using some objective criterion as for example stated in the algorithmic outline above. While many other criteria are possible and many algorithms exists in the literature, for practical purposes we will use the approach based on Jacobi rotations \cite{CardosoSouloumiac1996} in search for the matrices $\bo W_{TSIR}$ and $\bo W_{TSAVE}$ that maximize \eqref{eq::TSIRmax} and \eqref{eq::TSAVEmax}, respectively. This algorithm was recommended in \cite{IllnerMiettinenFuchsTaskinenNordhausenOjaTheis2015} and for more details about  joint diagonalization in BSS see for example \cite{TheisInouye2006,ChabrielKleinsteuberMoreauShenTichavskyYeredor2014,
MiettinenNordhausenOjaTaskinen2014b,
MiettinenTaskinenNordhausenOja2015,
MatilainenNordhausenOja2015,
MiettinenIllnerNordhausenOjaTaskinenTheis2016,
MiettinenNordhausenTaskinen2017} and the references therein. In the following we will however not distinguish between joint diagonalization and joint approximate diagonalization.

Given a solution $\bo W=(\bo w_1,...,\bo w_k)'$, the maximum value of the criterion function is $\sum_{j\in S}\sum_{i=1}^k \lambda_{ij}$, where
\[
\lambda_{ij}=\left( \bo w_i^\top \E[(\bo I_p - \cov(\bo x_t^{st}| y_{t+j}))^2] \bo w_i\right)^2,
\]
in a sense that it measures the contribution of the $s$th lag of the  $i$th linear combination to this maximum value, $i=1,\ldots,k$; $j\in S$.
$\bs \Gamma_{TSAVE}(\bo x;y)$ is unique if $\lambda_{i\cdot}=\sum_{j\in
\mathcal{S}} \lambda_{ij}$, $i=1,\ldots,k$ are distinct. The components
$\bs \Gamma_{TSAVE}(\bo x;y)\bo x$ are standardized and can be naturally ordered so that $\lambda_{1\cdot}\ge \ldots \ge \lambda_{k\cdot}$.
The large value $\lambda_{i\cdot}$ indicates  a strong dependence between the time series $(\bo\Gamma(\bo x;y)\bo x)_i$ and  $y$. The higher the value of $\lambda_{ij}$,
the stronger is the dependence between $(\bo\Gamma(\bo x;y)\bo x)_{it}$ and $y_{t+j}$. Identifying the relevant sources and lags is however difficult due to the possible serial correlations in $\bo x$ which means that $\lambda_{ij}$ might vanish only slowly to zero
with $s$ for irrelevant lags. 

As for SAVE, the unmixing matrix estimate is obtained for the sliced version $\bs \Gamma_{TSAVE}(\bo x;y^{sl})$, where $y^{sl}$ is a discrete time series such that
$y^{sl}_t=h\Leftrightarrow y_t\in S_h$, $h=1,...,H$.
As with SAVE it can be also here concluded that TSAVE will be more sensitive to the number of slices $H$ as also TSAVE, just like SAVE, estimates a higher order moment and therefore needs more information (see Section \ref{sec::H}).

Consider the following important property of TSAVE.

\begin{result}\label{RES4}
Let $\bo x^* = \bo A \bo x + \bo b$, where $\bo A$ is a full rank $p \times p$ matrix and $\bo b$ a $p$-vector.
TSAVE is affine equivariant in the sense that, for all transformed time series $\bo x^*$, $\bs \Gamma_{TSAVE}(\bo x^*;y)\bo x^*=\bs \Gamma_{TSAVE}(\bo x;y)\bo x$ up to the signs and the location of the component series.
\end{result}

Result \ref{RES4} also means that $\bs \Gamma_{TSAVE}(\bo x^*;y) = \bo J\bs \Gamma_{TSAVE}(\bo x;y) \bo A^{-1}$, where $\bo J$ is a $p \times p$ diagonal matrix with diagonal elements $\pm 1$, up to the location.
The proof is straightforward and hence is omitted from here.

To derive asymptotic properties of  TSAVE the challenge consists of deriving  the joint limiting distributions of $\sqrt{T}(\widehat{\cov}(\bo x) - \cov(\bo x))$ and $\sqrt{T}(\hat{ G}_{1,j}(\bo x, y) -  G_{1,j}(\bo x, y))$ for all lags $j \in S$ for which probably stronger assumptions on the process $\bo x_t$ need to be made. Therefore this is beyond the scope of this paper and we just outline how the asymptotics could be derived given the joint distribution of these matrices and that the signal dimension $k$ would be known.
   
Write $\bo G_j := G_{1,j}(\bo x, y)$, for all $j = 1, \ldots, S$. The maximization \eqref{eq::TSAVEmax} can also be written as 
\begin{align}\label{eq::TSAVEmaxALT}
&\sum_{j \in S}\sum_{i=1}^k \left[\bo w_i^\top \bo G^*_j \bo w_i  \right]^2,
\end{align}
where $\bo G^*_j = \cov(\bo x)^{-1/2} \bo G_j \cov(\bo x)^{-1/2}$. Denote then $\bo \Gamma :=\bo \Gamma_{TSAVE}$ and $\bo M = M(\bo W) = ((m(\bo w_i))_{i = 1, \ldots, {k}}^{\top}$, where $m(\bo w_i) = \sum_{j = 1}^{\hat{s}} \left[\bo w_i^\top \bo G^*_j \bo w_i  \right]\bo G^*_j \bo w_i$, for $i = 1, \ldots, {k}$.
Now we can use the Lagrange multiplier technique, which yields $\bo W \bo M = \bo M \bo W^{\top}$ and $\bo W \bo W^{\top} = \bo I_{{k}}$. These equations lead to a fixed-point algorithm (see e.g. \citep{CardosoSouloumiac1996}) with a step 
$\bo W \leftarrow (\bo M \bo M^{\top})^{-1/2}\bo M$.

Using this we can search for the limiting distributions of $\hat{\bo \Gamma}$ and $\hat{\bo W}$, with known dimension ${k}$ and lags in $S$. As the estimate is also affine equivariant, we can wlog consider here the case, where $\cov(\bo x) = \bo I_p$ and $\bo W = (\bo I_k, \bo 0)$. 

Let $T$ be the length of the time series. We also need to assume here that the joint limiting distribution of $\sqrt{T}(\widehat{\cov}(\bo x) - \bo I_p)$ and $\sqrt{T}(\hat{\bo G}_j - \bo G_j)$ is known. Assuming the distribution is known, the joint limiting distribution of $\hat{\bo W}$ and $\hat{\bo M}$ satisfies the conditions
\[
\sqrt{T}(\hat{\bo W} - \bo W)\bo M' - \bo M\sqrt{T}(\hat{\bo W} - \bo W)' = \sqrt{T}(\hat{\bo M} - \bo M)\bo W' - \bo W\sqrt{T}(\hat{\bo M} - \bo M)' + o_P(1)
\]
and
\[\sqrt{T}(\hat{\bo W} - \bo W)\bo W' = -\bo W\sqrt{T}(\hat{\bo W} - \bo W)' + o_P(1),
\]
which then can be used in finding the joint limiting distribution of $\sqrt{T}(\hat{\bo W} - \bo W)$ and $\sqrt{T}(\widehat{\cov}(\bo x) - \bo I_p)$.

Finally we get $\sqrt{T}(\hat{\bo \Gamma} - \bo W)= \sqrt{T}(\hat{\bo W} - \bo W) - \frac{1}{2}\bo W \sqrt{T}(\widehat{\cov}(\bo x) - \bo I_p) + o_P(1)$. 
For similar derivations based on the Langrangian multiplier technique, see for example \citep{MiettinenIllnerNordhausenOjaTaskinenTheis2016}.

\subsection{A hybrid of TSIR and TSAVE}\label{sec::TSSH}

As both SIR and SAVE have their advantages and drawbacks, a hybrid of SIR and SAVE using a convex combination of the two supervised matrices was proposed in \cite{ZhuOhtakiLi2007}.
As the time series versions of SIR and SAVE show similar behaviour to the original SIR and TSAVE, respectively, we similarly propose here a convex combination of TSIR and TSAVE methods. We call this time series SIR and SAVE hybrid method TSSH.
In TSSH we are searching for a $k\times p$  matrix $\bo W_{TSSH}=(\bo w_1,\ldots,\bo w_k)^\top$ with orthonormal rows $\bo w_1,\ldots,\bo w_k$ that maximizes
\begin{align*}
\sum_{j \in S}\sum_{i=1}^k \left(\bo w_i^{\top} \left((1-a) * G_{1,j}(\bo x^{st}, y) +
  a * G_{2,j}(\bo x^{st}, y)\right)\bo w_i \right)^2,
\end{align*}
where $a \in [0,1]$ and $S$ a set of chosen lags as before. Then $a=0$ gives TSIR and $a=1$ gives TSAVE.

Note that the results similar to Results \ref{RES2} and \ref{RES3} and the Result \ref{RES4} apply to TSSH as well.
Also the search for the number of latent sources and lags goes as for TSIR and TSAVE.

In addition to the issues that TSIR and TSAVE have, a proper value for the coefficient $a$ needs to be found. This is discussed in Section \ref{sec::a}.

Extending the $SIR|SAVE$ method by \cite{ShakerPrendergast2011} for time series is challenging, as we need to search not only for sources but also lags corresponding to each of the sources.

\subsection{Identification of $k$ and the lags of interest}\label{sec::lagdir}

In practice the number of sources, i.e. the value of $k$, is not known and needs to be estimated as well. Also the important lags regarding the sources need to be found.
We can choose these by using the quantities $\lambda_{ij}$.
However, at this stage of the development of TSAVE, formal testing is not possible yet and we suggest to use the same strategies as suggested for TSIR in \cite{MatilainenCrouxNordhausenOja2017}.

For that purpose consider the matrix $\bo L = l_{ij}$ where
\[
l_{ij} = \frac{\lambda_{ij}}{\sum_{i=1}^k \sum_{j = 1}^s \lambda_{ij}}, \ \ \ i=1,\ldots,k;\ j = 1, \ldots, s,
\]
contains the scaled pseudo eigenvalues and the scaling is chosen such that the elements of $\bo L$ add up to 1. Note that we have assume here that we use the first $s$ lags, as currently we do not have information that would suggest to use some other set of lags.
Row and column sums of $\bo L$ will be again denoted as $l_{i\cdot}=\sum_{j=1}^s l_{ij}$, as before, and $l_{\cdot j}=\sum_{i=1}^k l_{ij}$.

Assume then that $\Gamma_{TSAVE}(\bo x;y)$ is defined such that the latent sources are ordered according to their magnitudes of $\lambda_{1\cdot} \geq \ldots \geq \lambda_{k\cdot}$ and $k=p$. Then \cite{MatilainenCrouxNordhausenOja2017} suggested different strategies to find the appropriate amount of sources and the lags corresponding to those sources, by trying to explain similar as in principal component analysis (PCA) $100 \cdot P\%$ of the dependence between the latent sources and the response series.

The suggested strategies can be summarized as:

\begin{description}
  \item[ALL LAGS:] keep all $s$ lags and find the smallest value $\hat{k}$ such that $\sum_{i=1}^{\hat{k}}\l_{i\cdot}\ge P$.
  \item[ALL SOURCES:] keep all $k$ sources and find the smallest $\hat{s}$ in such way that $\sum_{j=1}^{\hat{s}}\l_{\cdot j}\ge P$.
  \item[RECTANGLE:] find $\hat{k}$ and $\hat{s}$ with the smallest product $\hat{k}\hat{s}$ in such way that $\sum_{i=1}^{\hat{k}}\sum_{j=1}^{\hat{s}}\l_{ij}\ge P$.
  \item[BIGGEST VALUES:] find the smallest number $\hat{r}$ of elements $(i_1,j_1),\ldots,(i_{\hat{r}},j_{\hat{r}})$ of $\bo L$ in such way that $\sum_{k=1}^{\hat{r}} l_{i_k j_k}\ge P$.
\end{description}

While the first two strategies assume some prior knowledge about the number of lags or sources respectively, the last two methods seem suitable for general use.
As in the iid case, the `real' amount of sources $k$ may not be found due to the slicing (value of $H$) and/or the method used, and hence it is possible that $\hat{k} < k$.

Natural values for P are then for example $0.5$ or $0.8$. How the different strategies perform will also be considered in the example and simulation section.

\section{Examples and simulations}\label{sec::ExSim}

In this section the differences between TSIR and TSAVE are first visualized. Then the simulation settings and the prediction models are presented. In Section \ref{sec::H} we search for the best values for the number of slices $H$ in TSIR and TSAVE and in Section \ref{sec::a} the appropriate values for the coefficient $a$ for TSSH. In both cases we aim to give some guidelines how to choose them in practice. Finally we compare TSAVE with TSIR, SIR and SAVE (applied to time series case) in Section \ref{sec::comparison}.

Note that TSIR, TSAVE and TSSH, together with the different selection strategies described in the previous section, are implemented in the R package tsBSS \cite{tsBSS}
and are used together with the R package JADE \cite{MiettinenNordhausenTaskinen2017} in this section.

\subsection{Visualization of the differences of TSIR and TSAVE}\label{sec::visual}

It is already well established that the regular SIR works efficiently with linear relationships, but not when the relationship in $y = g(\bo z^{(1)}) + \epsilon$ is specified by a symmetric function $g$ \cite{Li1991, CookWeisberg1991}.
On the other hand, the regular SAVE works with a symmetric function $g$.

\begin{figure}[h]
\includegraphics[scale=0.48]{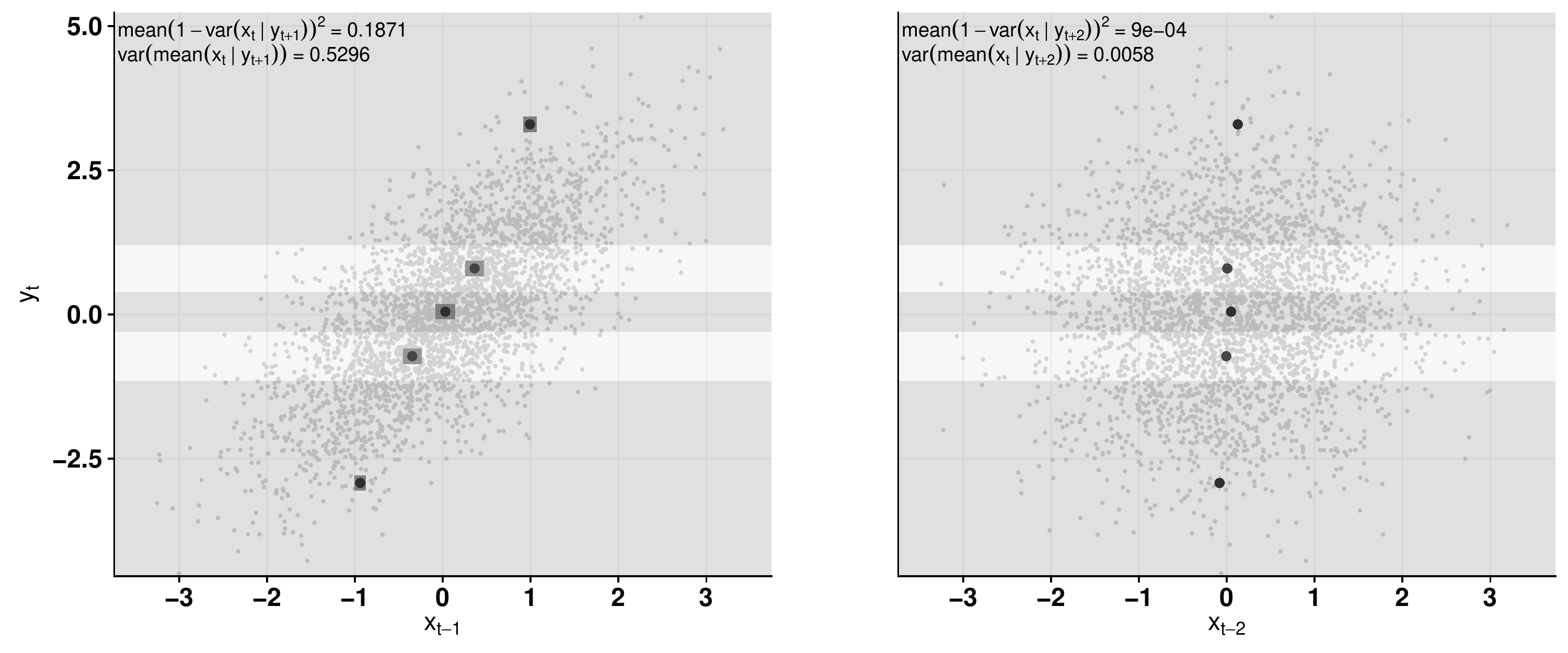}
\includegraphics[scale=0.48]{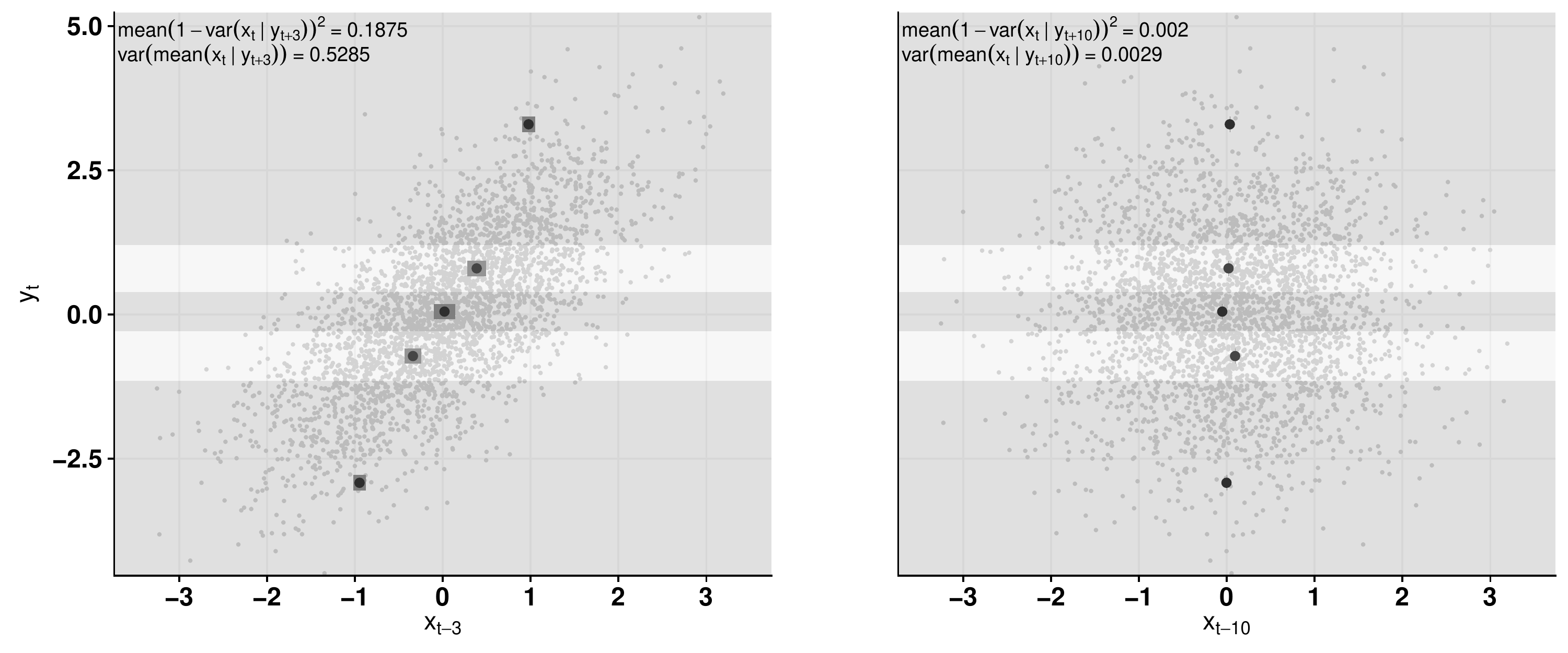}
\caption{Model {\it M1}: Scatterplot of $y_t$ and $x_{t - j}$, $j = 1, 2, 3, 10$ with slices of $y_t$ as the shaded areas}\label{fig::A}
\end{figure}

\begin{figure}[h]
\includegraphics[scale=0.48]{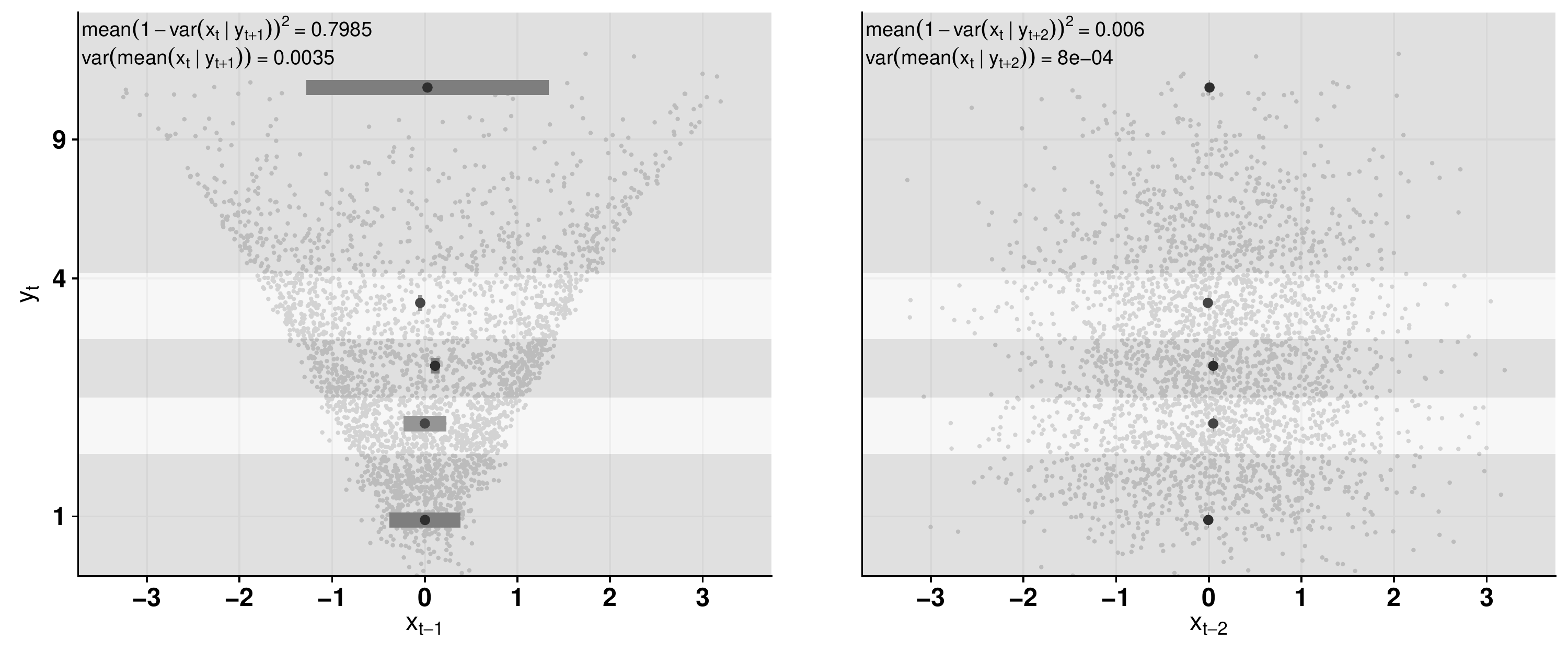}
\includegraphics[scale=0.48]{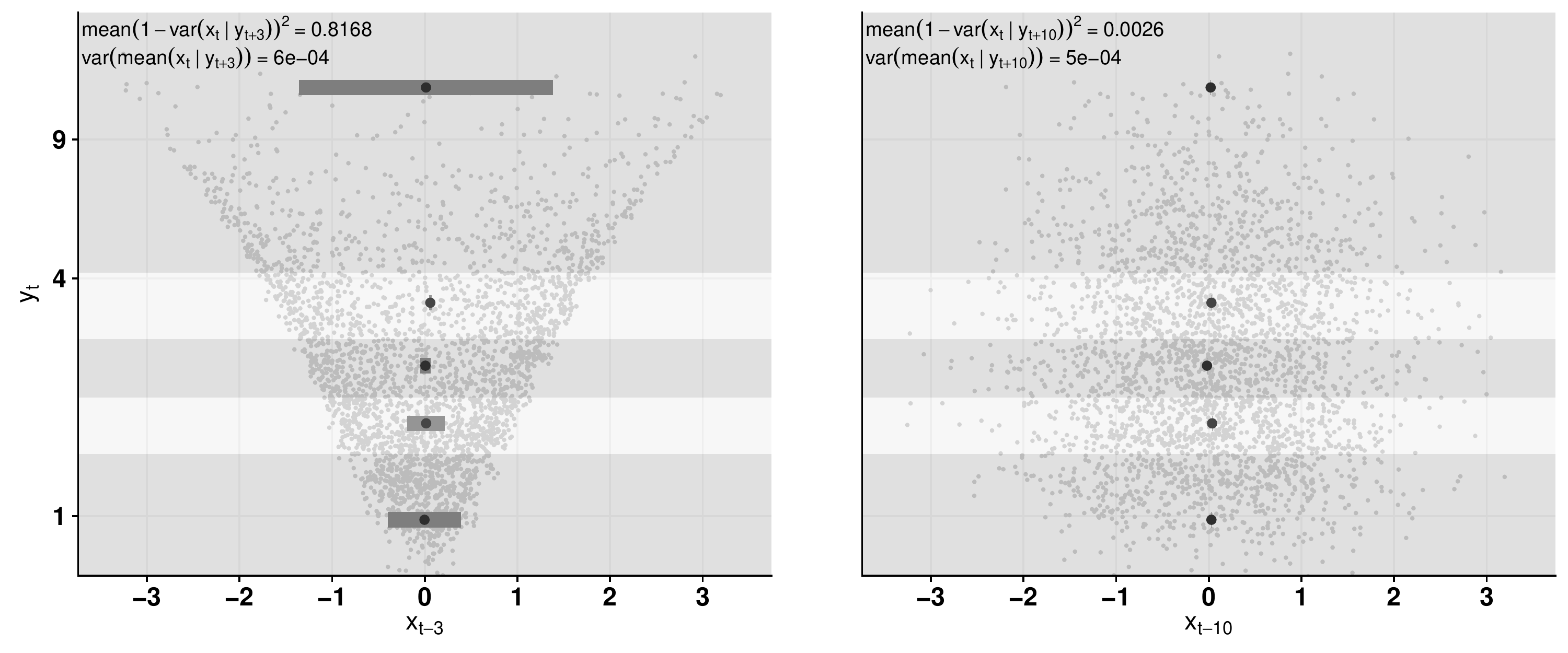}
\caption{Model {\it M2}: Scatterplot of $y_t$ and $x_{t - j}$, $j = 1, 2, 3, 10$ with slices of $y_t$ as the shaded areas. The $y$-axis is in a logarithmic scale.}\label{fig::B}
\end{figure}

To consider the differences between the time series versions of both methods, consider the examples where the response $y$ at time $t$ be
\begin{enumerate}
\item[\it M1:] $y_t = x_{t - 1} + x_{t - 3} + \epsilon_t$
\item[\it M2:] $y_t = 1 + x_{t - 1}^2 + x_{t - 3}^2 + \epsilon_t$
\end{enumerate}
where $\epsilon_t \sim N(0, 0.2)$ and $x_t$ follows an AR(1) model with $\phi = 0.1$.
To illustrate how and where TSIR and TSAVE work, we plot the values of $y_t$ against the values of $x_{t - j}$, where $j = 1, 2, 3$ or $10$ for all the models (see Figures \ref{fig::A} and \ref{fig::B}).

The larger black dots in the figures denote the sample values of $\E(x_{t}|y_{t + j})$ in each slice. Also the variance of these values is added to each figure as text. A non-zero variance value indicates that TSIR is able to find a relationship between $y_t$ and $x_{t - j}$.

The width of the dark gray bars added around the black dots corresponds to the sample value of $(1 - $Var$(x_{t}|y_{t + j}))^2$ in each slice.
When TSAVE cannot find a relationship between $y_t$ and $x_{t - j}$, all these values are so close to zero that the bars are hardly visible.
Also the means of these values are added to each figure as text. A non-zero mean indicates that TSAVE is able to find a relationship between $y_t$ and $x_{t - j}$.

From Figure~\ref{fig::A} and \ref{fig::B} it can be seen that TSIR cannot find any relationship between $y_t$ and $x_{t - j}$, when $j = 2$ or $10$, as $y_t$ did not depend on $x_t$ with those lags.
However, the results are different with $j = 1$ and $3$. As seen on the left side panels of Figure~\ref{fig::A}, $y_t$ and $x_{t-j}$ have a strong linear relationship. The variance of the slice means is clearly non-zero and indicates that TSIR finds the relationship. Also a non-zero value for the mean of the conditional variances indicates that also TSAVE works with linear relationships.

On the left side panels of Figure~\ref{fig::B}, $y_t$ and $x_{t-j}$ have a strong quadratic relationship. It can easily be seen that TSAVE finds the quadratic relationship. However, as the mean values for each slice are close to zero, TSIR fails to find the quadratic relationship, similar to the regular SIR in iid regression.

Based on the figures and the values in them, it can be concluded that TSAVE finds both the linear and the quadratic relationship between $y_t$ and $x_t$ with lags 1 and 3 in the models {\it M1} and {\it M2}.

Next we illustrate how the appropriate lags and the amount of the latent sources are chosen in TSAVE and TSIR using the strategies mentioned in Section~\ref{sec::lagdir}.

Consider a $4$-variate time series $\bo x = (x_1, x_2, x_3, x_4)^{\top}$, where $x_1$ and $x_2$ are AR(1) processes with $\phi = 0.2$, $x_3$ is ARMA(1,1) with $\phi = 0.3$ and $\theta = -0.4$, and $x_4$ is MA(1) model with $\theta = -0.4$.  As both methods are affine equivariant, we have chosen $\bo \Omega = \bo I_4$ as the mixing matrix and therefore $\bo x = \bo z$ for all $t \in \mathbf{Z}$.
The time series $\bo z$ are standardized and the length of the time series is $T = 10000$.

In choosing the number of sources and the lags, we use $P=0.8$ as the threshold value, and $H = 5$ as the number of slices. Tables are constructed as the average values of the  elements $l_{ij}$ over 100 repetitions, and we have used lags $1, \ldots, 12$ for both methods.
Assume now that the response $y$ at a time $t$ depends on the predictors as follows.
$$
y_t = z_{1,t-1}^2 + 3z_{2,t-5} + \epsilon_t,
$$
where $\epsilon_t \sim N(0,1)$. Table~\ref{table::ARMAModelBlow} includes the $\bo L$ matrices for both TSAVE and TSIR based on this model. It can be seen that TSAVE finds two latent sources and five lags. Also already the values of the elements $l_{12}$ and $l_{51}$ are clearly bigger than others and together they already explain more than 80 \% of the dependence between the response $y$ and the predictors. On the other hand TSIR finds only one source (and five lags), which seems to explain by far the most of the dependence, but fails to find the other source with the quadratic relationship.

\begin{table}
{\small
\centering
\begin{minipage}{\textwidth}
\begin{minipage}[b]{0.49\textwidth}
\begin{tabular}{p{0.8cm}|p{0.7cm}p{0.7cm}:p{0.7cm}p{0.7cm}|p{0.7cm}}
 & $\bo w'_1 \bo x^{st}$ & $\bo w'_2 \bo x^{st}$ & $\bo w'_3 \bo x^{st}$ & $\bo w'_4 \bo x^{st}$ & Sum \\
  \hline
  $t - 1$ & 0.002 & \multicolumn{1}{p{0.7cm}|}{\cellcolor{gray!25} 0.317} & 0.002 & 0.002 & 0.323 \\
  $t - 2$ & 0.002 & \multicolumn{1}{p{0.7cm}|}{ 0.003} & 0.002 & 0.002 & 0.010 \\
  $t - 3$ & 0.002 & \multicolumn{1}{p{0.7cm}|}{ 0.002} & 0.003 & 0.002 & 0.009 \\
  $t - 4$ & 0.003 & \multicolumn{1}{p{0.7cm}|}{ 0.002} & 0.003 & 0.002 & 0.010 \\
  $t - 5$ & \cellcolor{gray!25} 0.576 & \multicolumn{1}{p{0.7cm}|}{ 0.002} & 0.002 & 0.002 & 0.582 \\
\hdashline
\cline{2-3}
  $t - 6$ & 0.003 & 0.002 & 0.003 & 0.002 & 0.010 \\
  $t - 7$ & 0.002 & 0.002 & 0.003 & 0.002 & 0.009 \\
  $t - 8$ & 0.002 & 0.002 & 0.003 & 0.002 & 0.009 \\
  $t - 9$ & 0.002 & 0.002 & 0.002 & 0.002 & 0.009 \\
  $t - 10$ & 0.002 & 0.002 & 0.003 & 0.002 & 0.009 \\
  $t - 11$ & 0.002 & 0.002 & 0.003 & 0.002 & 0.009 \\
  $t - 12$ & 0.002 & 0.002 & 0.002 & 0.002 & 0.009 \\
     \hline
  Sum & 0.603 & 0.342 & 0.029 & 0.025 & 1.000 \\
\end{tabular}
\end{minipage}
\begin{minipage}[b]{0.49\textwidth}
\centering
\begin{tabular}{p{0.8cm}|p{0.7cm}:p{0.7cm}p{0.7cm}p{0.7cm}|p{0.7cm}}
 & $\bo w'_1 \bo x^{st}$ & $\bo w'_2 \bo x^{st}$ & $\bo w'_3 \bo x^{st}$ & $\bo w'_4 \bo x^{st}$ & Sum \\
  \hline
  $t - 1$ & \multicolumn{1}{p{0.7cm}|}{ 0.001} & 0.001 & 0.001 & 0.001 & 0.004 \\
  $t - 2$ & \multicolumn{1}{p{0.7cm}|}{ 0.001} & 0.001 & 0.001 & 0.001 & 0.004 \\
  $t - 3$ & \multicolumn{1}{p{0.7cm}|}{ 0.002} & 0.001 & 0.001 & 0.001 & 0.006 \\
  $t - 4$ & \multicolumn{1}{p{0.7cm}|}{ 0.035} & 0.001 & 0.001 & 0.001 & 0.038 \\
  $t - 5$ & \multicolumn{1}{p{0.7cm}|}{\cellcolor{gray!25} 0.878} & 0.001 & 0.001 & 0.001 & 0.881 \\
\hdashline
\cline{2-2}
  $t - 6$ & 0.036 & 0.001 & 0.001 & 0.001 & 0.039 \\
  $t - 7$ & 0.002 & 0.001 & 0.001 & 0.001 & 0.006 \\
  $t - 8$ & 0.001 & 0.001 & 0.001 & 0.001 & 0.004 \\
  $t - 9$ & 0.001 & 0.001 & 0.001 & 0.001 & 0.004 \\
  $t - 10$ & 0.001 & 0.001 & 0.001 & 0.001 & 0.004 \\
  $t - 11$ & 0.001 & 0.001 & 0.001 & 0.001 & 0.004 \\
  $t - 12$ & 0.001 & 0.001 & 0.001 & 0.001 & 0.004 \\
  \hline
  Sum & 0.961 & 0.015 & 0.013 & 0.011 & 1.000 \\
  \end{tabular}
\end{minipage}
\end{minipage}
}
\caption{The matrix $\bo L$ with row sums and column sums: TSAVE (left panel) and TSIR (right panel)\label{table::ARMAModelBlow}
   }
\end{table}

\subsection{Models and prediction}\label{sec::model}

The results presented here are based on the following ARMA models, where the four $\bo z$ component series are as follows.
\begin{eqnarray*}
\mbox{Components 1 and 2:} && \mbox{ $AR(1)$ with $\phi=0.2$ (or $0.8$).}\\
\mbox{Component 3:} && \mbox{ $ARMA(1,1)$ with $ \phi = 0.3$ and $\theta= 0.4$.}\\
\mbox{Component 4:} && \mbox{ $MA(1)$ with $\theta= -0.4$, respectively.}
\end{eqnarray*}

Note that for the first two components two different $\phi$ values are used to compare how the level of the autocorrelation affects the results.
The response series $y$ depends then on the first two components $z_1$ and $z_2$, in the following different ways:
\begin{align*}
\mbox{Model {\it A}:}& \ y_t = 2z_{1,t-1} + 3z_{2,t-1} + \epsilon_t \\
\mbox{Model {\it B}:}& \ y_t = z_{1,t-1}^2 + 3z_{2,t-5} + \epsilon_t \\
\mbox{Model {\it C}:}& \ y_t = (2 z_{1,t-1} + 3z_{2,t-1})^2 + \epsilon_t \\
\mbox{Model {\it D}:}& \ y_t = z_{1,t-1}^2 + 3z_{2,t-5}^2 + \epsilon_t \\
\mbox{Model {\it E}:}& \ y_t = 2z_{1,t-1}^3 + 3z_{2,t-5}^2 + \epsilon_t
\end{align*}
All the models have iid $N(0,1)$-distributed innovations $\epsilon_t$. As before $\bo\Omega=\bo I_4$ is used as the mixing matrix. 

Note that we have also performed additional simulations which evaluated what happens if there are almost non-stationary components, stochastic volatility components or components with heavy-tailed innovations. The exact settings are detailed in the appendix together with the corresponding results. While there are maybe minor differences, we believe that the guidelines derived on the settings specified above suffice in practice.

For prediction we use the prediction model \eqref{eq::predmodel}. As an approximation of the function $f$ we use both simple linear regression and also regression with quadratic $B$-splines (for model {\it E} cubic $B$-splines, as it includes a factor of the form $z^3$).
The size of the testing set is 100, i.e. we predict the last 100 values of the data. To predict the value for $T-100 + i$, $i = 1, 2, \ldots, 100$, we use the observations $i, \ldots, T-100 + (i-1)$ as a training set (`rolling window approach').

We estimate the accuracy of the prediction by calculating the root mean square error (RMSE) based on the one-step-ahead prediction errors $\hat{\epsilon}_t$ of the last 100 observations (testing set).
The lags used to create the matrix $\bo L$ are $1, \ldots, 12$ and the number of repetitions is 500.

\subsection{On the number of slices for TSIR and TSAVE}\label{sec::H}

In \cite{MatilainenCrouxNordhausenOja2017} simulations for TSIR are conducted only with value $H = 10$, which is a common value used with SIR. Here we aim to go a bit deeper and provide some guidelines for choosing the value $H$ for TSAVE as well as for TSIR.

To find the optimal number of slices $H$, we conduct an experiment with different strategies and with two different threshold values $P = $ 0.5 and 0.8 to find the important lags and the number of sources.
We use time series lengths $T = 500, 1000, 2000$ and 3000.

First we predict the values using all the strategies with linear and spline predictions. With $H = 2, 5, 10, 20$ and $40$ and models {\it A} -- {\it D}, we calculate the RMSE values compared to value $H = 10$, which is commonly used in e.g. SIR and has been used in TSIR in \cite{MatilainenCrouxNordhausenOja2017}.
The results are for the components with low and high autocorrelation as well as for time series lengths $T = 500$, 1000, 2000 and 3000.

The choices $H = 20$ and $H = 40$ do not work that well in any of the settings in TSAVE. We conclude from this that there are then simply too few observations per slice.
Thus we show here results only concerning values $H = 2, 5$ and 10. The choices $H = 2$ and $H = 5$ are compared to $H = 10$. If the choice is better than $H = 10$, then the relative RMSE values will generally be lower than one.

In model {\it A} the linear predictions are the most efficient and spline predictions may add a little bit of additional noise.
In models {\it B} -- {\it D}, however, the linear predictions are not very good at determining the best value of $H$. Only with $T = 500$ the choice $H = 2$ seems to be a bit better than others, while in longer time series any possible difference is barely visible. This is expected, since the relationship between the predictors and the response is not linear.
Thus the spline predictions is preferred for determining the optimal value for $H$. 

In both low and high dependence settings the threshold value $P = 0.5$ seems to be working well for models {\it A} and {\it C}, where only one source is expected to be found, and choosing $P = 0.8$ has only a very small effect on results. 
For models {\it B} and {\it D} it seems that $P = 0.5$ might be enough in short time series ($T = 500$), but in longer time series, especially when we have sources with high autocorrelation, using $P = 0.8$ is crucial for the second source to be found.
Thus $P = 0.8$ is seems to be a safe choice in general.

\begin{figure}[!htb]
\includegraphics[scale=0.5]{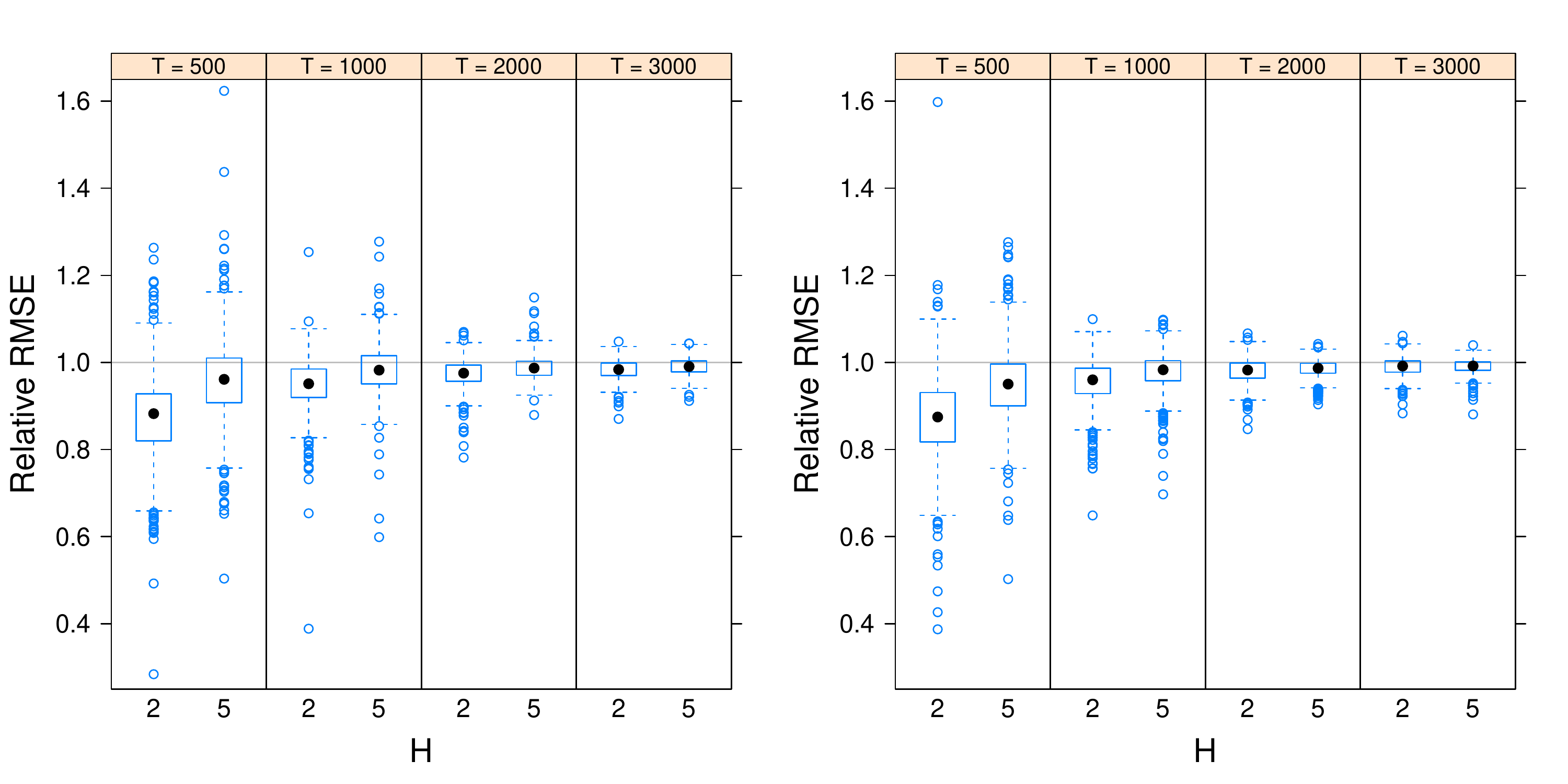}
\caption{TSAVE. Model {\it A} with the biggest values strategy. Relative RMSE values compared to $H = 10$ with $\phi = 0.2$ (left panel) and $\phi = 0.8$ (right panel).}\label{fig::AH_RMSE3000}
\end{figure}

\begin{figure}[!htb]
\includegraphics[scale=0.5]{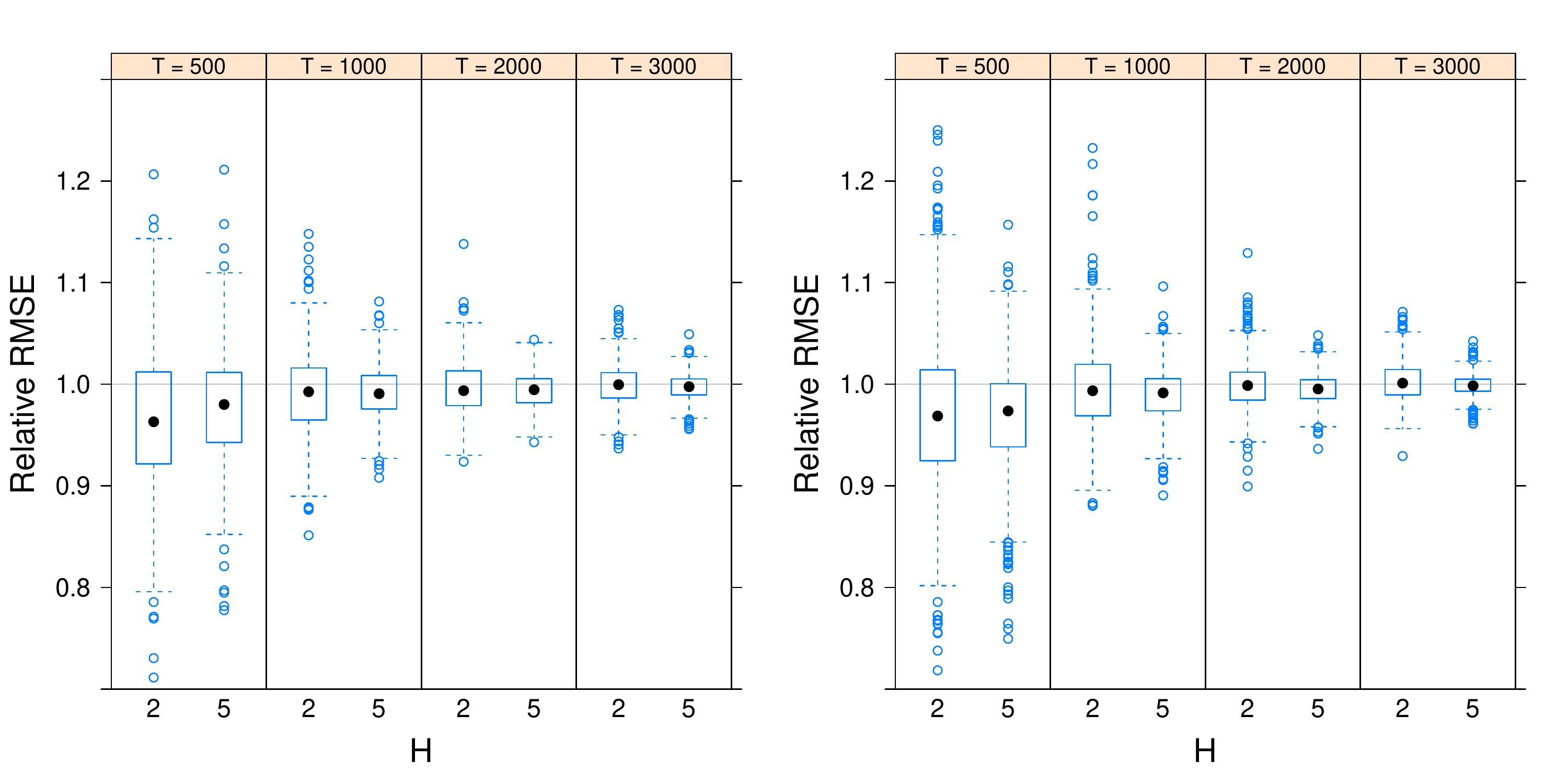}
\caption{TSAVE. Model {\it B} with the biggest values strategy. Relative RMSE values compared to $H = 10$ with $\phi = 0.2$ (left panel) and $\phi = 0.8$ (right panel).}\label{fig::BH_RMSE3000}
\end{figure}

\begin{figure}[!htb]
\includegraphics[scale=0.5]{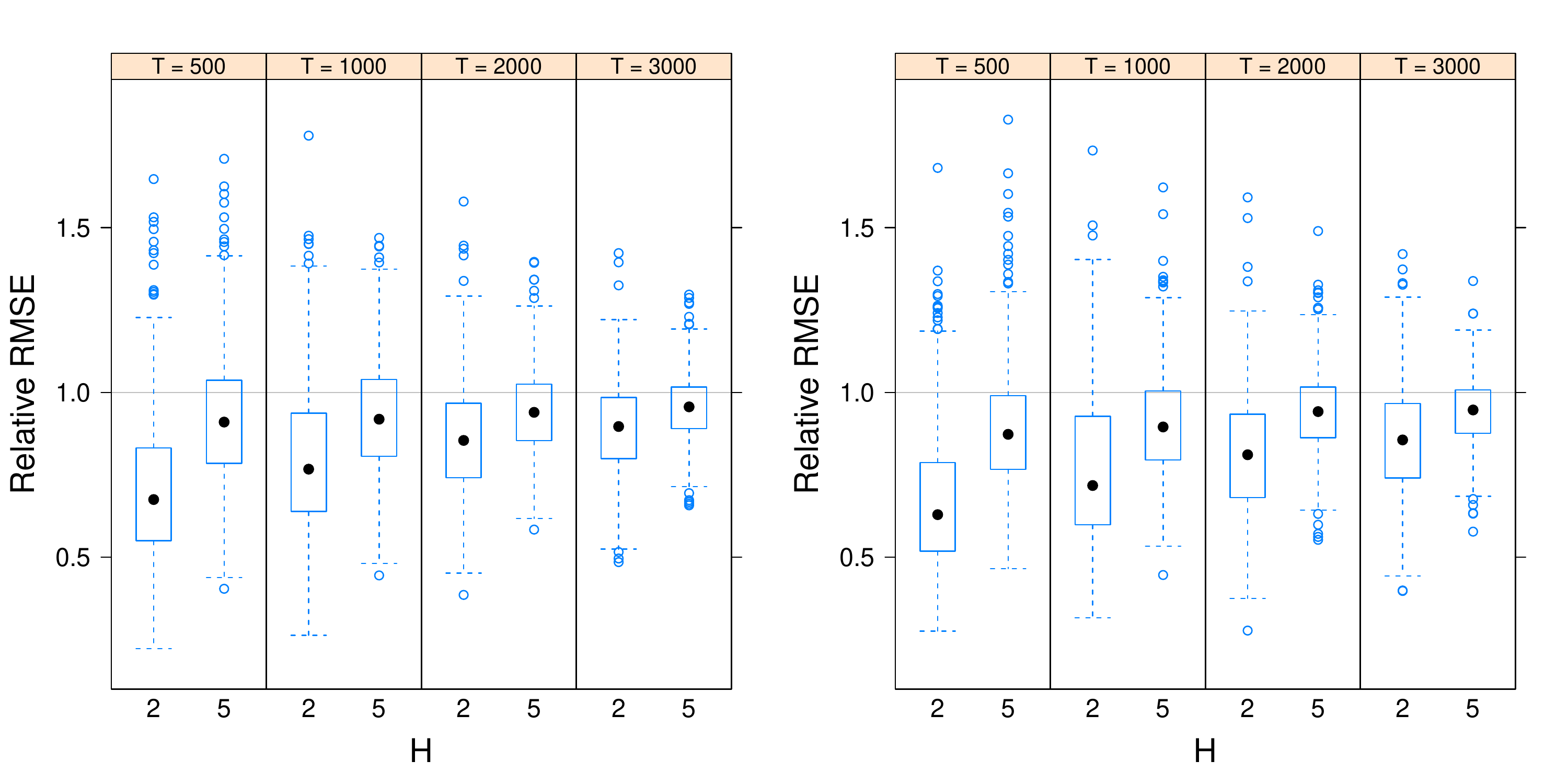}
\caption{TSAVE. Model {\it C} with the biggest values strategy. Relative RMSE values compared to $H = 10$ with $\phi = 0.2$ (left panel) and $\phi = 0.8$ (right panel).}\label{fig::CH_RMSE3000}
\end{figure}

\begin{figure}[!htb]
\includegraphics[scale=0.5]{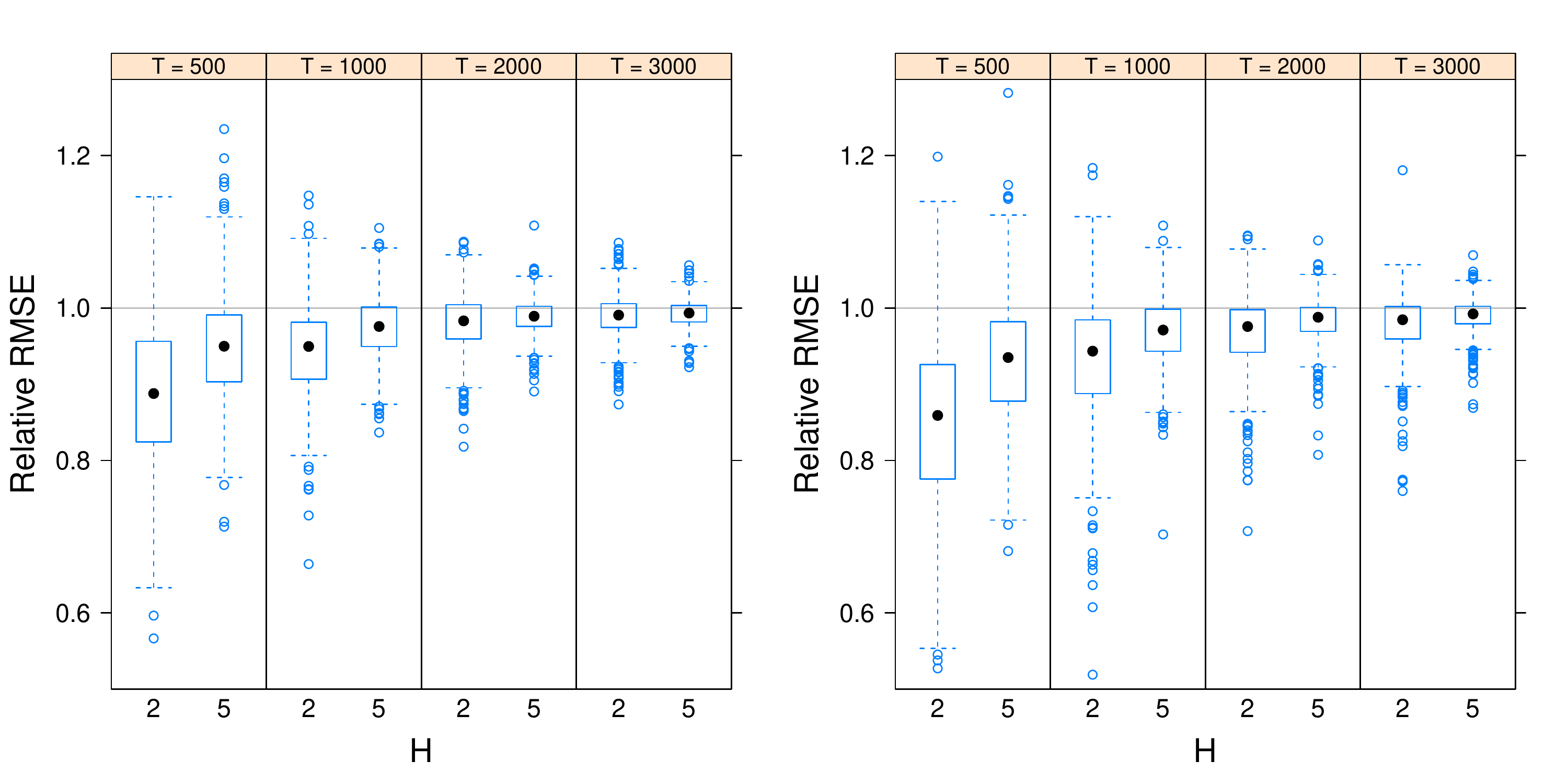}
\caption{TSAVE. Model {\it D} with the biggest values strategy. Relative RMSE values compared to $H = 10$ with $\phi = 0.2$ (left panel) and $\phi = 0.8$ (right panel).}\label{fig::DH_RMSE3000}
\end{figure}

When comparing the different strategies to select the number of sources and lags, the biggest values strategy seems to produce almost always the best results, and in the remaining few cases it is very close to the best. 
 Figures \ref{fig::AH_RMSE3000}--\ref{fig::DH_RMSE3000} show then based on that strategy the relative RMSE for models {\it A} -- {\it D} and the different sample sizes. And these figures clearly indicate that using only 2 or 5 slices for TSAVE is clearly better than 10 slices. Only with increasing sample size the differences vanish which means that then all slices contain enough observations.
The same can also be observed using other strategies for the selection (not shown here) although then in rare cases can be 5 slices better than 2 slices.

\begin{figure}[!htb]
\includegraphics[scale=0.5]{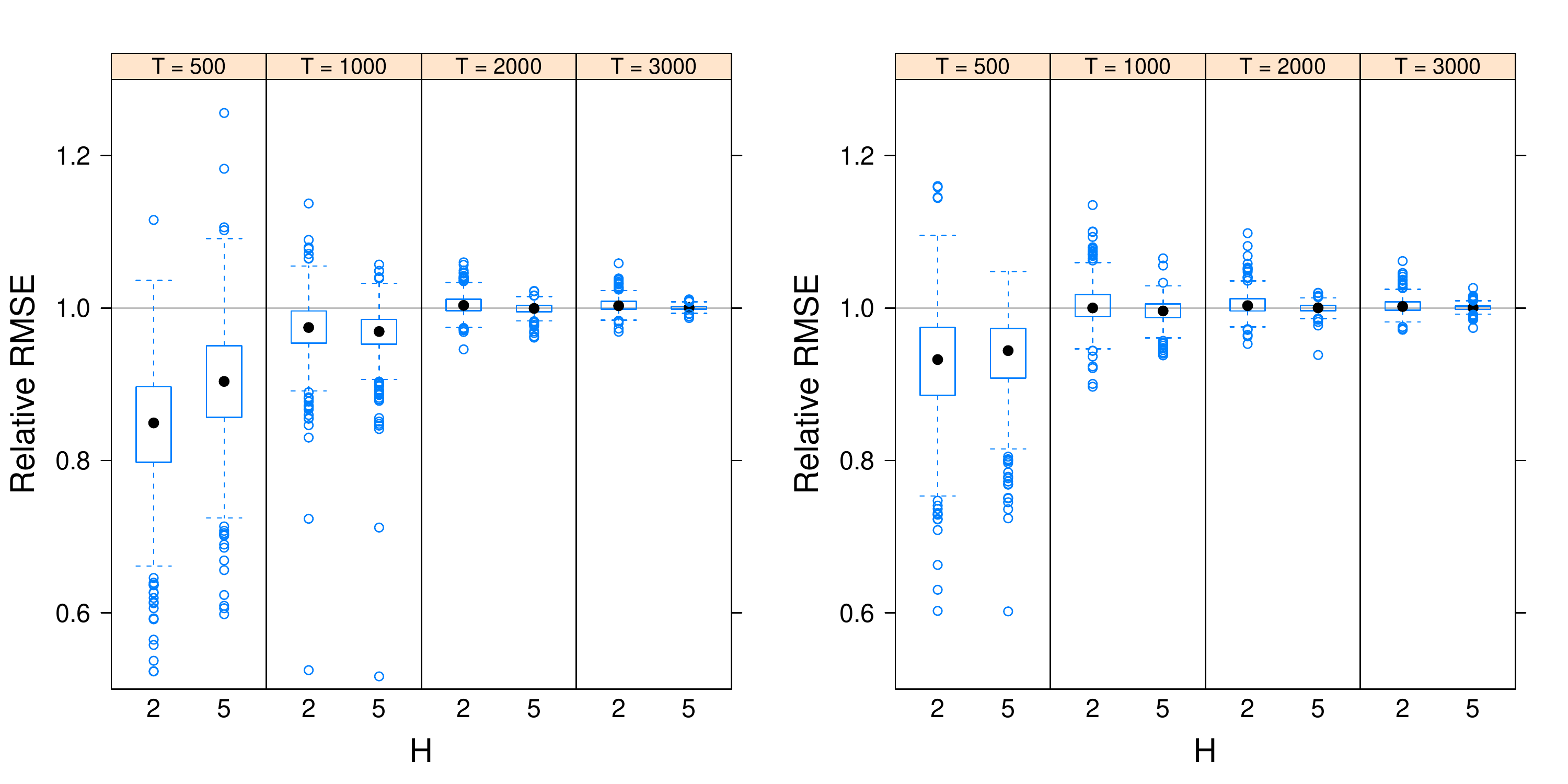}
\caption{TSIR. Model {\it A} with the biggest values strategy. Relative RMSE values compared to $H = 10$ with low (left panel) and high (right panel) depencency of the sources.}\label{fig::AH_RMSE3000_SIR}
\end{figure}

For TSIR the choice $H=10$ seems to be the safest when linear predictions are used.
However, when using spline predictions, $H = 2$ and $H = 5$ may be better choices with shortest time series, i.e. with $T = 500$.
Figure \ref{fig::AH_RMSE3000_SIR} includes results from spline predictions using the biggest values strategy.
Note that with the models {\it B} -- {\it D}, TSIR does not work well. Thus the evaluation of the value of $H$ for TSIR is not considered in those models.

We could also look for the best $H$ with the $\bo L$ matrix. For the models {\it A} and {\it C} we can check, if only one source with at least lag 1 is found, as only one is expected.
For the models {\it B} and {\it D} we can first check that if one source with lag 1 and one source with lag 5 are found. If that is true, then we can check that if only the two sources are found.

\begin{figure}[!htb]
\includegraphics[scale=0.7]{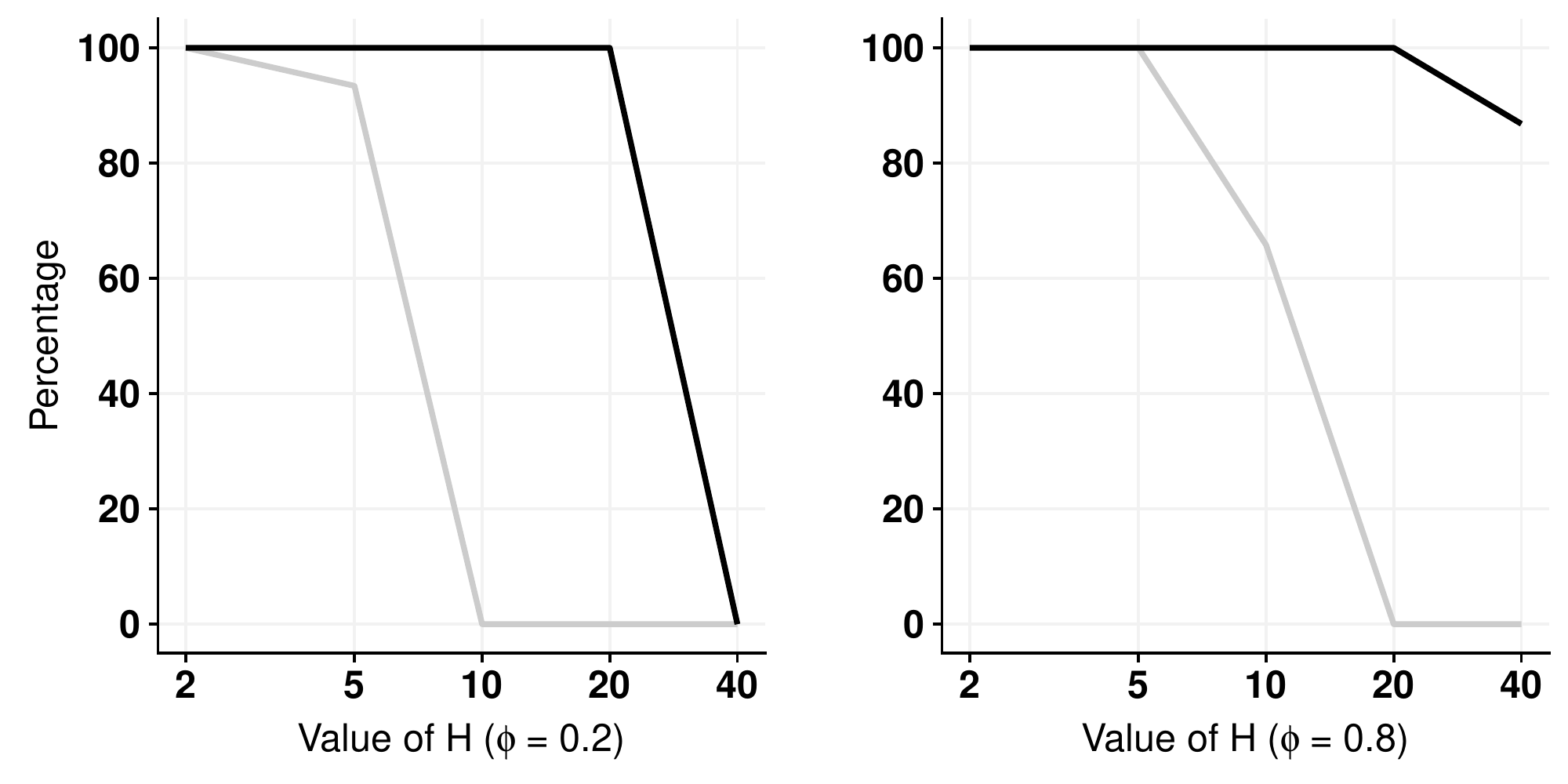}
\caption{TSAVE. Model {\it C} with $T = 3000$ using the biggest values strategy. Percentage of cases that finds the correct lags and correct amount of sources. Black line: $P = 0.5$, gray line: $P = 0.8$.}\label{fig::CP3000}
\end{figure}

With the biggest values strategy with $T = 3000$ for model {\it A}, generally $H = 2, 5$ and 10 are good choices when $P = 0.8$, and $H = 2$ is the best when $P = 0.5$.
In all the other models also $H = 2$ seems to be the most efficient choice when $P = 0.8$.
For model {\it C} the choices $H = 2$ and 5 are generally safe, while with $P = 0.8$ also $H = 10$ and 20 seem to be good enough. As an example, Figure \ref{fig::CP3000} has the results for model {\it C} with $T = 3000$.
In model {\it D} the threshold value $P = 0.5$ seems to be generally too low for efficiently finding the right amount of sources, while with $P = 0.8$ choices $H = 2$ and $H = 5$ seem to produce the expected results.
For TSIR any value $H \leq 10$ seems to be safe for model {\it A}.
For a short time series ($T = 500$), $H = 2$ seems mostly the safest choice for TSAVE and $H = 2, 5$ and 10 for TSIR.

To conclude this section, we can say that for TSIR $H=10$ is generally a reliable choice, however, with short time series a lower $H$ might be beneficial, depending on the prediction method used. For TSAVE the number of observations per slice is more important and depends also on the data generation process. Based on our simulations we recommend to have at least 100 observations per slice and for the time series lengths considered here our preferred choice is $H=2$, but also $H=5$ seems good. For a shorter time series $H=2$ might be the only choice, but the longer the time series the smaller differences there are between the values of H and then also a larger $H$ would be reliable.

Furthermore, the simulations suggest that the value of $P$ has a big influence on the number of selected components and $P=0.5$ is more restrictive than $P=0.8$, which is however also very intuitive. The biggest values strategy to choose the amount of sources and the lags corresponding to them is also recommended.

\subsection{On the TSSH method and the choice of coefficient $a$}\label{sec::a}

In model {\it A}, the relationship between the response and the predictors  is linear. Already \cite{MatilainenCrouxNordhausenOja2017} show that TSIR works in such models efficiently. On the other hand, the models {\it B}-{\it E} have symmetric parts. As seen in Section \ref{sec::visual}, TSIR in unable to find the relationship in such case (see Figure \ref{fig::B}). This is also seen later in the simulation results of Section \ref{sec::comparison}. Also from Figure \ref{fig::B} it can be seen that TSAVE still works with the linear relationships, but not as efficiently as TSIR.

This preference for different structures of the different methods was the main motivation for the introduction of the hybrid in Section~\ref{sec::H}. Now we consider the optimal value of $a$ for model {\it E}. This model is similar to the model 4 in \cite{ZhuOhtakiLi2007}, for which the hybrid of iid SIR and SAVE shows clearly better performance than SIR or SAVE separately.
Therefore we could expect here that the TSIR part uncovers the asymmetric part of the dependence $2z_{1,t-1}^3$ efficiently and TSAVE the symmetric part $3z_{2,t-5}^2$, and  hopefully both together work even better. The question how much weight should be given to which method, i.e. the proper value for $a$. For the iid hybrid method \cite{ZhuOhtakiLi2007} recommend as general rule of thumb to use $a = 0.5$.

\begin{figure}[!htb]
\includegraphics[scale=0.5]{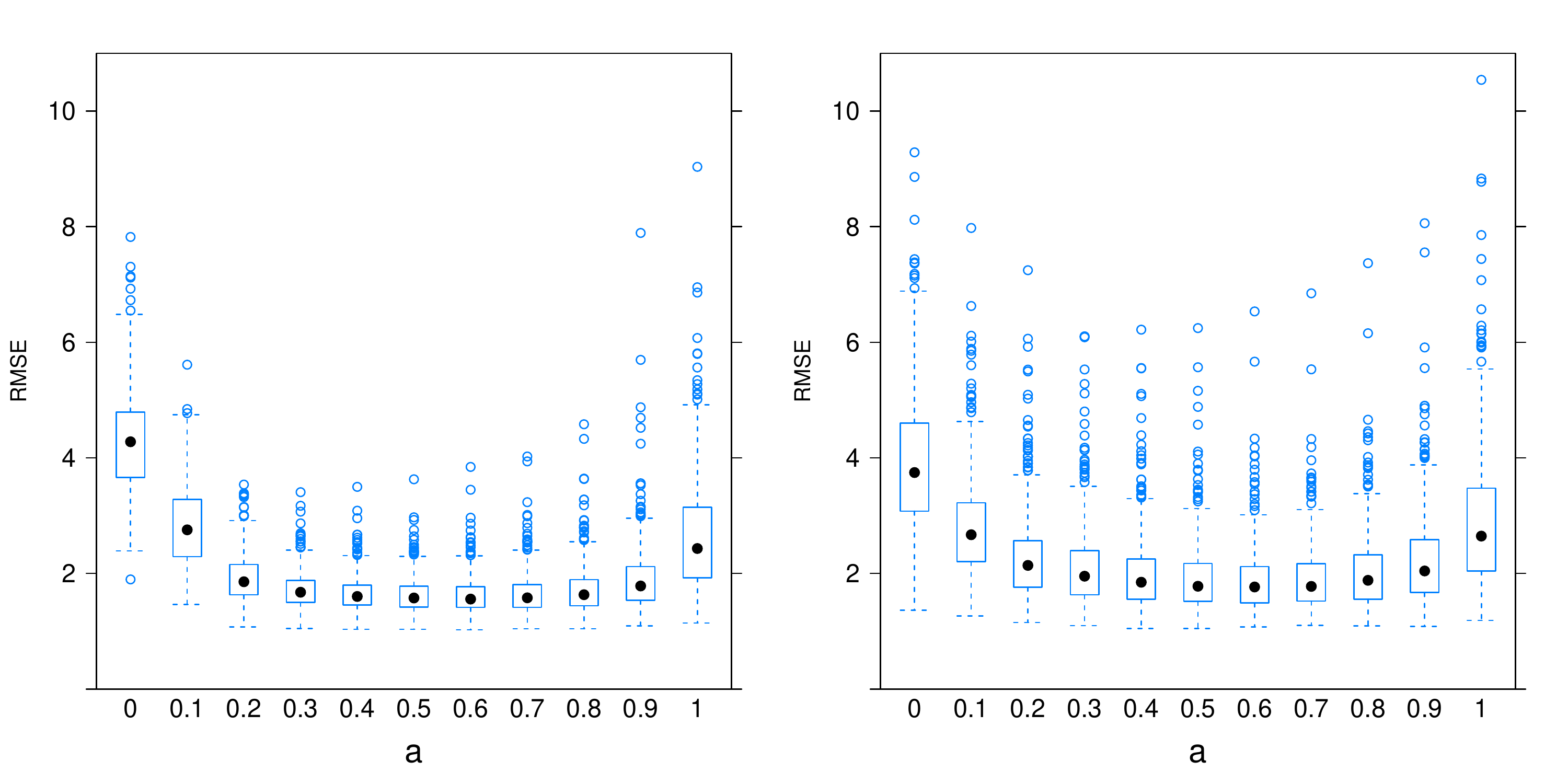}
\caption{TSSH. Model {\it E} with the biggest values strategy:
$T = 500$ and $H = 2$ for TSAVE part and $H = 10$ for TSIR part. RMSE values with $\phi = 0.2$ (left panel) and $\phi = 0.8$ (right panel).}\label{fig::HP500_RMSE2}
\end{figure}

\begin{figure}[!htb]
\includegraphics[scale=0.5]{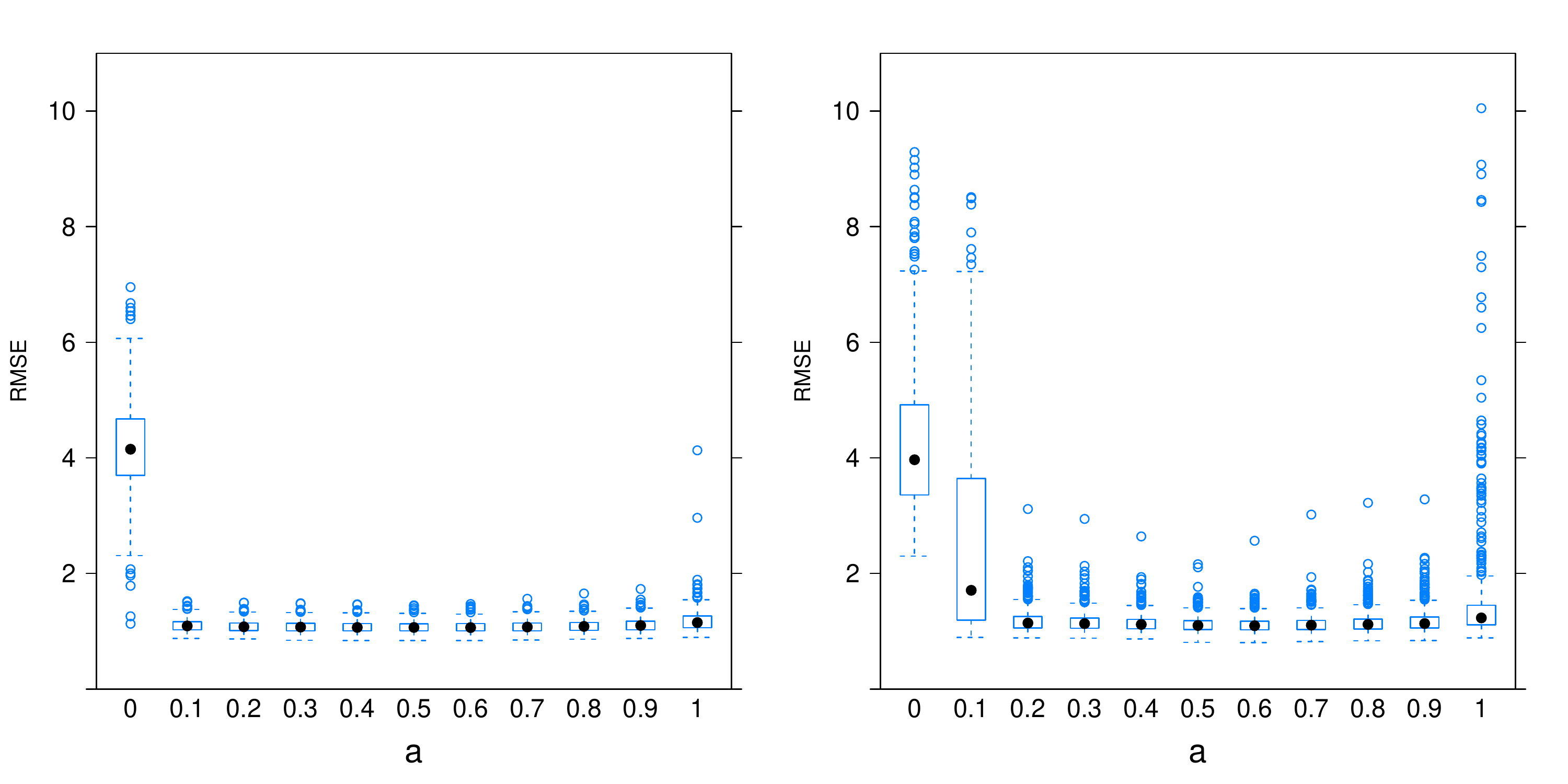}
\caption{TSSH. Model {\it E} with the biggest values strategy: $T = 3000$ and $H = 2$ for TSAVE part and $H = 10$ for TSIR part. RMSE values with $\phi = 0.2$ (left panel) and $\phi = 0.8$ (right panel).}\label{fig::HP3000_RMSE2}
\end{figure}

Following our general guidelines from the previous section, in the following presentation the results are based on using $H=10$ slices for TSIR part and $H=2$ slices for TSAVE part. To select the components, $P$ is set to 0.8 and cubic $B$-splines are used for the prediction. The RMSE values are then computed for the values $a = 0, 0.1, 0.2, \ldots, 1$, where $a=0$ refers to the pure TSIR method and $a=1$ to the pure TSAVE method.
We show here the RMSE for the time series lengths $T = 500$ and 3000 in Figure~\ref{fig::HP500_RMSE2} and Figure~\ref{fig::HP3000_RMSE2}.
 It seems that there are not so big differences as long as not all or almost all of the weight is given to TSIR or all the weight given to TSAVE. The central values of $a$ seem to be a little bit better.

\begin{figure}[!htb]
\includegraphics[scale=0.7]{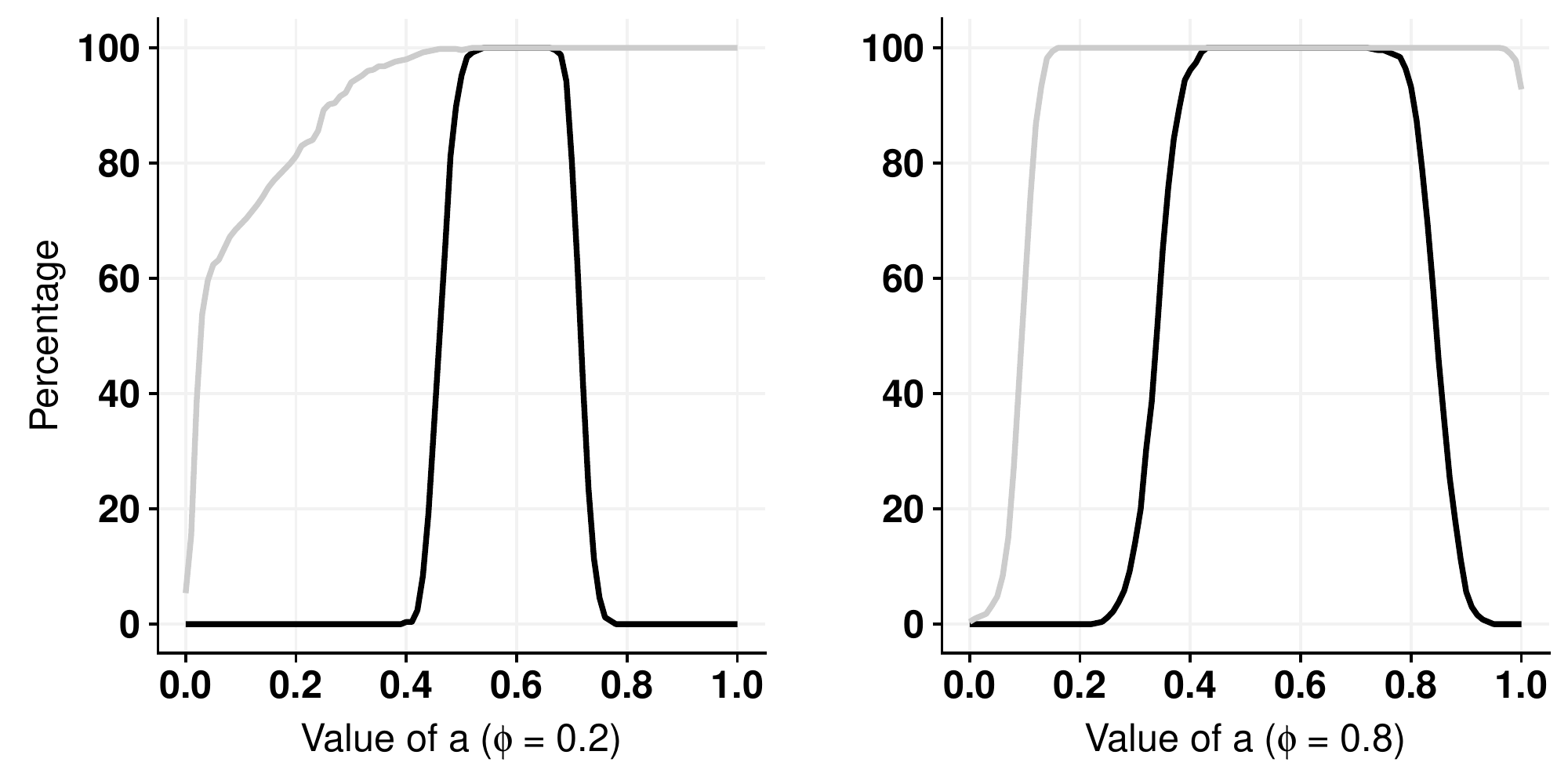}
\caption{TSSH. Model {\it E} with the biggest values strategy: $T = 3000$ and $H = 2$ for TSAVE part and $H = 10$ for TSIR part. Percentage of cases that finds the correct lags and correct amount of sources. Black line: $P = 0.5$, gray line: $P = 0.8$.}\label{fig::HP3000}
\end{figure}

To investigate this  a bit further, we also compare the choices of $P=0.5$ and $P=0.8$ using the $\bo L$ matrix. Figure~\ref{fig::HP3000} gives then for $T=3000$
the percentages of right number of sources and appropriate lags chosen, based on the biggest value strategy, for $a = 0, 0.01, 0.02, \ldots, 1$

From Figure~\ref{fig::HP3000} we can conclude that values for $a$ around 0.5 and 0.6 are reasonable choices. Considering results using other selection strategies not shown here we can in general recommend the value $a=0.5$, which coincides with the recommendation of the regular SIR and SAVE hybrid. 

The main feature we observed for the hybrid is that with values closer to $a=0.5$, the value of $P$ is less crucial and one often comes to the same conclusion. Quite different from what we have seen when using only TSIR or only TSAVE. 

\subsection{Comparison to other approaches}\label{sec::comparison}

To compare different methods, we simulate with time series length $T = 3000$ using $H=2$  for TSAVE and $H = 10$ for TSIR, $P = 0.8$ and the biggest values strategy, as recommended in Section \ref{sec::H}.
For models {\it A} -- {\it D} the relative RMSE values are compared to Oracle estimator, where the functional form of the relationship between the response and the predictors is known, but the coefficients are estimated.

\begin{figure}[!htb]
\includegraphics[scale=0.5]{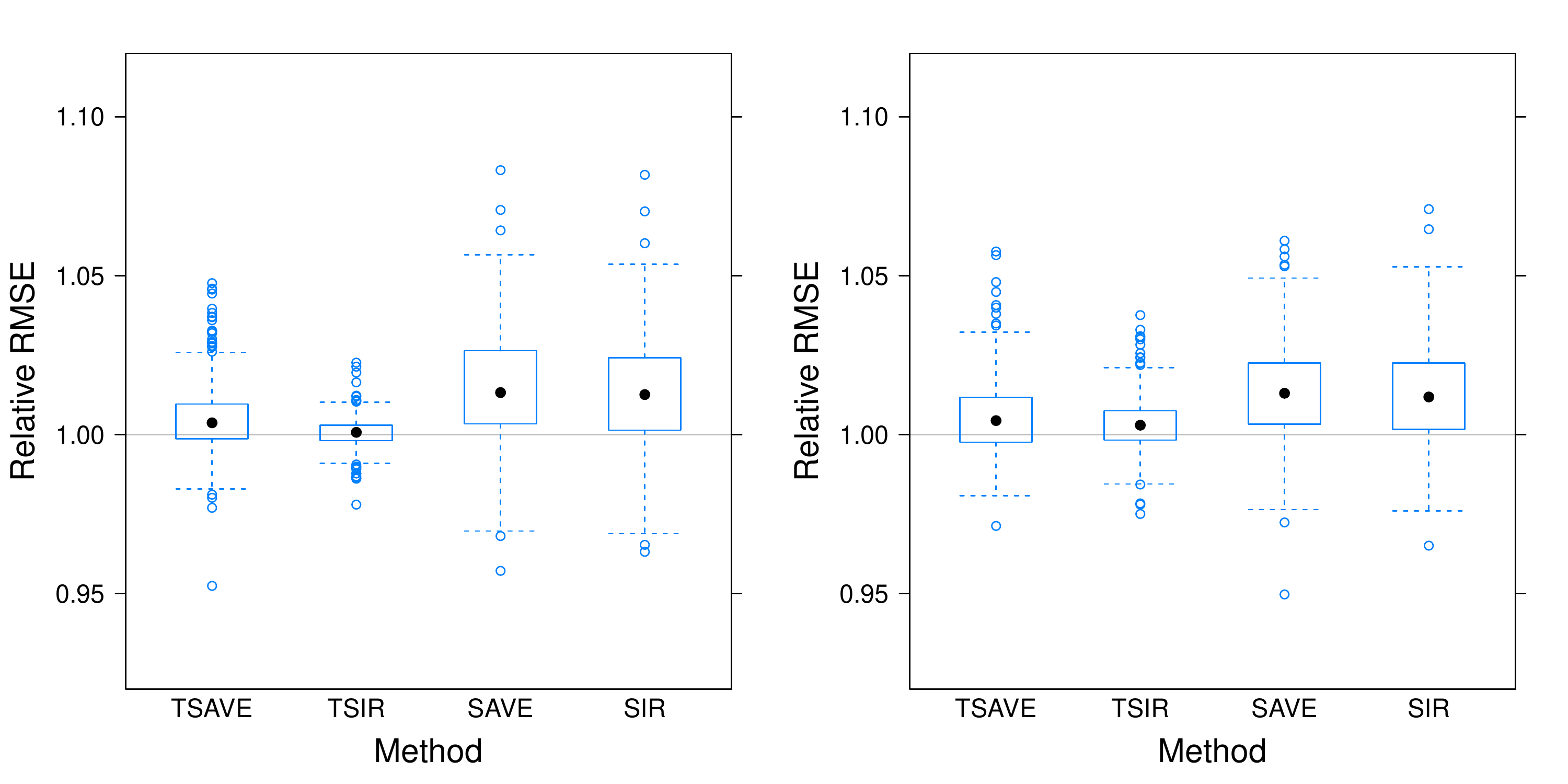}
\caption{Model {\it A} with biggest values strategy. Relative RMSE values compared to Oracle estimator with $\phi = 0.2$ (left panel) and $\phi = 0.8$ (right panel).}\label{fig::A_RMSE3000}
\end{figure}

\begin{figure}[!htb]
\includegraphics[scale=0.5]{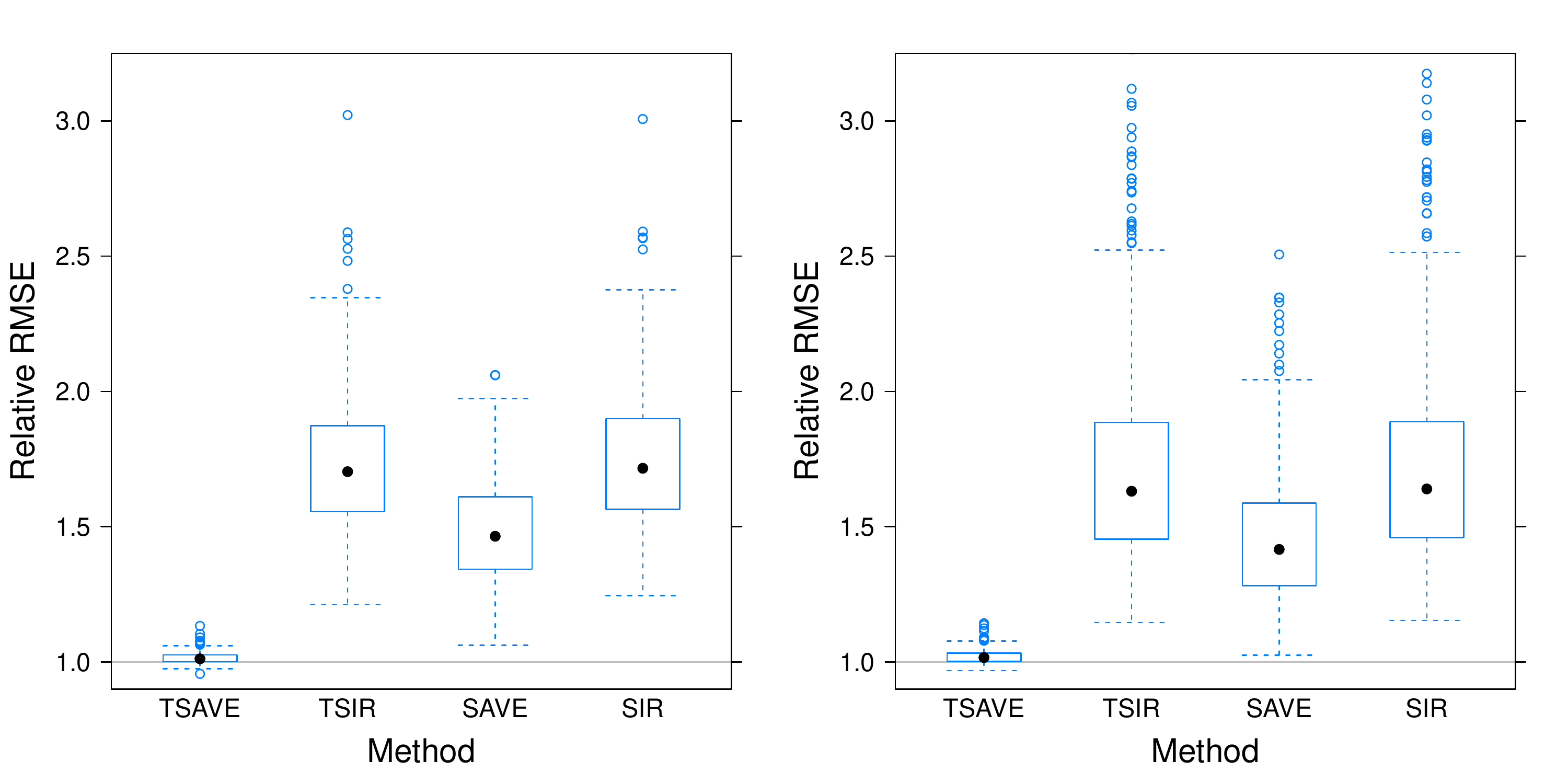}
\caption{Model {\it B} with biggest values strategy. Relative RMSE values compared to Oracle estimator with $\phi = 0.2$ (left panel) and $\phi = 0.8$ (right panel).}\label{fig::B_RMSE3000}
\end{figure}

\begin{figure}[!htb]
\includegraphics[scale=0.5]{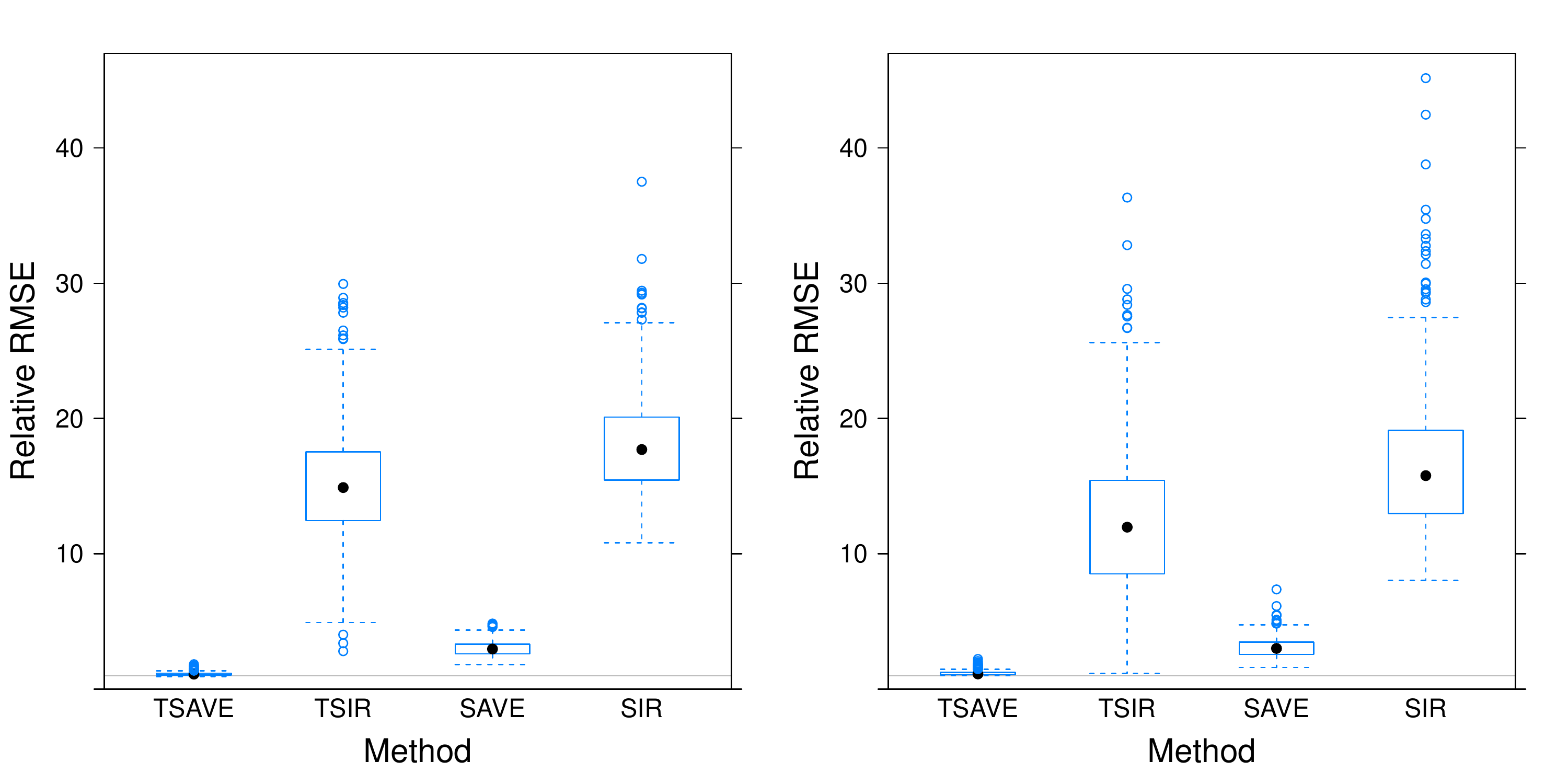}
\caption{Model {\it C} with biggest values strategy. Relative RMSE values compared to Oracle estimator with $\phi = 0.2$ (left panel) and $\phi = 0.8$ (right panel).}\label{fig::C_RMSE3000}
\end{figure}

\begin{figure}[!htb]
\includegraphics[scale=0.5]{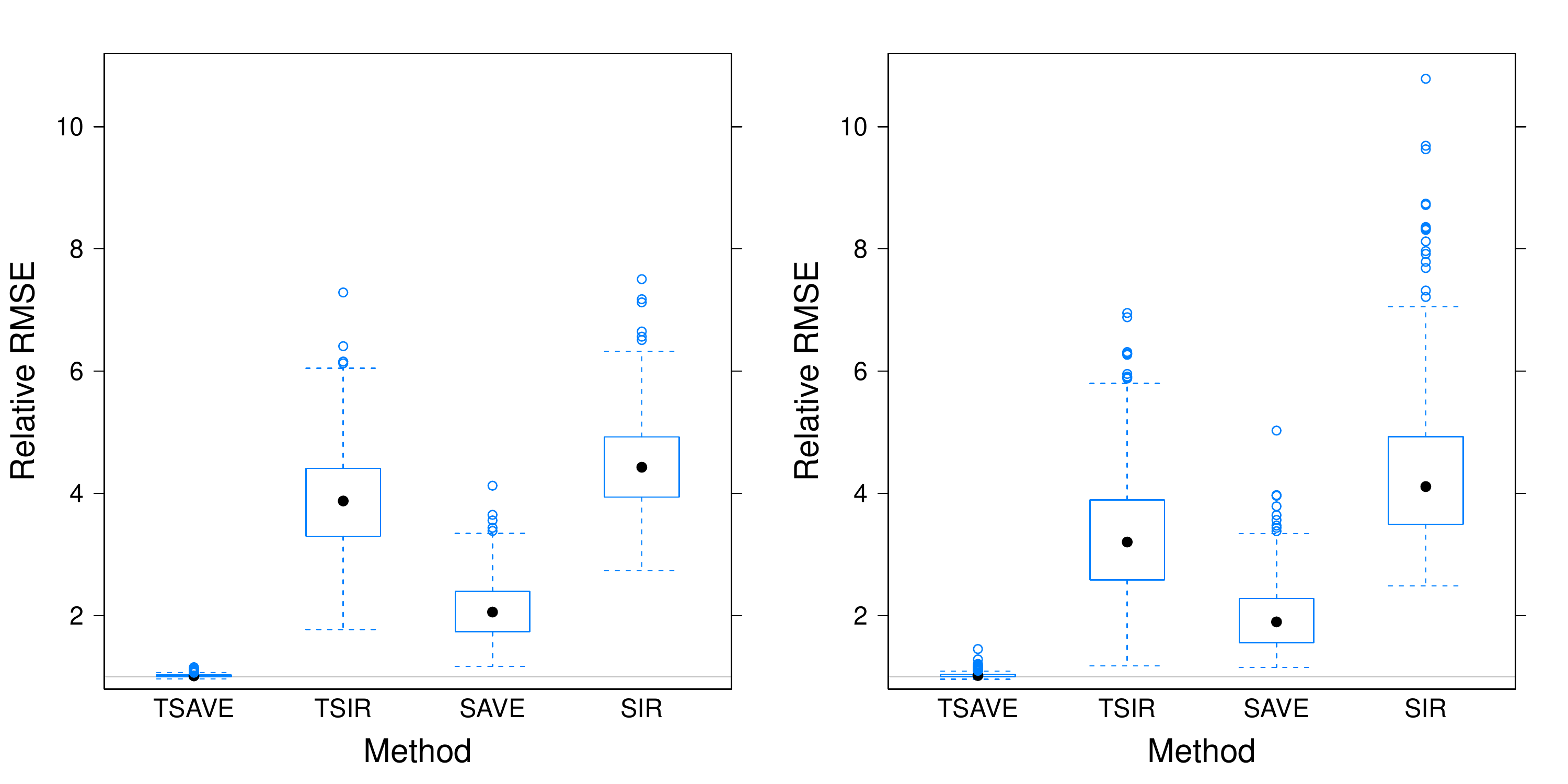}
\caption{Model {\it D} with biggest values strategy. Relative RMSE values compared to Oracle estimator with $\phi = 0.2$ (left panel) and $\phi = 0.8$ (right panel).}\label{fig::D_RMSE3000}
\end{figure}

Figures \ref{fig::A_RMSE3000} -- \ref{fig::D_RMSE3000} have the relative RMSE values based on different methods compared to the Oracle estimator.
The methods here are TSAVE and TSIR as well as the original SAVE and SIR (Becker \& Fried SIR \cite{BeckerFried2003}) with the lagged values of $\bo x$ as predictors, i.e. with $\bo x_t^* = (\bo x_{t-1}, \ldots, \bo x_{t-s})$. Here we have used $s = 12$.

For original SIR we have used $H = 10$ \cite{Li1991} and for original SAVE $H = 5$, as $H = 2$ may be too low for SAVE (see for example \cite{LiZhu2007}). To choose the number of sources, we use the ordered empirical eigenvalues $\lambda_i$, $i = 1, \ldots, s\cdot p$, of the supervised matrices $\cov[\E(\bo x^{*,st}|y^{sl})]$ in the original SIR and $\E[(\bo I_p - \cov(\bo x^{*,st}|y^{sl}))^2]$ in the original SAVE.
The chosen number of sources is the minimal $\hat{k}$ for which $\sum_{i=1}^{\hat{k}} \lambda_i / \sum_{i=1}^{s\cdot p} \lambda_i \ge P = 0.8$.

From Figure \ref{fig::A_RMSE3000} we can see that TSIR and TSAVE both work very well. From Figures \ref{fig::B_RMSE3000} -- \ref{fig::D_RMSE3000} we see that TSAVE clearly works the best in the models {\it B}, {\it C} and {\it D}, while also the original SAVE with lagged variables as predictors works better than TSIR and the original SIR.
The iid versions using the lagged variables as predictors do not work that well except in the linear case (Model {\it A}).
If we used $H = 5$ instead of $H = 2$, the results would be very similar.

To evaluate the effect of the dimension we included as a final setting also a setup, where we have $p = 10$ components and the `true' number of series that the response depends on is $k = 3$. The simulations were conducted using several different time series lengths ($T = 500, 1000, 2000, 3000$ and $5000$), but here we show only the results based on $T = 3000$ and other results can be found in the appendix.

The innovations are standard normal unless otherwise stated. Components included here are

\begin{enumerate}
\item[$z_1$:] AR(1) with $\phi = -0.2$
\item[$z_2$:] AR(1) with $\phi = 0.8$ with heavy-tailed $t_4$ innovations
\item[$z_3$:] GARCH(1,1) with $\alpha = 0.05$ and $\beta = 0.93$
\item[$z_4$:] AR(1) with $\phi = 0.6$ with light-tailed $U(-1,1)$ innovations
\item[$z_5$:] AR(1) with $\phi = 0.98$
\item[$z_6$:] ARCH(2) with $\alpha_1 = 0.3$ and $\alpha_2$ = 0.4
\item[$z_7$:] GARCH(1,1) with $\alpha = 0.1$ and $\beta = 0.8$
\item[$z_8$:] ARMA(1,1) with $\phi = 0.3$ and $\theta = -0.6$
\item[$z_9$:] iid $N(0,1)$
\item[$z_{10}$:] iid $t_4$
\end{enumerate}

\begin{figure}[!htb]
\centering
\includegraphics[scale=0.48]{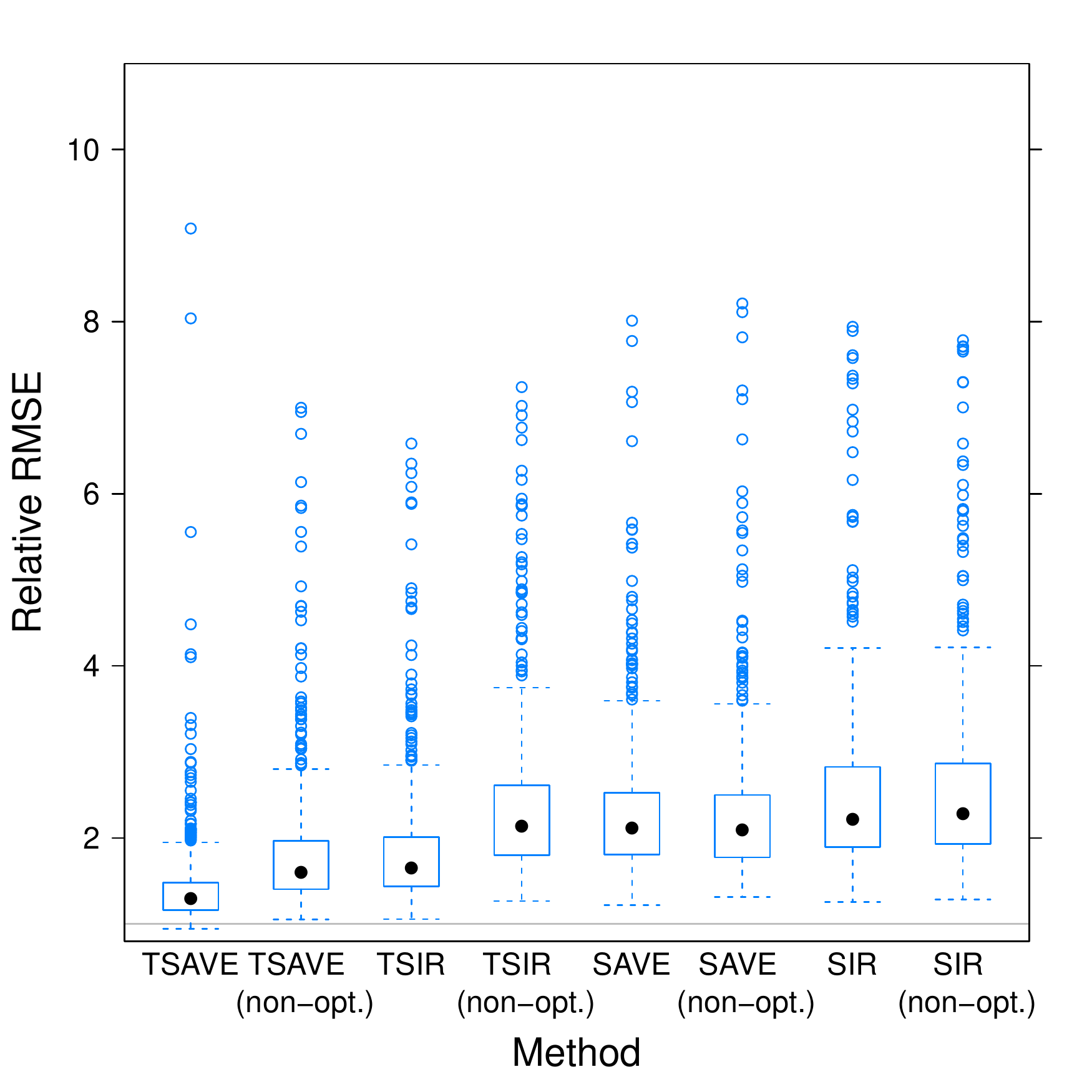}
\caption{Big setting with biggest values strategy. Relative RMSE values compared to Oracle estimator.}\label{fig::BIG_RMSE3000}
\end{figure}

All are standardized to meet the requirements for $\bo z$. The response is created as
$$
y_t = z_{1, t-1} + z_{2, t-2} + 0.5 z_{3, t - 4} + \epsilon_t,
$$
where $\epsilon_t \sim N(0,1)$. In order to examine what happens if $y_t$ is not well approximated in the respective spline class, we have used predictions based both on cubic (`optimal') and quadratic (`non-optimal') splines for all the methods used. The RMSE values with $T = 3000$ in Figure \ref{fig::BIG_RMSE3000} reveal that TSAVE with optimal prediction models produces clearly the best results. Also TSAVE with non-optimal prediction model gives very slightly better results than TSIR with optimal prediction model, and much better than TSIR with non-optimal predictions. All the time series versions behave better than the vectorized iid versions.

For $T = 2000$ and $T = 5000$ the above is also clearly true. For $T = 1000$ both choices are about equally good. With $T = 500$ it seems that the `non-optimal' prediction is better. It might be safer not to use `too accurate' prediction models when time series length are short and the number of dimensions large.
Also in shorter time series the vectorized SIR starts to produce better and better results compared to vectorized SAVE.
Figures are included in the appendix.

\section{Final comments}

SAVE and hybrids that include SIR and SAVE parts are well established for the iid case. After \cite{MatilainenCrouxNordhausenOja2017} introduced TSIR as a time series extension for SIR, we suggested in this paper TSAVE and TSSH as time series extensions of SAVE and the hybrid of SIR and SAVE, respectively. 
We demonstrated that these methods are superior for supervised dimension reduction in a time series context than applying their iid counterparts to the explaining variables and their lagged values.
We furthermore explored further the strategies for components and lag selection in the time series case suggested in \cite{MatilainenCrouxNordhausenOja2017} and could now give some general recommendations how to apply TSIR, TSAVE and TSSH in practice.
Not so surprisingly many of the recommendations of iid methods apply also in the time series context.
Also, while in \cite{MatilainenCrouxNordhausenOja2017} only $H = 10$ was used as the number of slices, here we conducted a simulation study to give some guidelines for choosing $H$ for TSIR and TSAVE.

The popularity of the iid versions of SIR, SAVE and hybrids of them also led to the introduction of modified versions of SAVE, like CSAVE \cite{LiZhu2007} and ESAVE \cite{ZhuOhtakiLi2007} and hybrids between these and SIR, or other versions of hybrids with SIR like $SIRII_a$ (see the rejoinder of \cite{Li1991}). Future work will be to investigate how these modified versions can be moved to a time series framework and if they are improvements compared to TSIR, TSAVE and TSSH.

\section*{Funding}
The work of M. Matilainen and H. Oja was supported by the Academy of Finland under Grant 268703; and the work of K. Nordhausen was supported by the CRoNoS COST Action IC1408.

\bibliographystyle{plainnat}

\clearpage

\section*{Appendix: Additional settings for `Examples and simulations'}

\subsection*{Settings}

The simulations from the main paper are continued here with some additional settings. These include having `almost non-stationary' components, stochastic volatility components as well as components with heavy-tailed innovations.

In these settings the components 1 and 2, mentioned in the paper, are
\begin{itemize}
\item $AR(1)$ processes with $\phi=0.97$ -- a setting where the series are close to the border of non-stationarity of the mean, i.e. there is `almost some kind of a trend',
\item GARCH(1,1) processes with $\alpha = 0.1$ and $\beta = 0.8$, and
\item components as in the main paper simulations, but innovations $\epsilon_t$ are $t_4$-distributed.
\end{itemize}

The components 3 and 4 are as in the main text, i.e. component 3 is $ARMA(1,1)$ with $ \phi = 0.3$ and $\theta= 0.4$ and component 4 is $MA(1)$ with $\theta= -0.4$, respectively. The response $y$ depends on the first two components $z_1$ and $z_2$,
\begin{align*}
\mbox{Model {\it A}:}& \ y_t = 2z_{1,t-1} + 3z_{2,t-1} + \epsilon_t \\
\mbox{Model {\it B}:}& \ y_t = z_{1,t-1}^2 + 3z_{2,t-5} + \epsilon_t \\
\mbox{Model {\it C}:}& \ y_t = (2 z_{1,t-1} + 3z_{2,t-1})^2 + \epsilon_t \\
\mbox{Model {\it D}:}& \ y_t = z_{1,t-1}^2 + 3z_{2,t-5}^2 + \epsilon_t \\
\mbox{Model {\it E}:}& \ y_t = 2z_{1,t-1}^3 + 3z_{2,t-5}^2 + \epsilon_t
\end{align*}
The models have iid $N(0,1)$-distributed innovations $\epsilon_t$ and $\bo\Omega=\bo I_4$ is used as the mixing matrix.

The last 100 values of the data are predicted using regression with quadratic $B$-splines for models {\it A -- \it D} and cubic $B$-splines for model {\it E}, as in the main text.
The accuracy is estimated with the average RMSE values of 500 repetitions. Twelve lags are used in the simulations.

\clearpage

\subsection*{The value of $H$ (number of slices)}

\begin{figure}[h]
\includegraphics[scale=0.48]{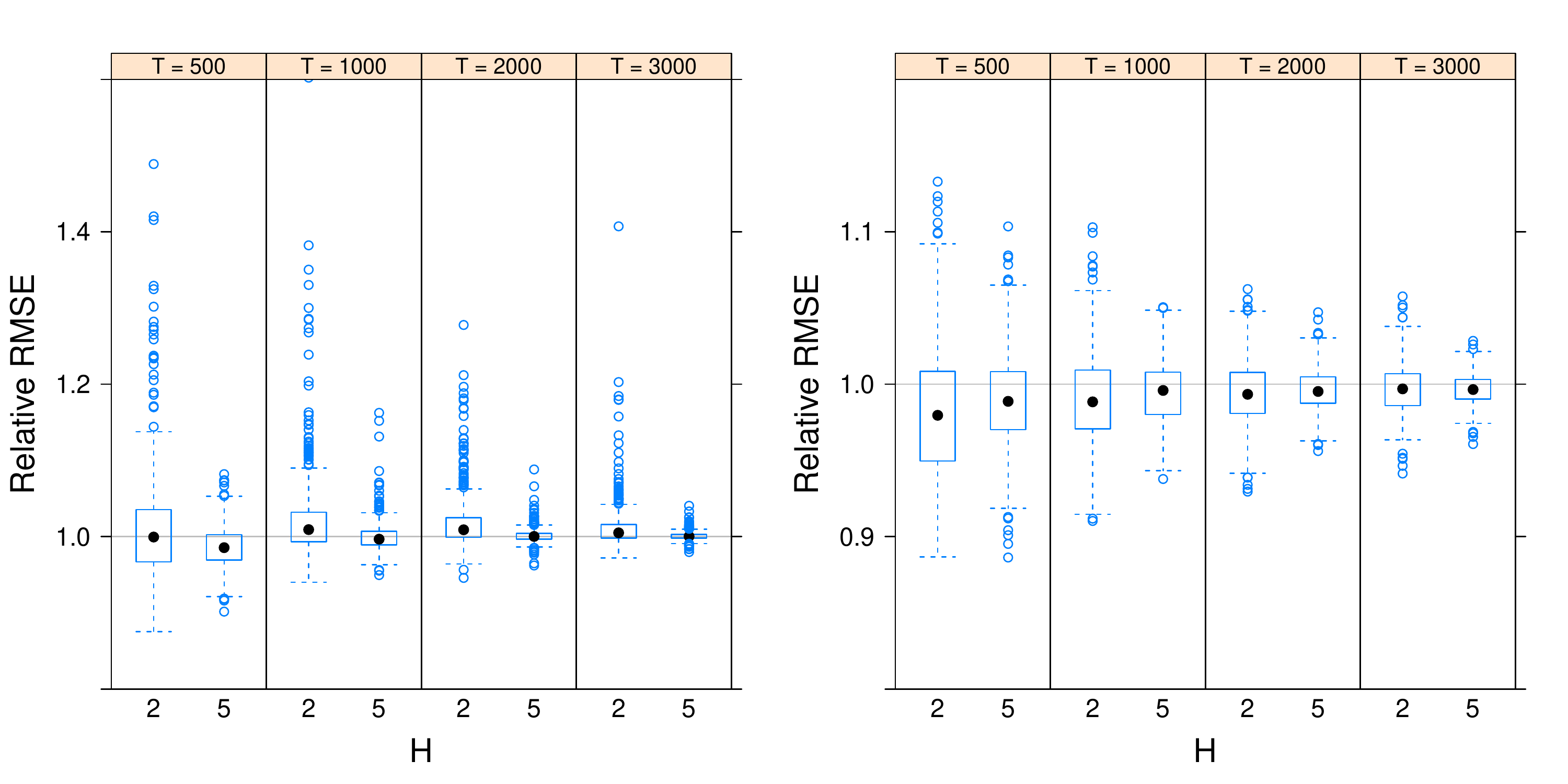}
\caption{TSAVE. Model {\it A} with the biggest values strategy. Relative RMSE values compared to $H = 10$ with borderline nonstationary setting (left panel) and GARCH setting (right panel).}\label{fig::Sup01}
\end{figure}

\begin{figure}[h]
\includegraphics[scale=0.48]{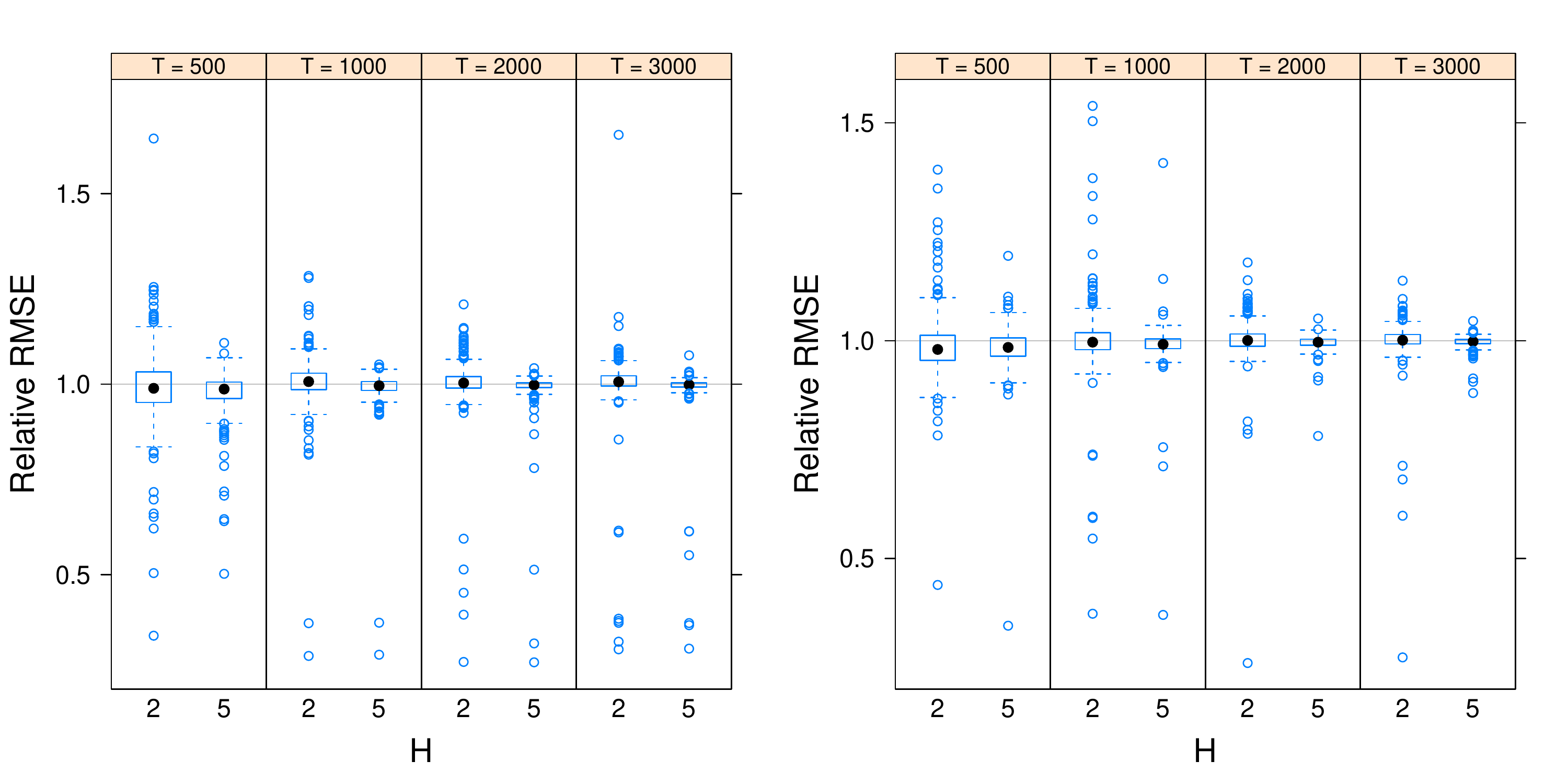}
\caption{TSAVE. Model {\it A} with the biggest values strategy. Relative RMSE values compared to $H = 10$ for $t_4$ innovations setting with $\phi = 0.2$ (left panel) and $\phi = 0.8$ (right panel).}\label{fig::Sup02}
\end{figure}

\begin{figure}[h]
\includegraphics[scale=0.48]{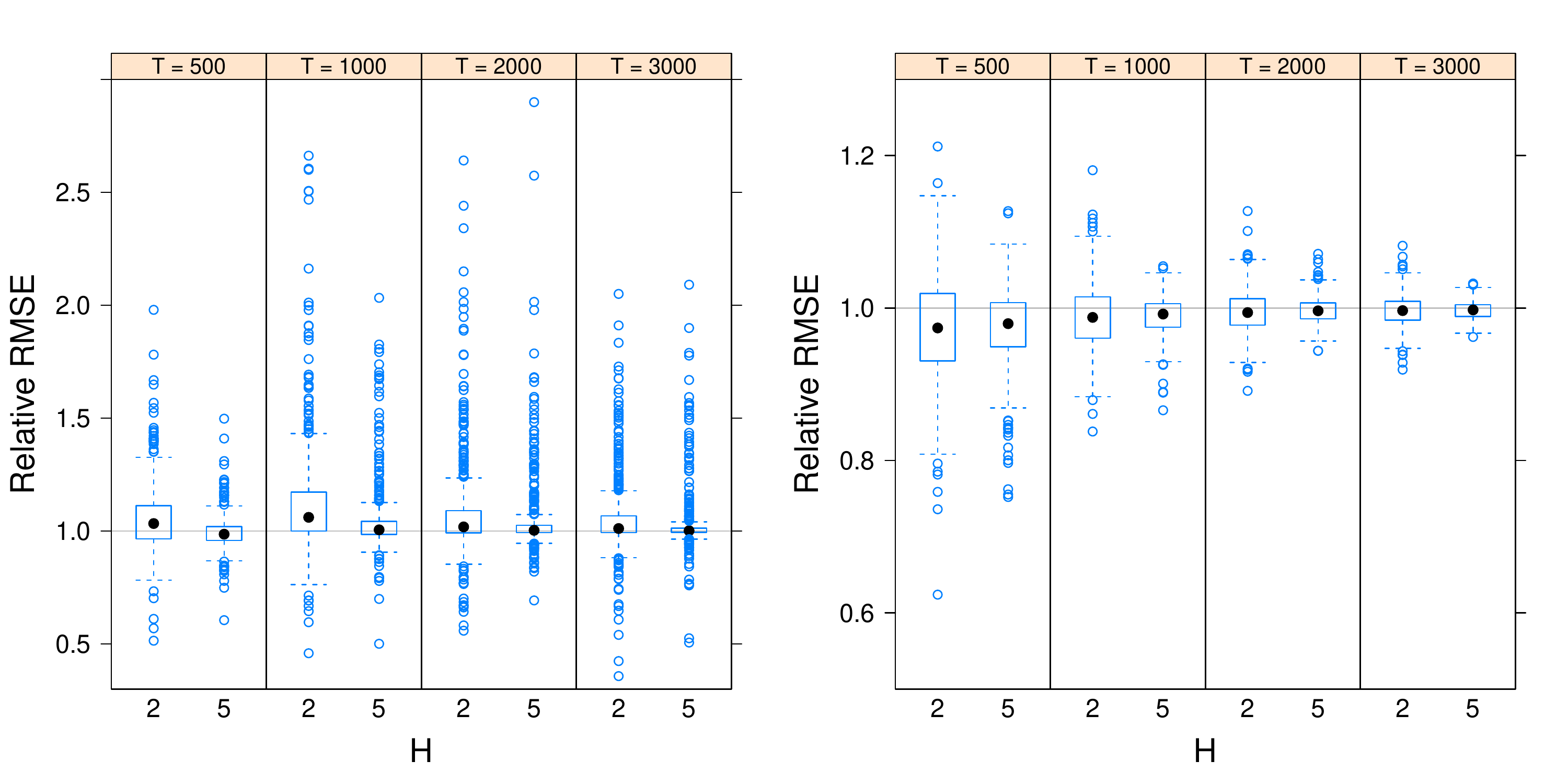}
\caption{TSAVE. Model {\it B} with the biggest values strategy. Relative RMSE values compared to $H = 10$ with borderline nonstationary setting (left panel) and GARCH setting (right panel).}\label{fig::Sup03}
\end{figure}

\begin{figure}[h]
\includegraphics[scale=0.48]{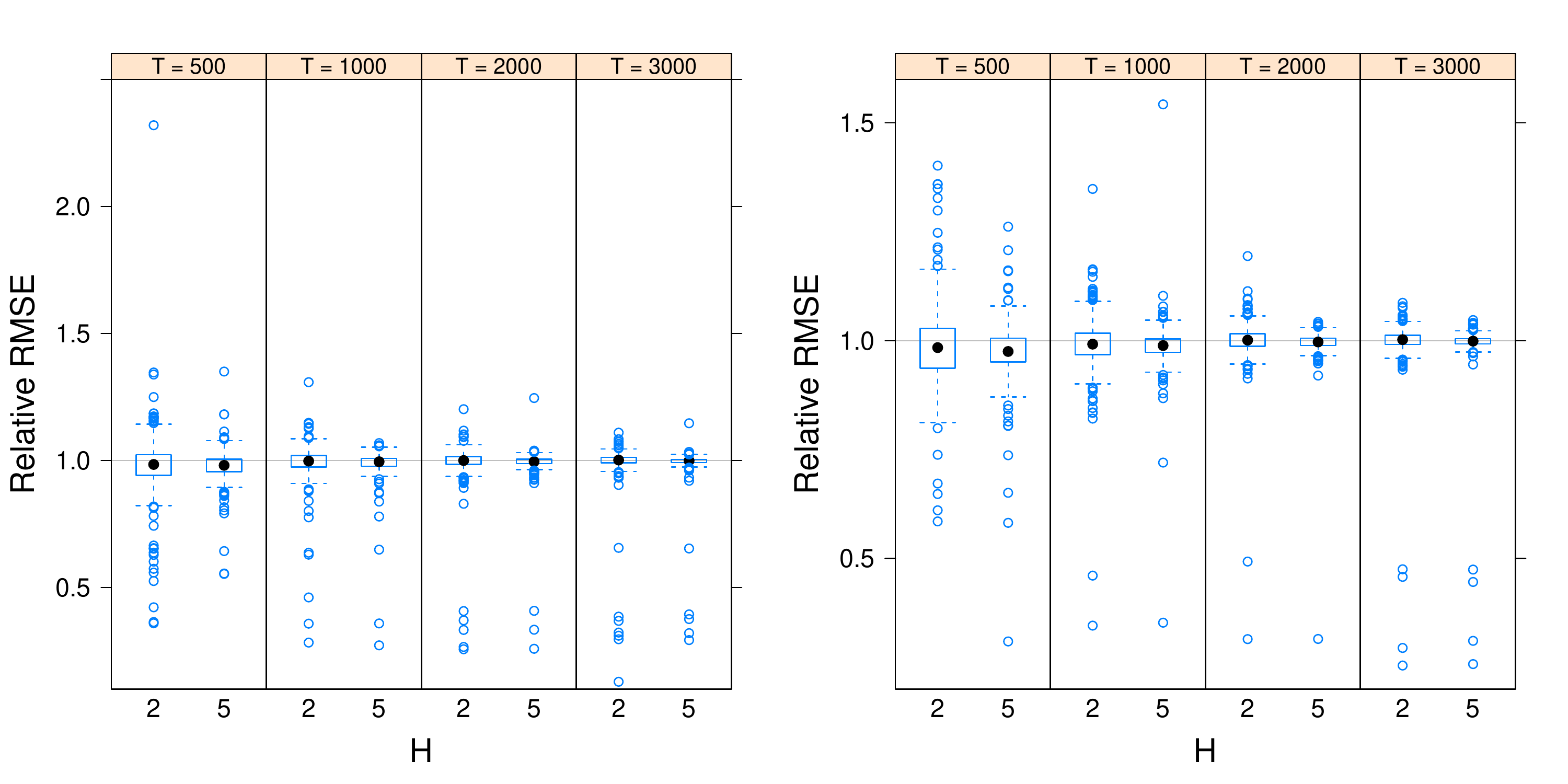}
\caption{TSAVE. Model {\it B} with the biggest values strategy. Relative RMSE values compared to $H = 10$ for $t_4$ innovations setting with $\phi = 0.2$ (left panel) and $\phi = 0.8$ (right panel).}\label{fig::Sup04}
\end{figure}

\begin{figure}[h]
\includegraphics[scale=0.48]{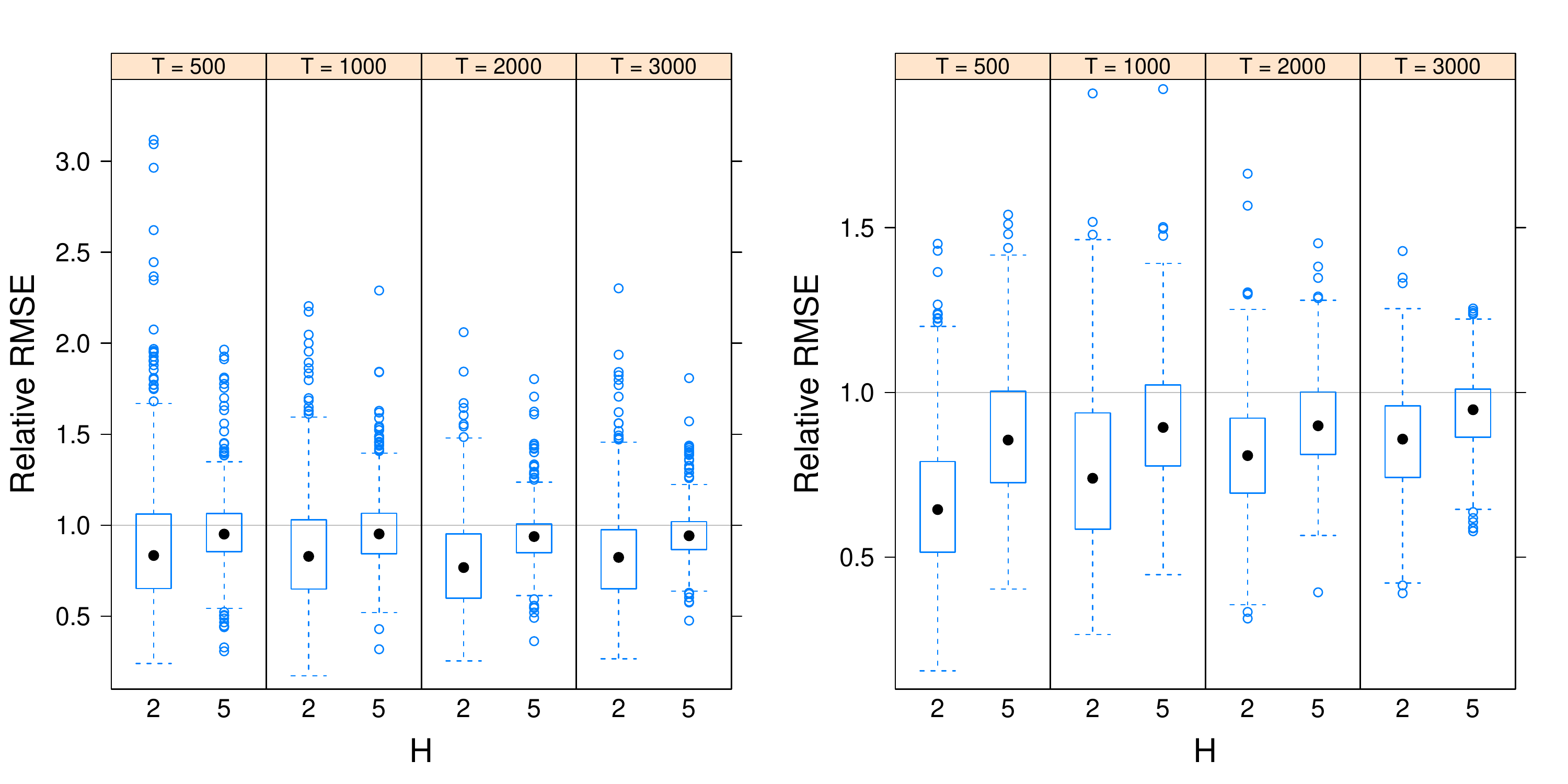}
\caption{TSAVE. Model {\it C} with the biggest values strategy. Relative RMSE values compared to $H = 10$ with borderline nonstationary setting (left panel) and GARCH setting (right panel).}\label{fig::Sup05}
\end{figure}

\begin{figure}[h]
\includegraphics[scale=0.48]{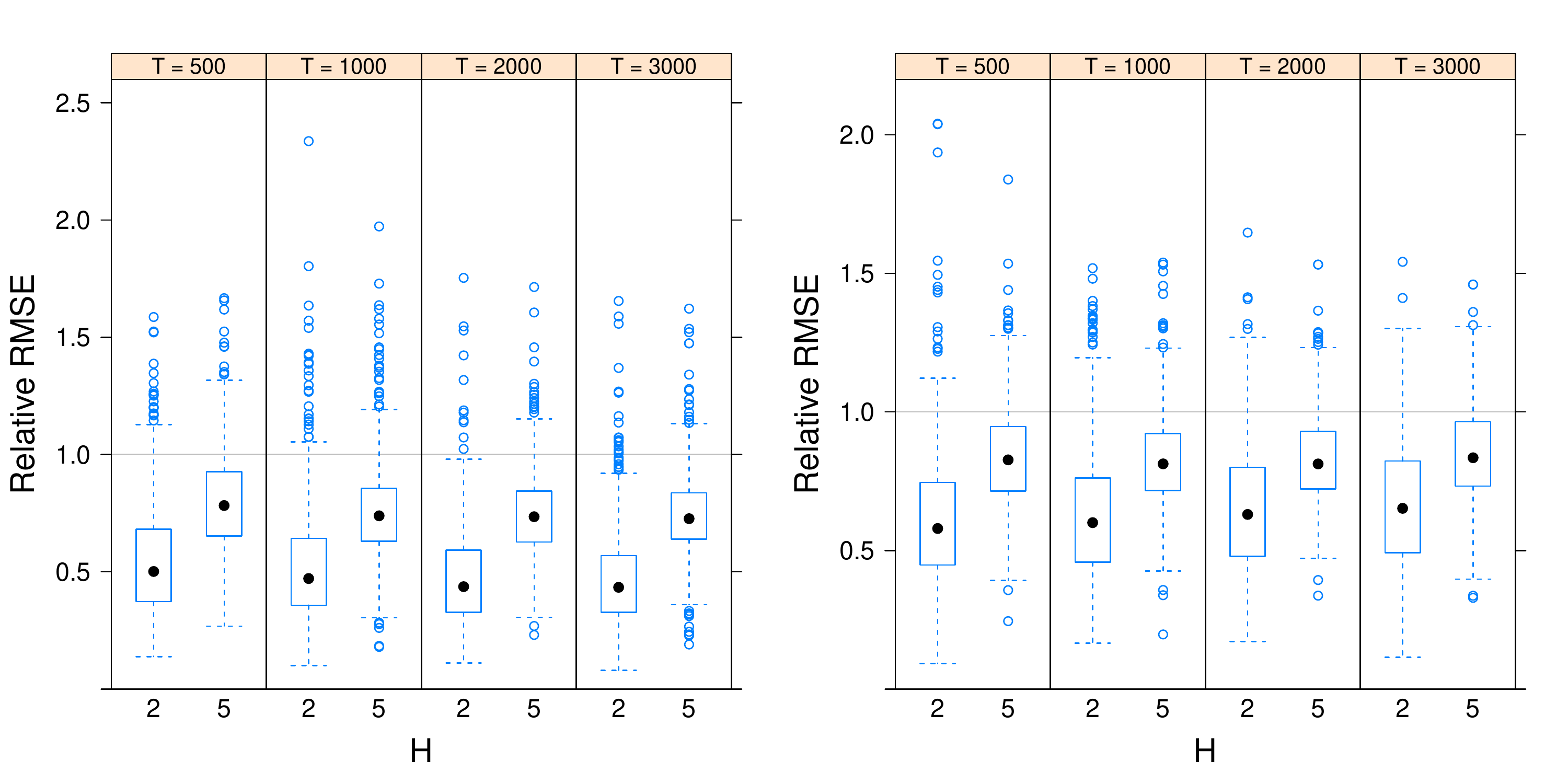}
\caption{TSAVE. Model {\it C} with the biggest values strategy. Relative RMSE values compared to $H = 10$ for $t_4$ innovations setting with $\phi = 0.2$ (left panel) and $\phi = 0.8$ (right panel).}\label{fig::Sup06}
\end{figure}

\begin{figure}[h]
\includegraphics[scale=0.48]{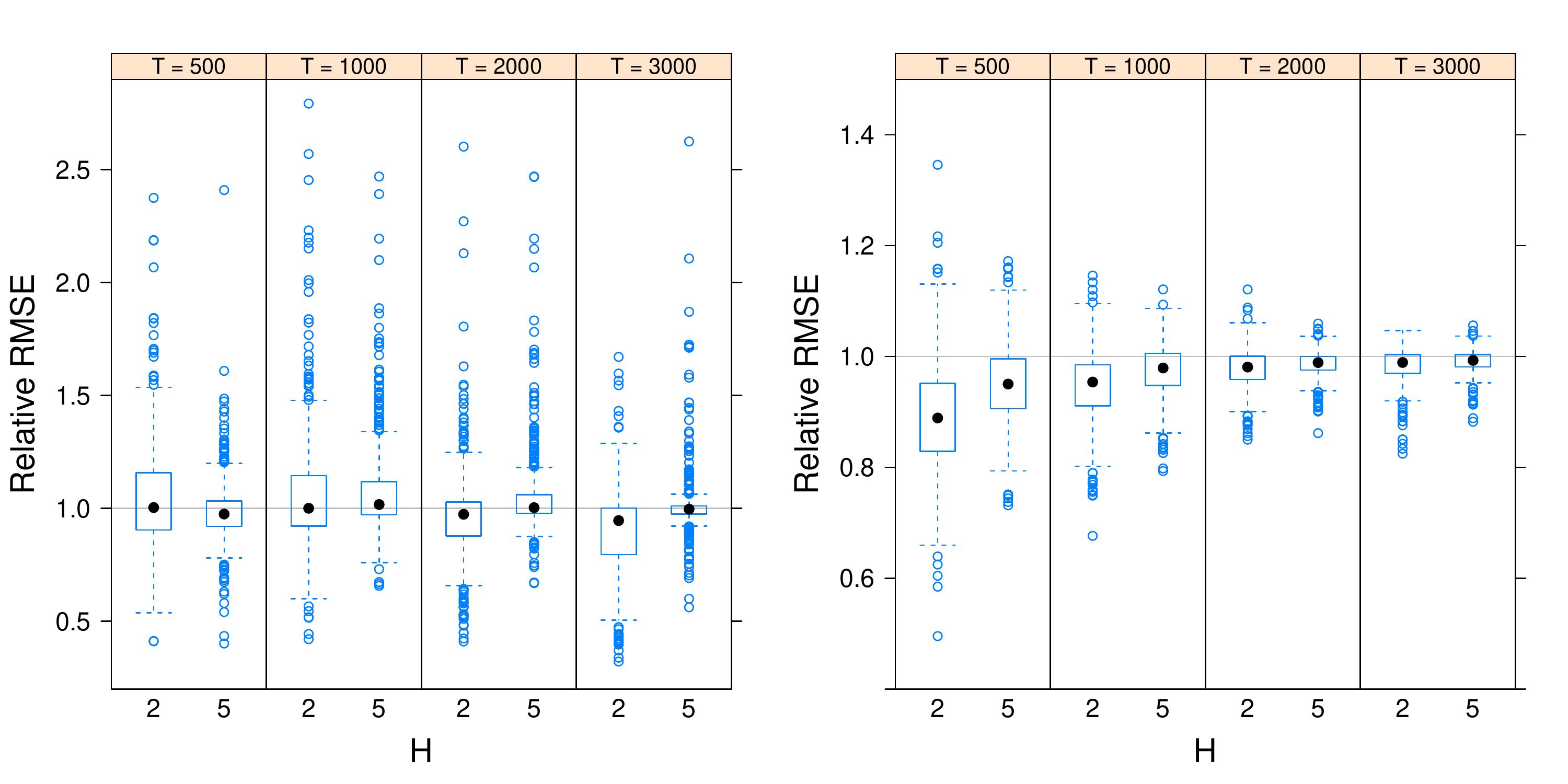}
\caption{TSAVE. Model {\it D} with the biggest values strategy. Relative RMSE values compared to $H = 10$ with borderline nonstationary setting (left panel) and GARCH setting (right panel).}\label{fig::Sup07}
\end{figure}

\begin{figure}[h]
\includegraphics[scale=0.48]{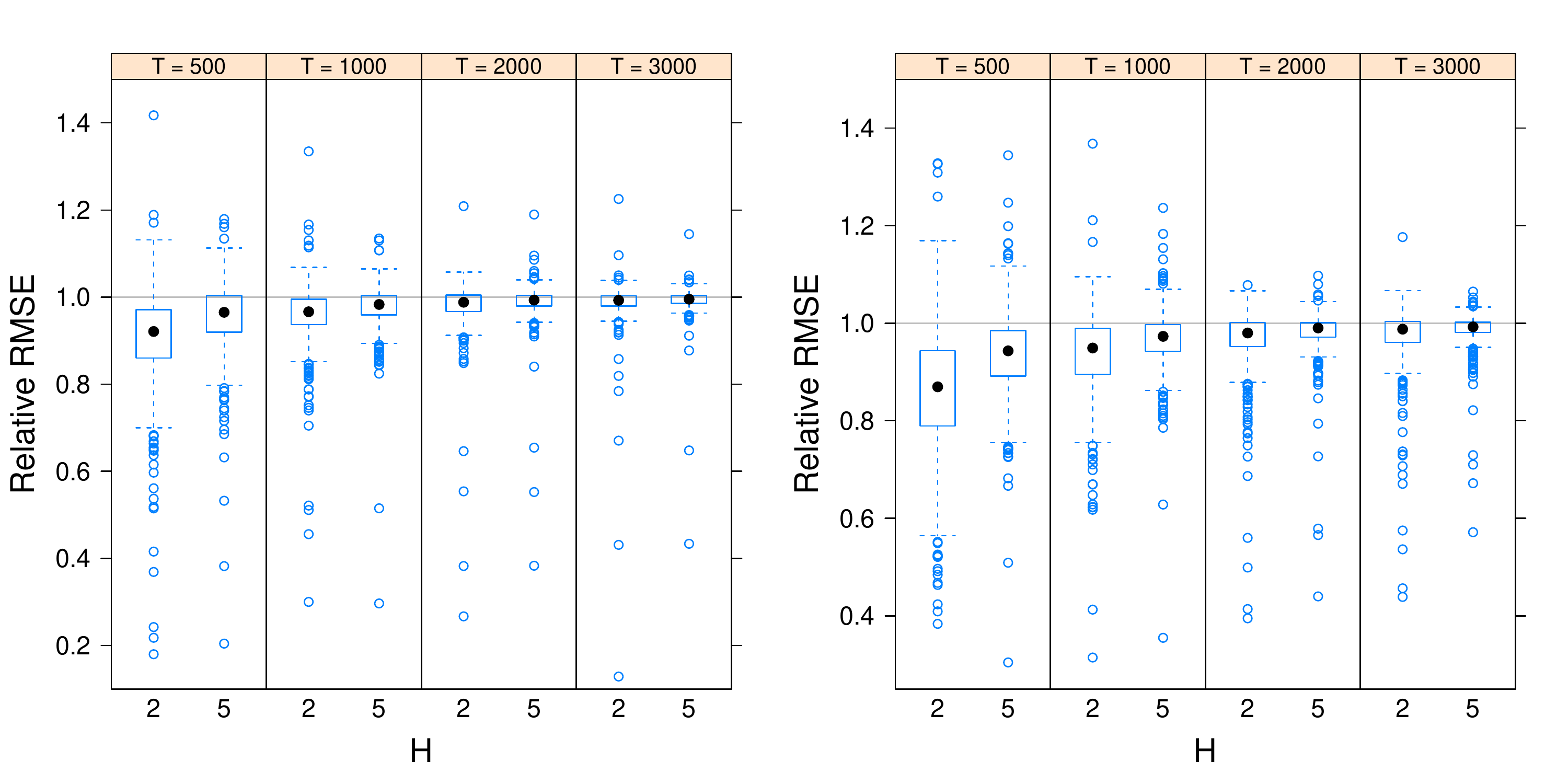}
\caption{TSAVE. Model {\it D} with the biggest values strategy. Relative RMSE values compared to $H = 10$ for $t_4$ innovations setting with $\phi = 0.2$ (left panel) and $\phi = 0.8$ (right panel).}\label{fig::Sup08}
\end{figure}

\begin{figure}[h]
\includegraphics[scale=0.48]{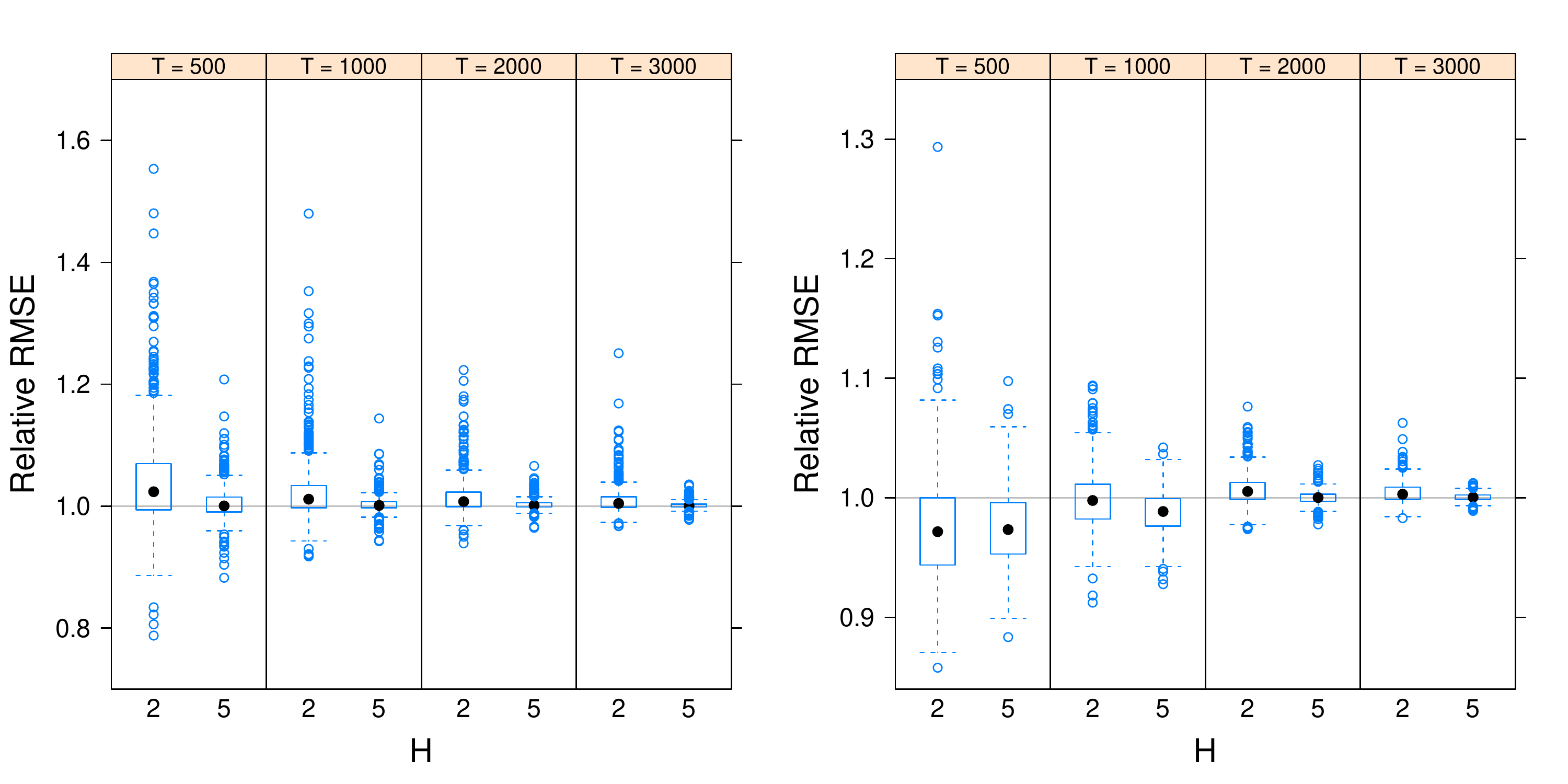}
\caption{TSIR. Model {\it A} with the biggest values strategy. Relative RMSE values compared to $H = 10$ with borderline nonstationary setting (left panel) and GARCH setting (right panel).}\label{fig::Sup09}
\end{figure}

\begin{figure}[h]
\includegraphics[scale=0.48]{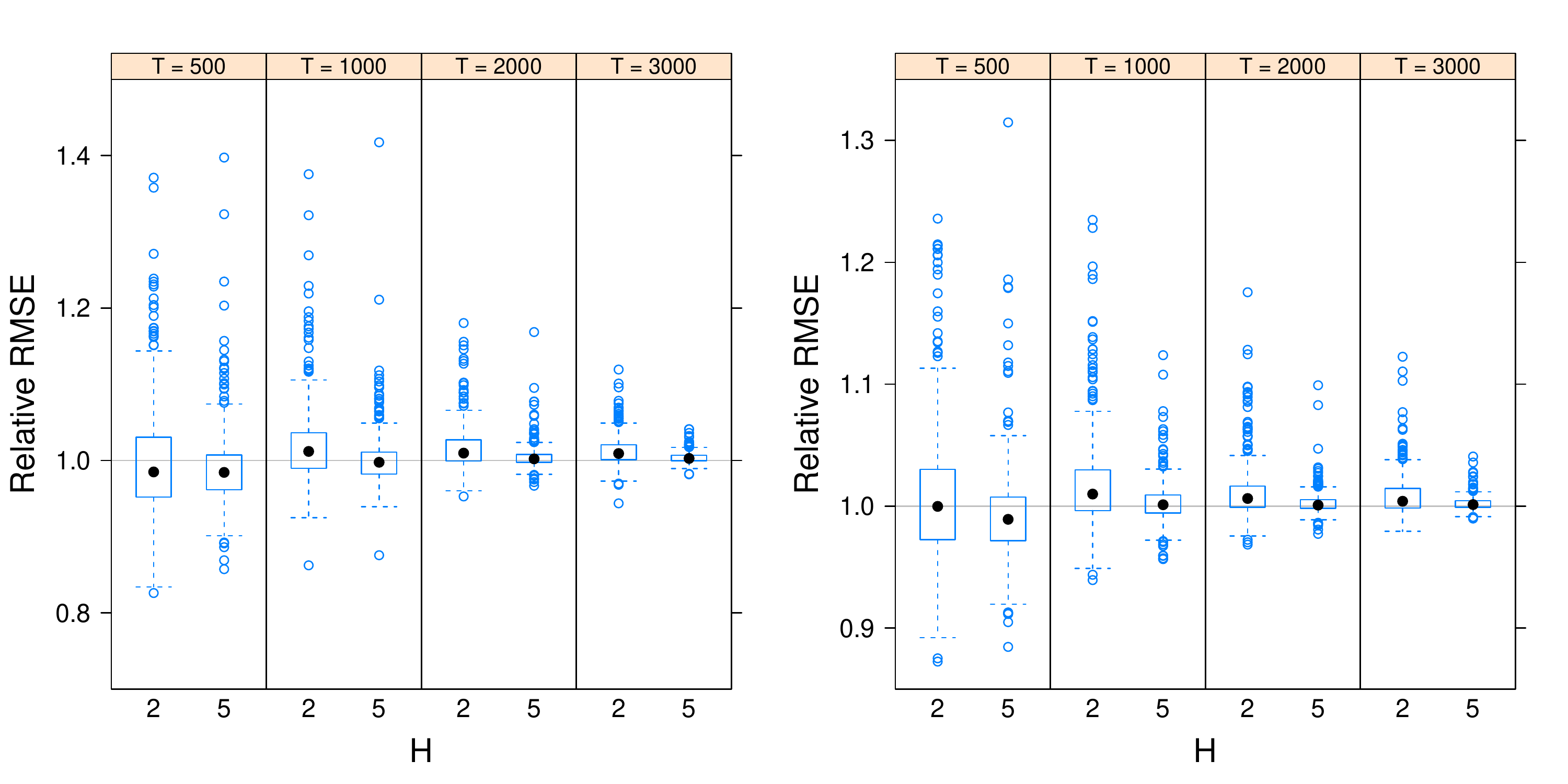}
\caption{TSIR. Model {\it A} with the biggest values strategy. Relative RMSE values compared to $H = 10$ for $t_4$ innovations setting with $\phi = 0.2$ (left panel) and $\phi = 0.8$ (right panel).}\label{fig::Sup10}
\end{figure}

Figures \ref{fig::Sup01} -- \ref{fig::Sup10} include the boxplots of the RMSE values for the search of an optimal $H$.
In the borderline non-stationarity setting in models {\it A} and {\it B} it seems that $H = 5$ is all-around the best, but in the other two models $H = 2$ seems generally the best, especially in longer time series. In GARCH setting $H = 2$ outperforms others in all models. The same is mostly true for both low and high dependence settings with $t_4$-innovations.


\clearpage

\subsection*{The value of $a$ (the coefficient for TSSH)}

\begin{figure}[h]
\includegraphics[scale=0.48]{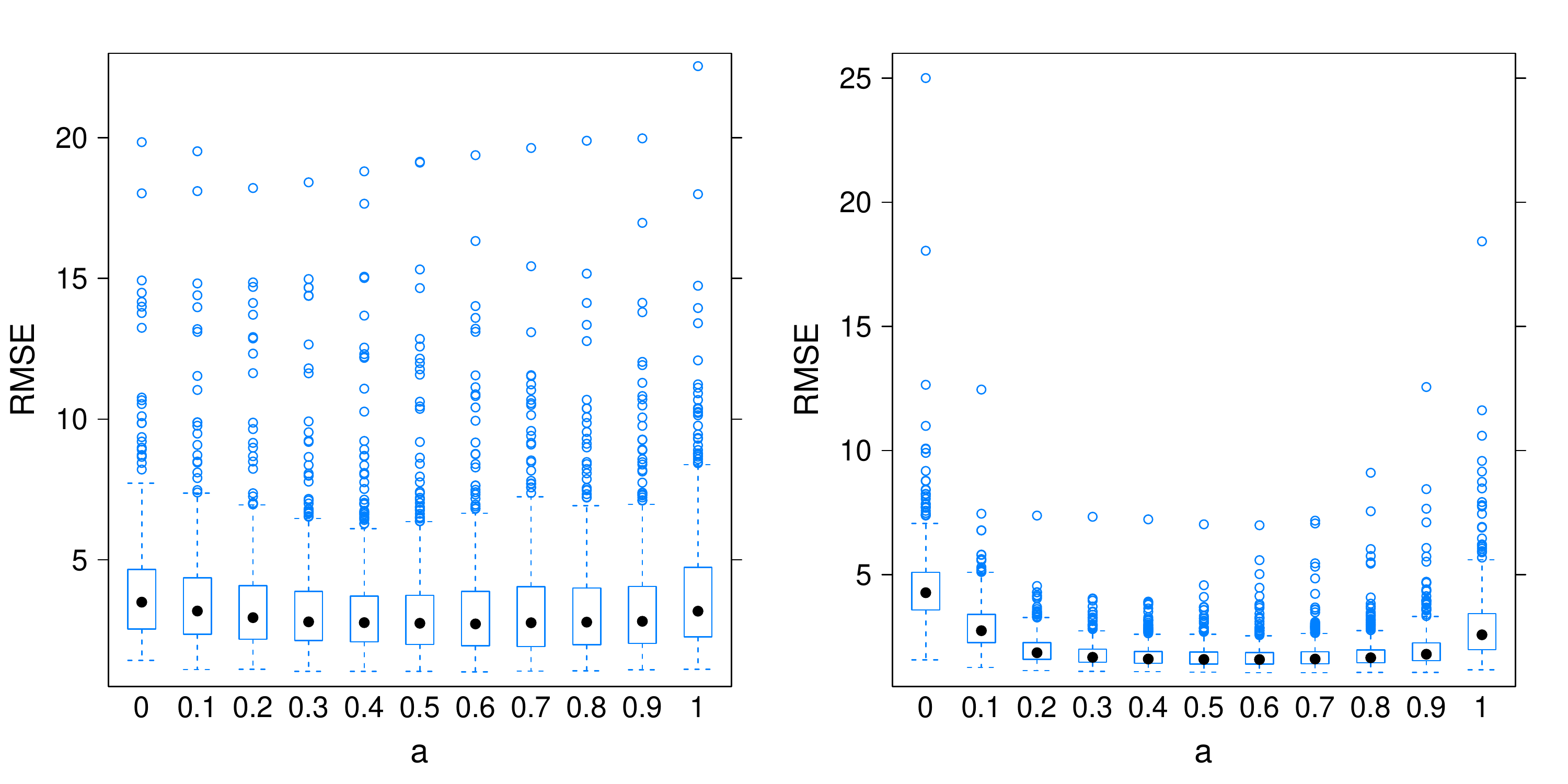}
\caption{TSSH. Model {\it E} with the biggest values strategy:
$T = 500$ and $H = 2$ for TSAVE part and $H = 10$ for TSIR part. RMSE values with borderline nonstationary setting (left panel) and GARCH setting (right panel).}\label{fig::Sup11}
\end{figure}

\begin{figure}[h]
\includegraphics[scale=0.48]{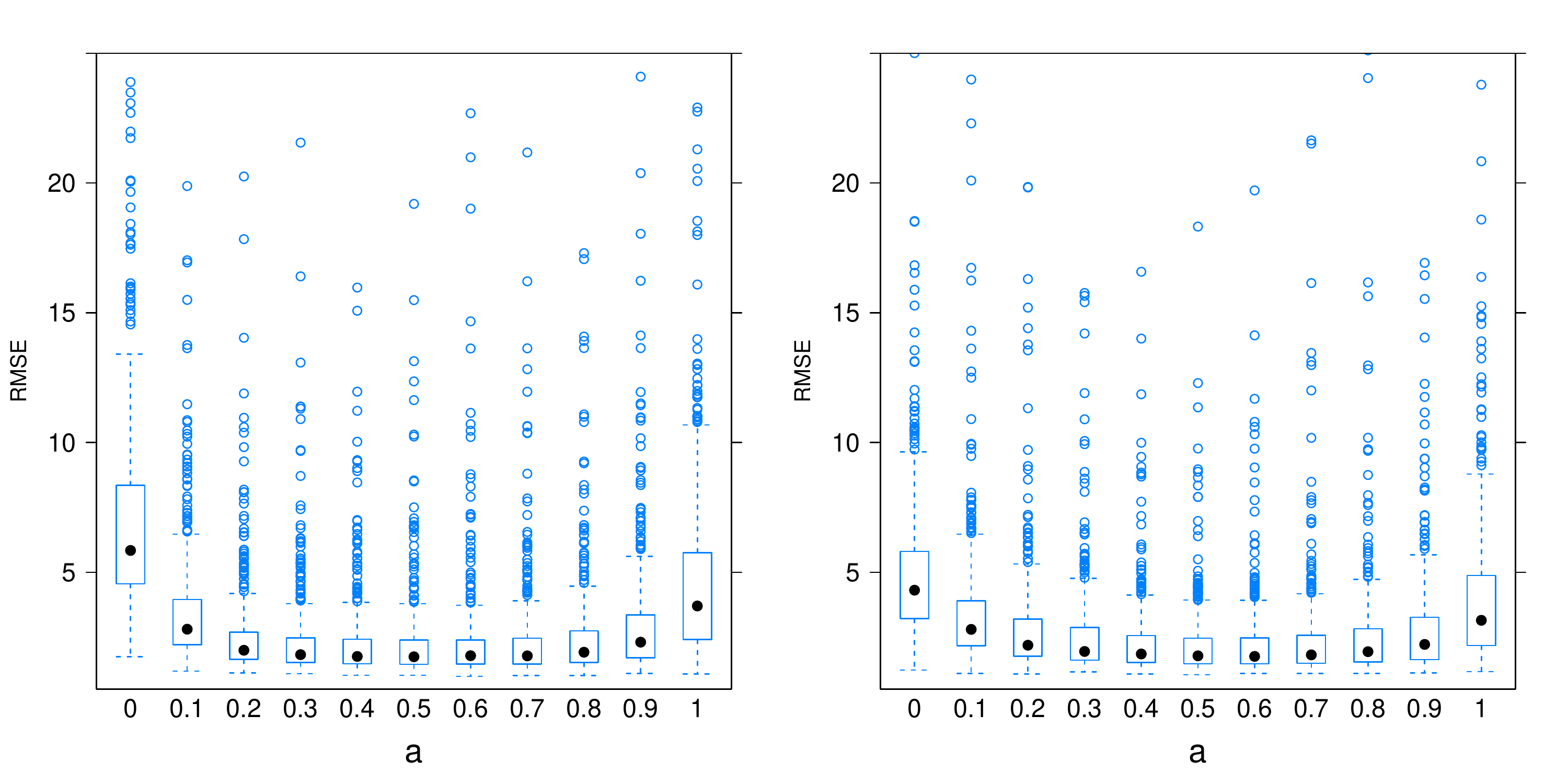}
\caption{TSSH. Model {\it E} with the biggest values strategy:
$T = 500$ and $H = 2$ for TSAVE part and $H = 10$ for TSIR part. RMSE values for $t_4$ innovations setting with $\phi = 0.2$ (left panel) and $\phi = 0.8$ (right panel).}\label{fig::Sup12}
\end{figure}

\begin{figure}[h]
\includegraphics[scale=0.48]{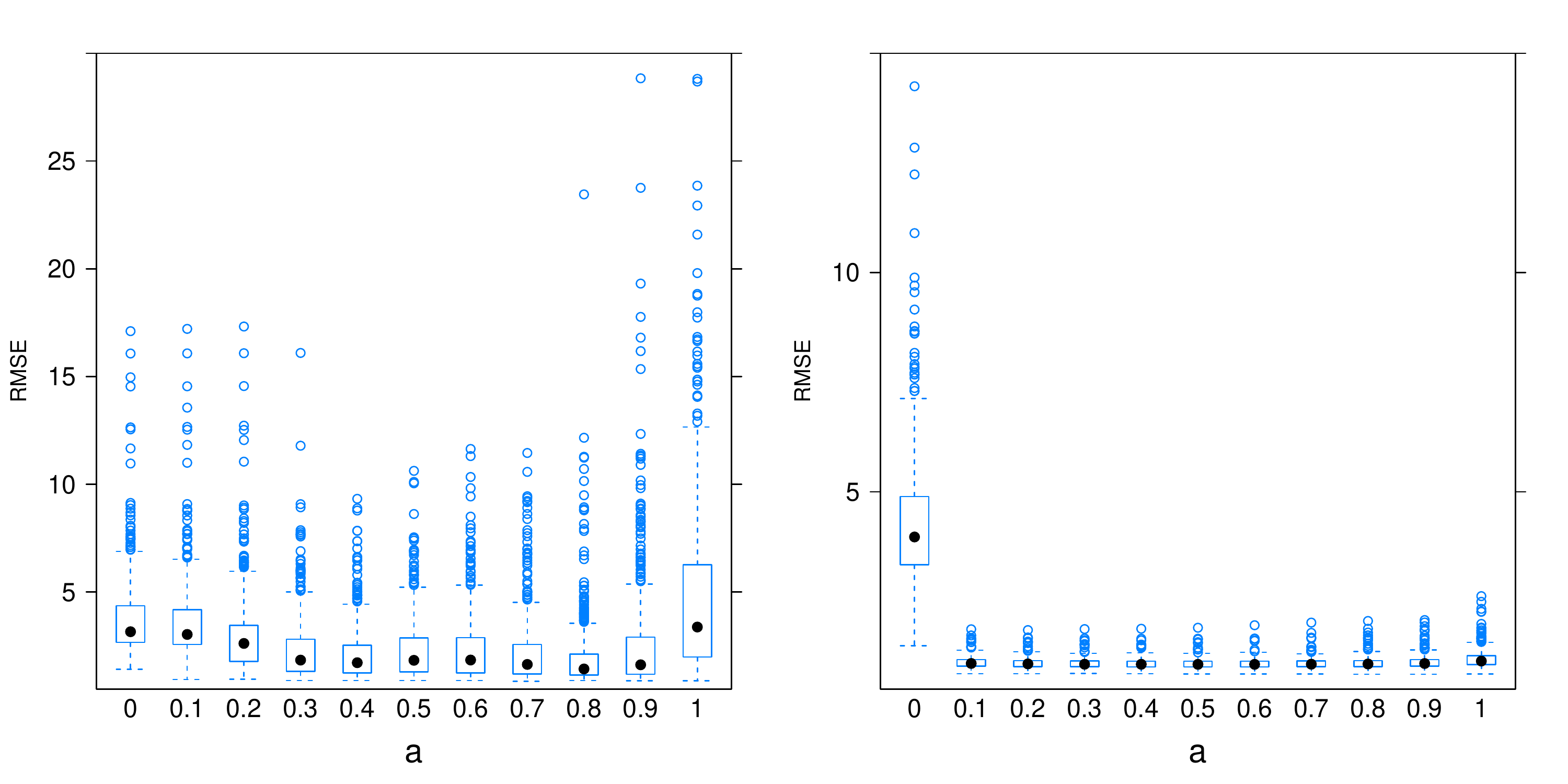}
\caption{TSSH. Model {\it E} with the biggest values strategy:
$T = 3000$ and $H = 2$ for TSAVE part and $H = 10$ for TSIR part. RMSE values with borderline nonstationary setting (left panel) and GARCH setting (right panel).}\label{fig::Sup13}
\end{figure}

\begin{figure}[h]
\includegraphics[scale=0.48]{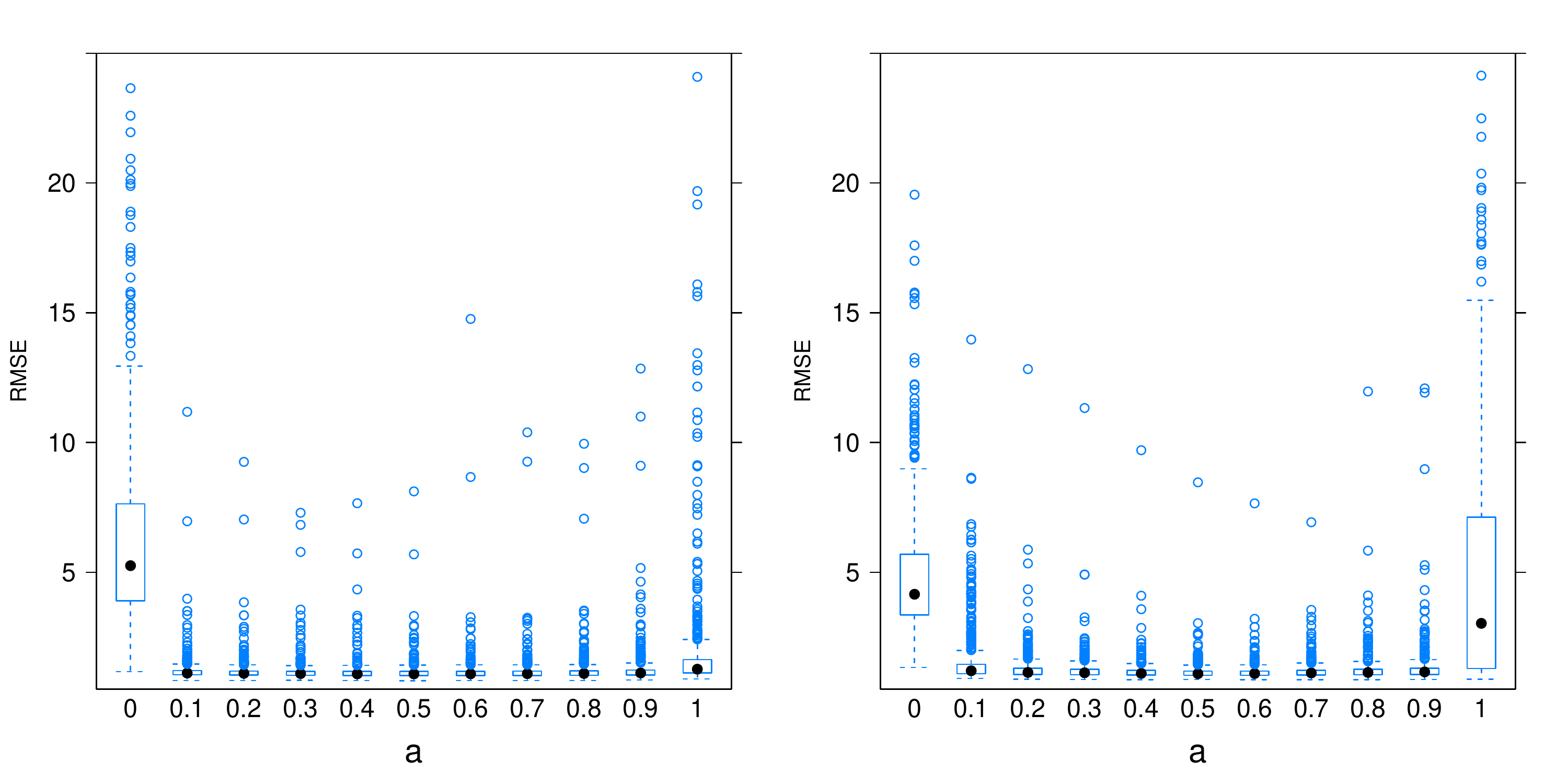}
\caption{TSSH. Model {\it E} with the biggest values strategy:
$T = 3000$ and $H = 2$ for TSAVE part and $H = 10$ for TSIR part. RMSE values for $t_4$ innovations setting with $\phi = 0.2$ (left panel) and $\phi = 0.8$ (right panel).}\label{fig::Sup14}
\end{figure}

Figures \ref{fig::Sup11} -- \ref{fig::Sup14} are the boxplots of the RMSE values regarding the choice of $a$.
The choices closer to the $a = 0.5$ seem to be the most reliable, and values close to $a = 0$ and $a = 1$ should be avoided. However, in the borderline non-stationary setting in longer time series $a = 0.8$ seems to produce surprisingly the best results, but choices closer to $a = 0.5$ are quite safe as well.

\clearpage

\subsection*{Comparison to other approaches}

\begin{figure}[h]
\includegraphics[scale=0.48]{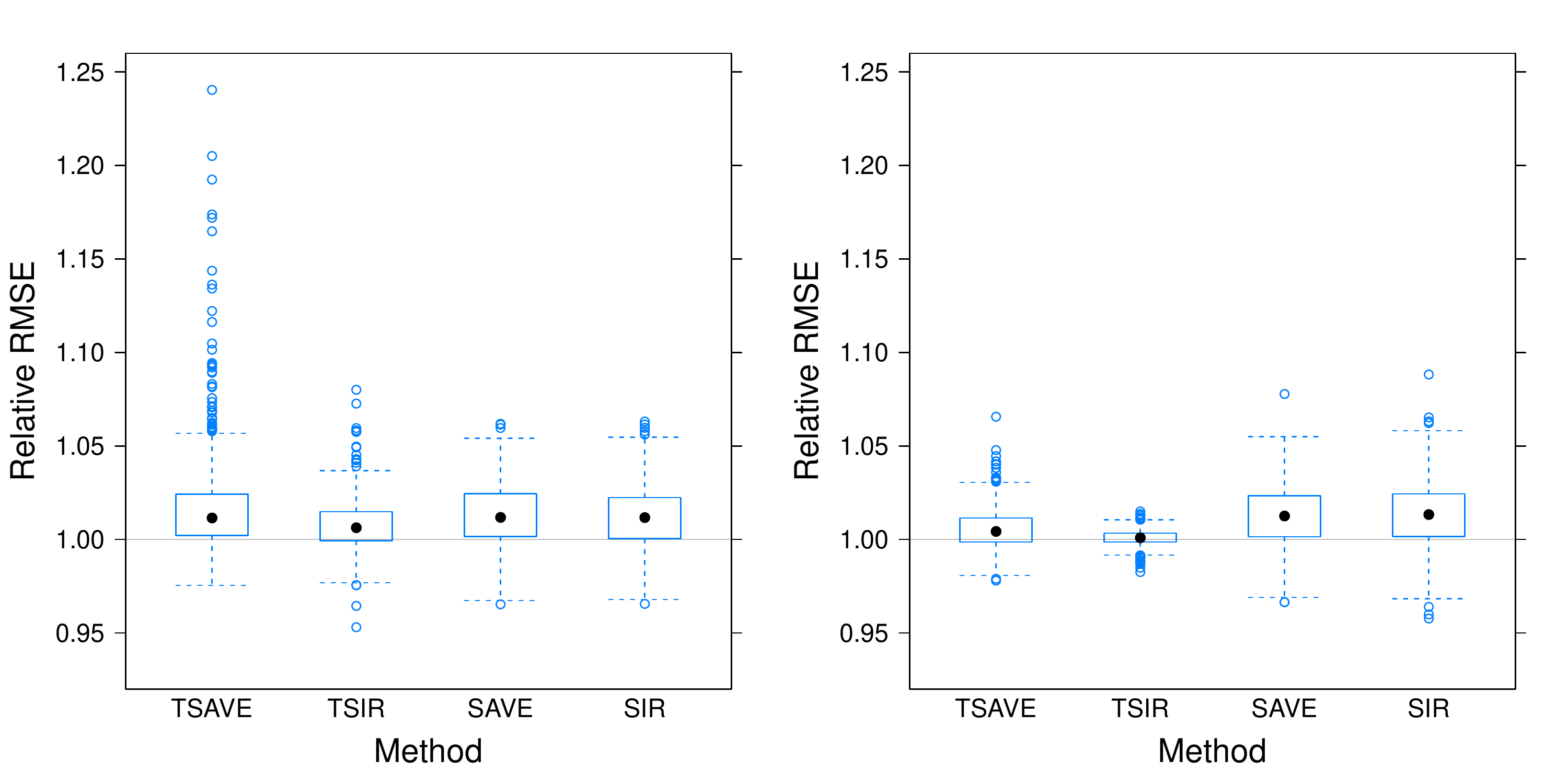}
\caption{Model {\it A} with biggest values strategy. Relative RMSE values compared to Oracle estimator borderline nonstationary setting (left panel) and GARCH setting (right panel).}\label{fig::Sup15}
\end{figure}

\begin{figure}[h]
\includegraphics[scale=0.48]{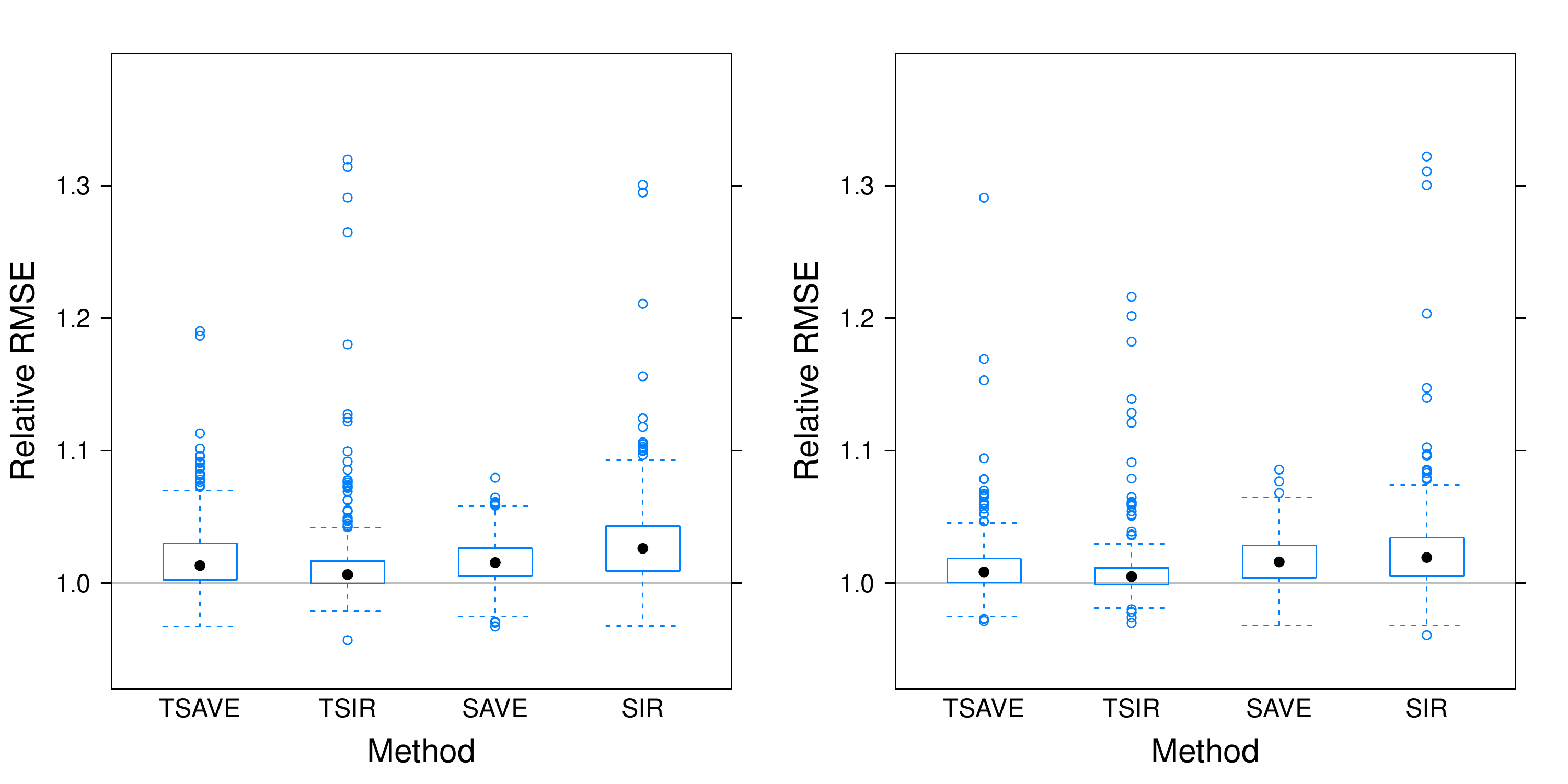}
\caption{Model {\it A} with biggest values strategy. Relative RMSE values compared to Oracle estimator for $t_4$ innovations setting with $\phi = 0.2$ (left panel) and $\phi = 0.8$ (right panel).}\label{fig::Sup16}
\end{figure}

\begin{figure}[h]
\includegraphics[scale=0.48]{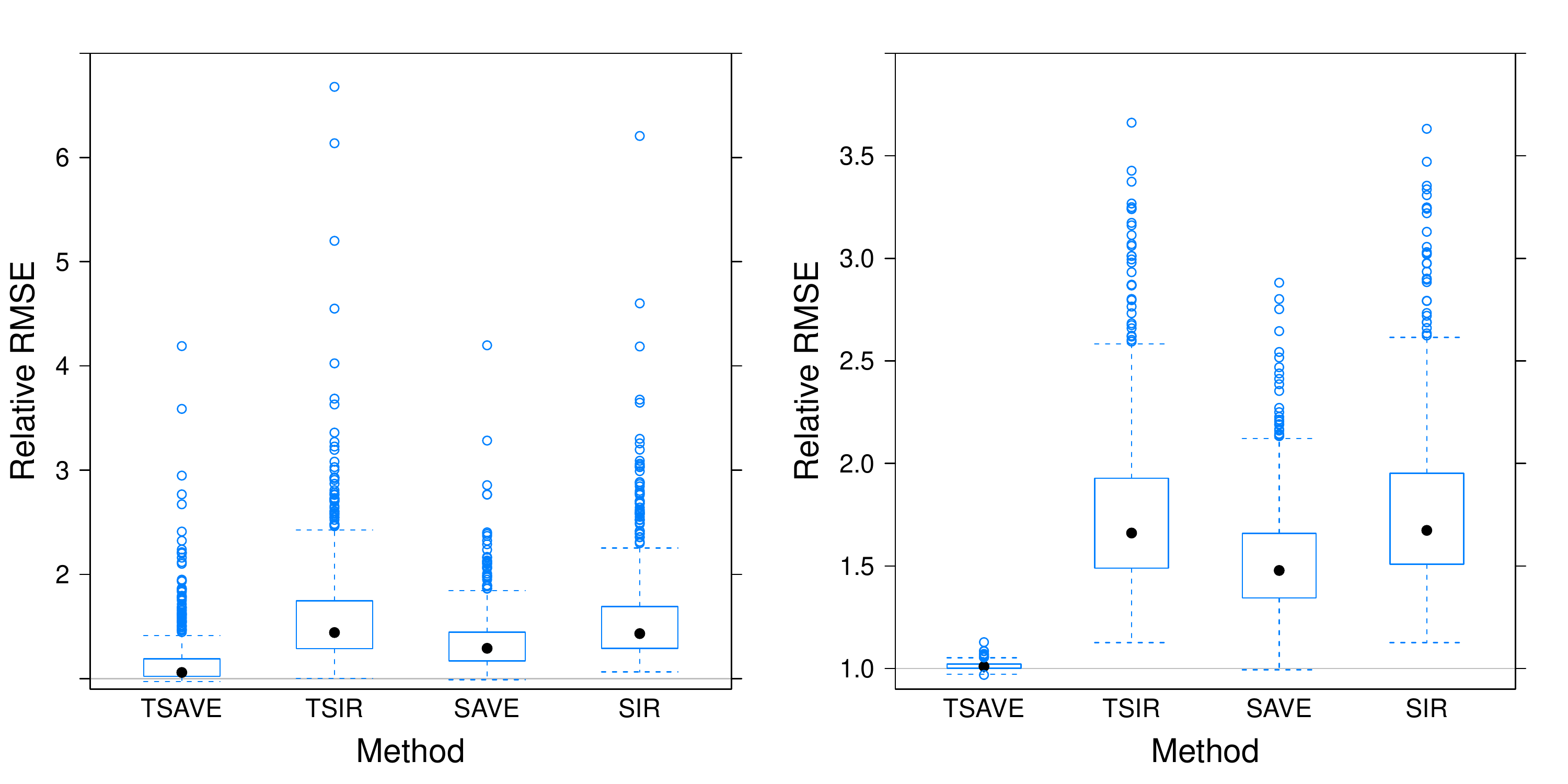}
\caption{Model {\it B} with biggest values strategy. Relative RMSE values compared to Oracle estimator borderline nonstationary setting (left panel) and GARCH setting (right panel).}\label{fig::Sup17}
\end{figure}

\begin{figure}[h]
\includegraphics[scale=0.48]{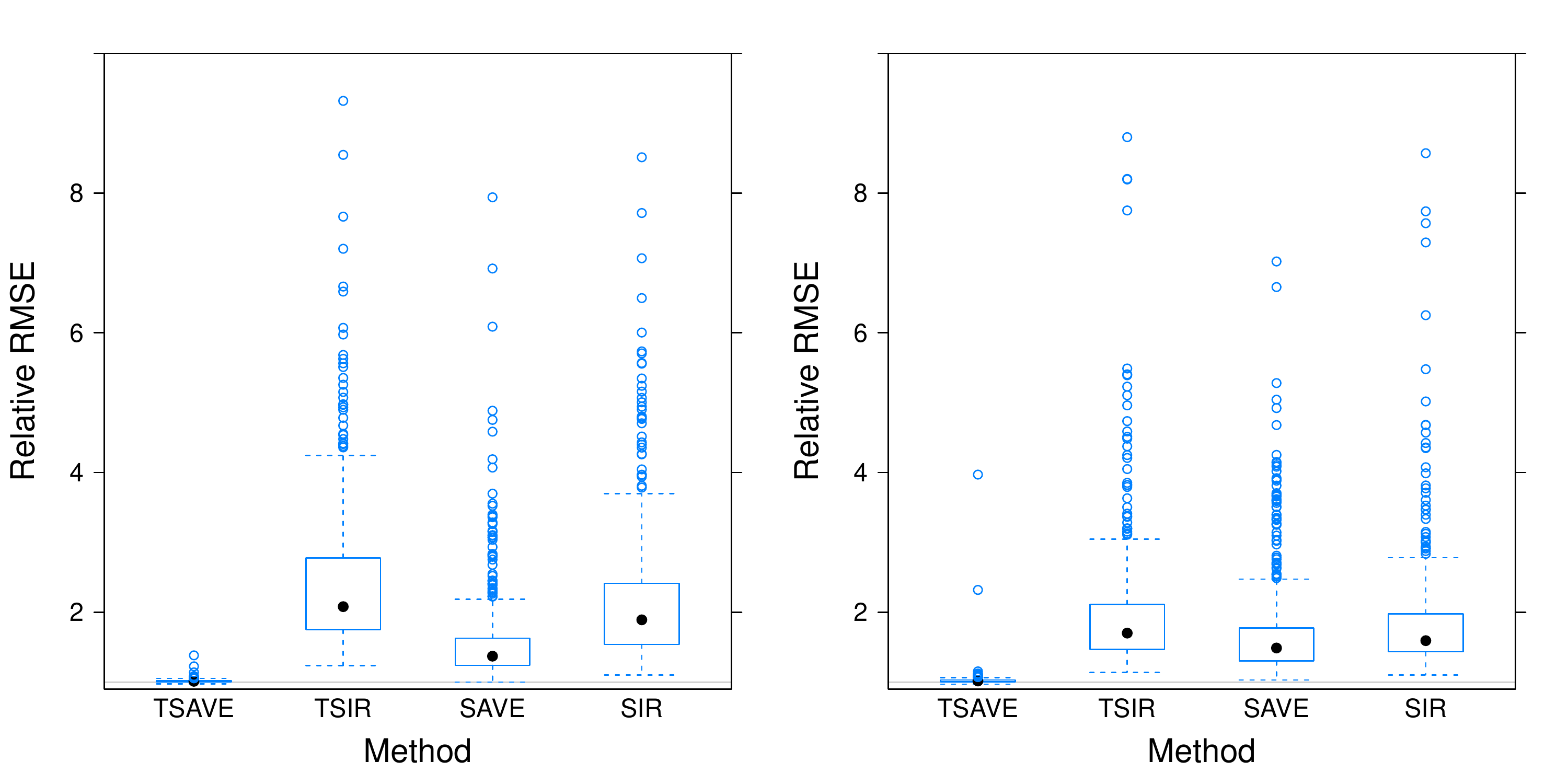}
\caption{Model {\it B} with biggest values strategy. Relative RMSE values compared to Oracle estimator for $t_4$ innovations setting with $\phi = 0.2$ (left panel) and $\phi = 0.8$ (right panel).}\label{fig::Sup18}
\end{figure}

\begin{figure}[h]
\includegraphics[scale=0.48]{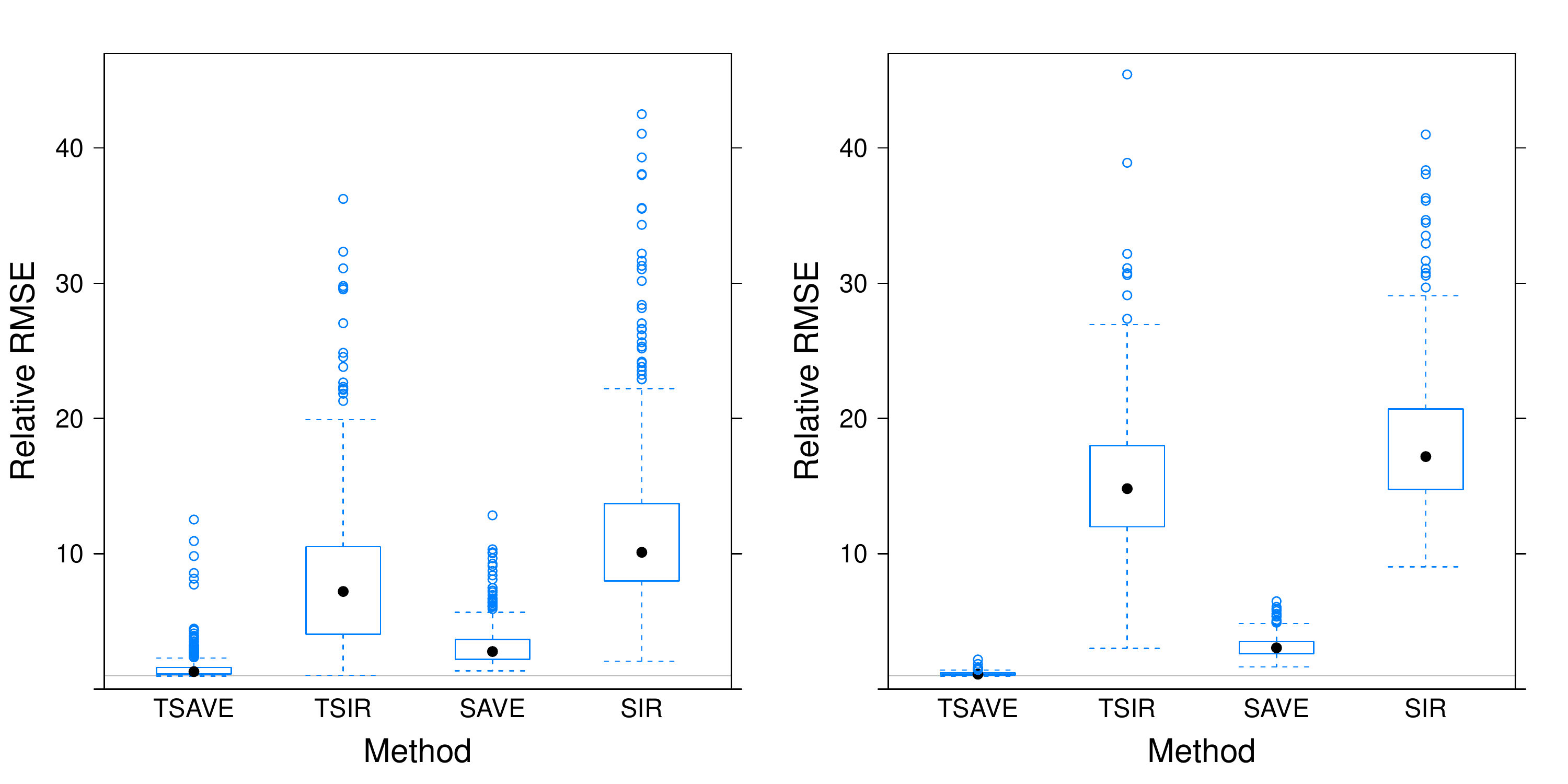}
\caption{Model {\it C} with biggest values strategy. Relative RMSE values compared to Oracle estimator borderline nonstationary setting (left panel) and GARCH setting (right panel).}\label{fig::Sup19}
\end{figure}

\begin{figure}[h]
\includegraphics[scale=0.48]{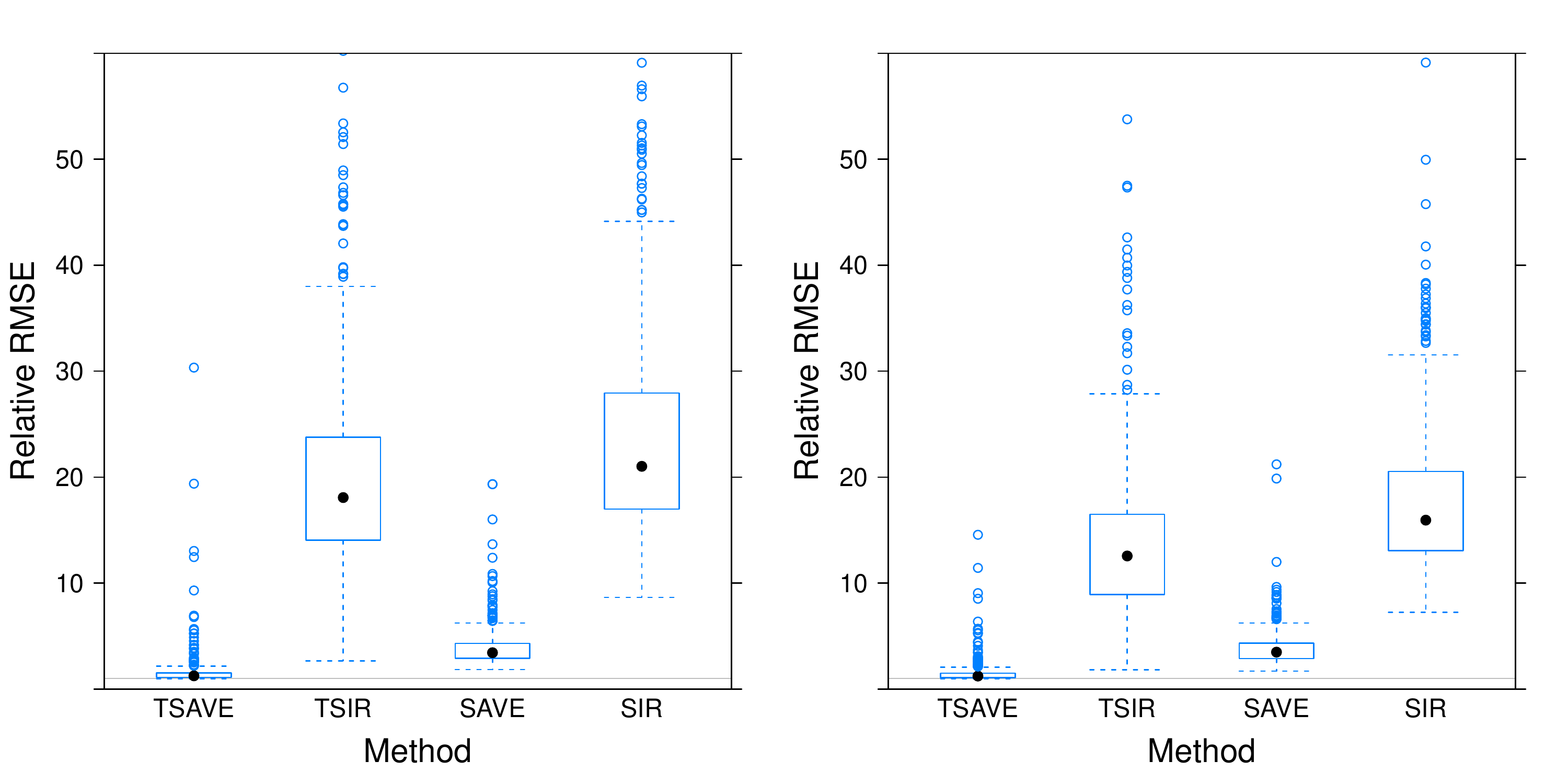}
\caption{Model {\it C} with biggest values strategy. Relative RMSE values compared to Oracle estimator for $t_4$ innovations setting with $\phi = 0.2$ (left panel) and $\phi = 0.8$ (right panel).}\label{fig::Sup20}
\end{figure}

\begin{figure}[h]
\includegraphics[scale=0.48]{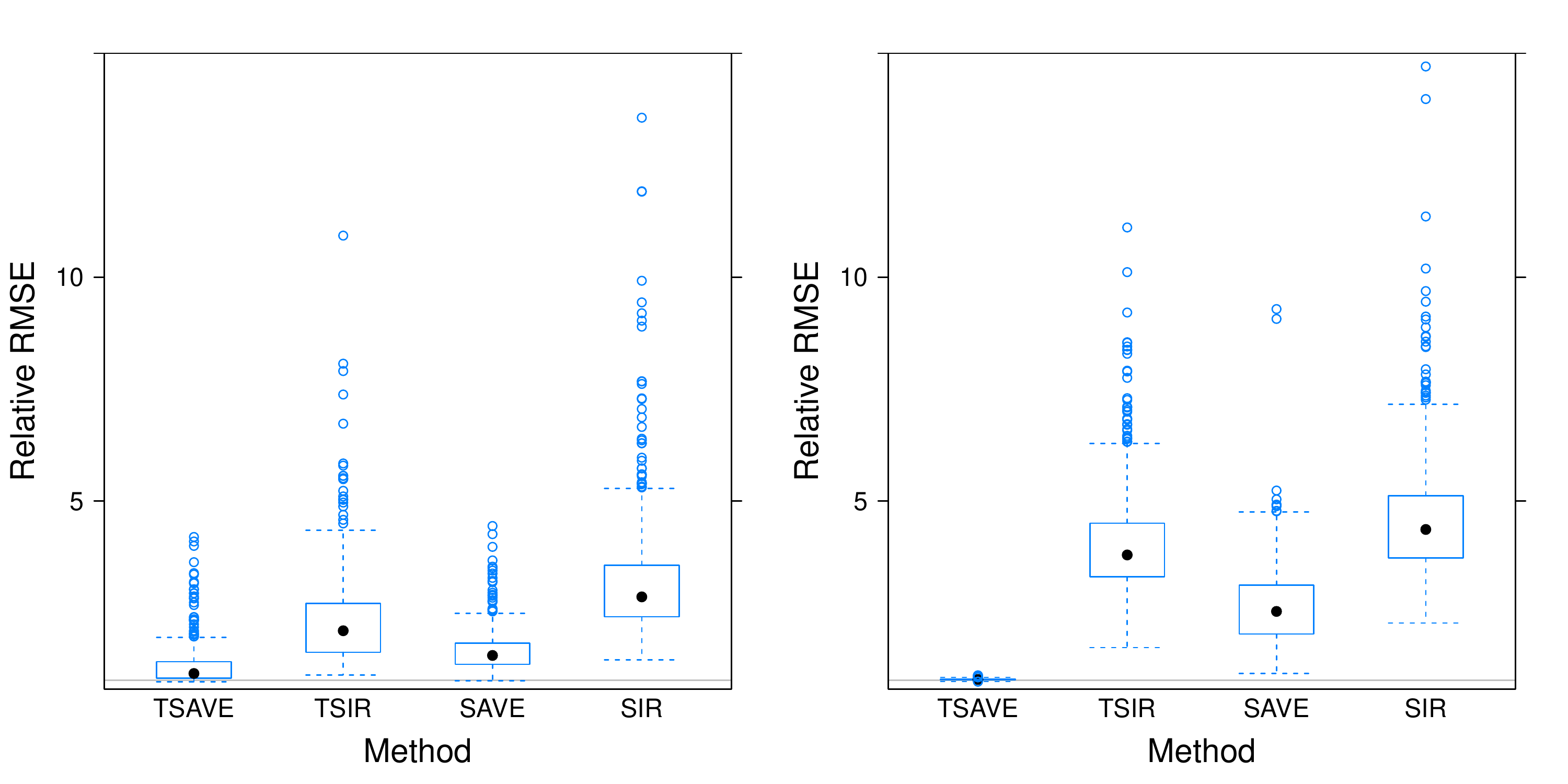}
\caption{Model {\it D} with biggest values strategy. Relative RMSE values compared to Oracle estimator borderline nonstationary setting (left panel) and GARCH setting (right panel).}\label{fig::Sup21}
\end{figure}

\begin{figure}[h]
\includegraphics[scale=0.48]{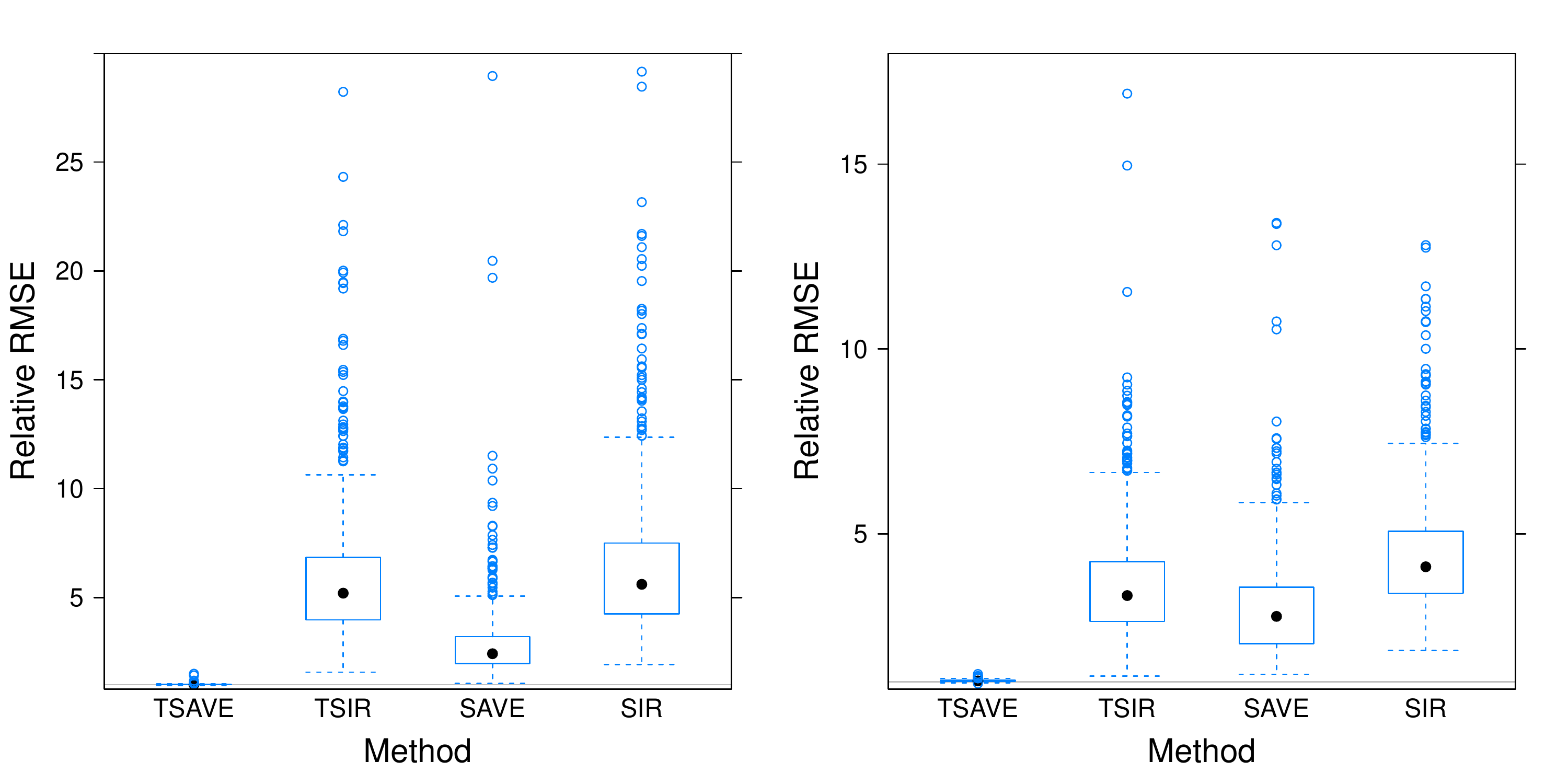}
\caption{Model {\it D} with biggest values strategy. Relative RMSE values compared to Oracle estimator for $t_4$ innovations setting with $\phi = 0.2$ (left panel) and $\phi = 0.8$ (right panel).}\label{fig::Sup22}
\end{figure}

Figures \ref{fig::Sup15} -- \ref{fig::Sup22} include the boxplots of the RMSE values compared to the Oracle model.
For models {\it B -- D} the results indicate that TSAVE is the better than  vectorized SAVE and TSIR better than vectorized SIR, and SAVE algorithms are better than the SIR algorithms. For model {\it A} TSIR produces the best results in all cases, as expected.


\clearpage

\section*{Additional results for $p = 10, k = 3$ setting: comparison to other approaches}

\begin{figure}[h]
\centering
\includegraphics[scale=0.44]{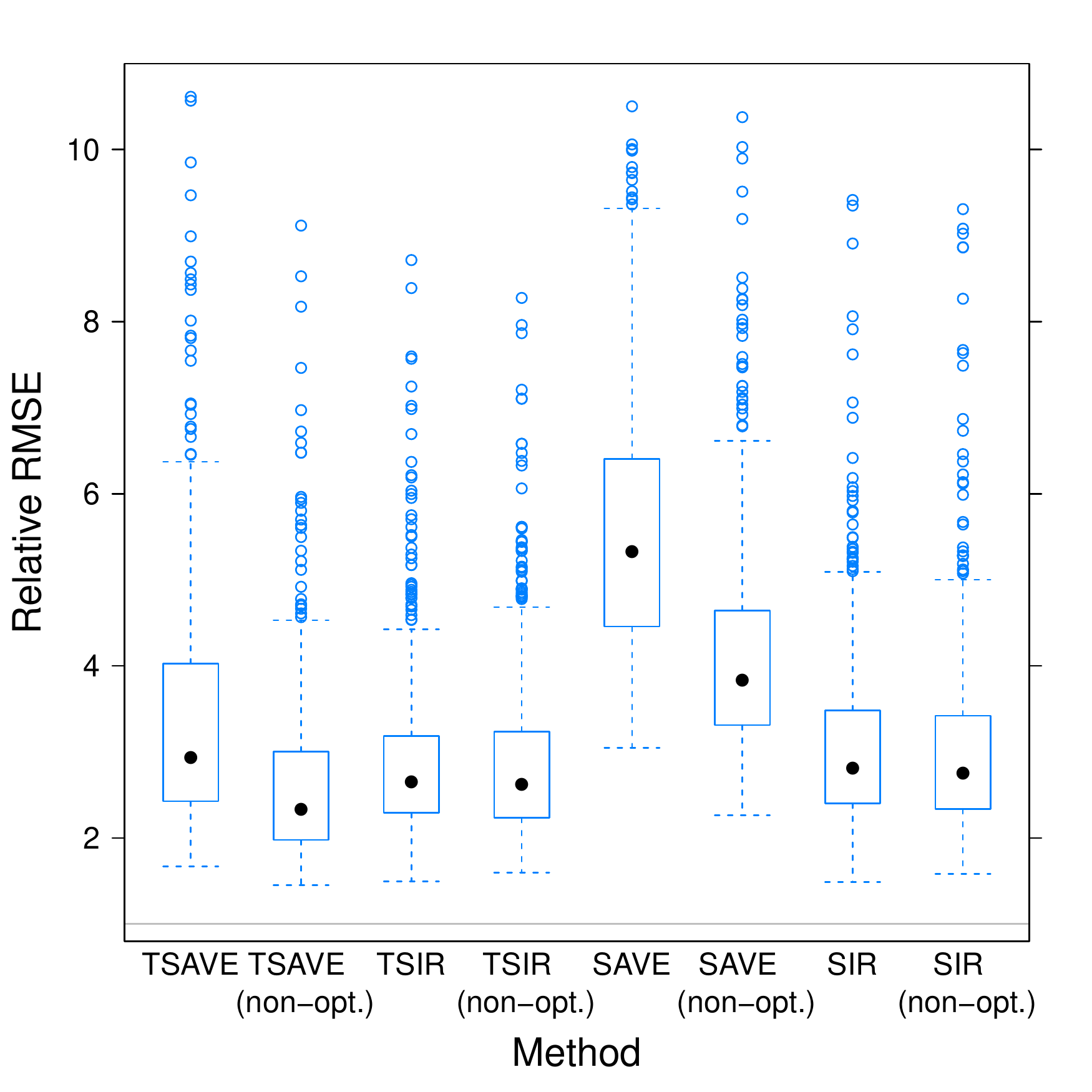}
\caption{$T = 500$: Big setting with biggest values strategy. Relative RMSE values compared to Oracle estimator.}\label{fig::Sup23}
\end{figure}
\begin{figure}[h]
\centering
\includegraphics[scale=0.44]{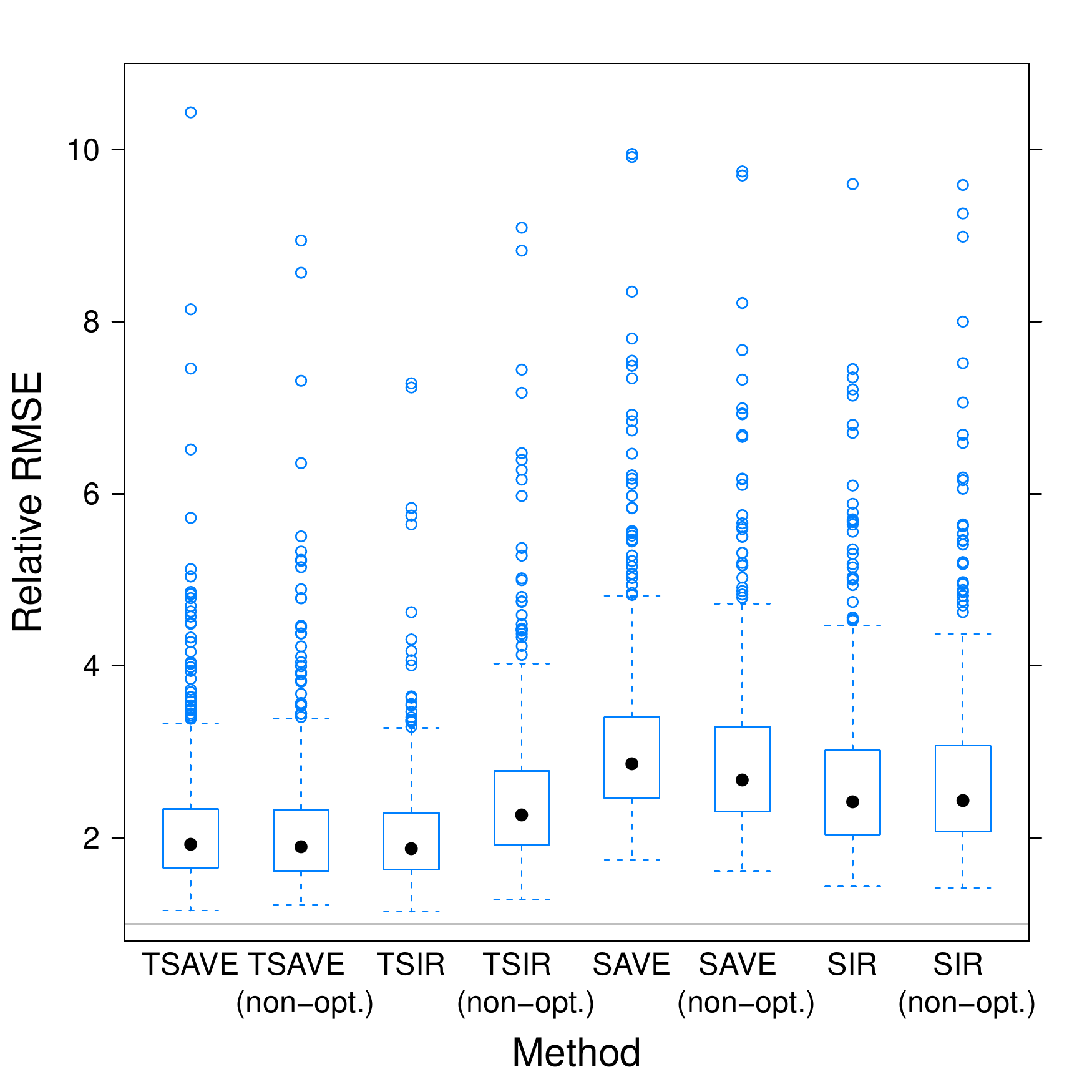}
\caption{$T = 1000$: Big setting with biggest values strategy. Relative RMSE values compared to Oracle estimator.}\label{fig::Sup24}
\end{figure}
\begin{figure}[h]
\centering
\includegraphics[scale=0.44]{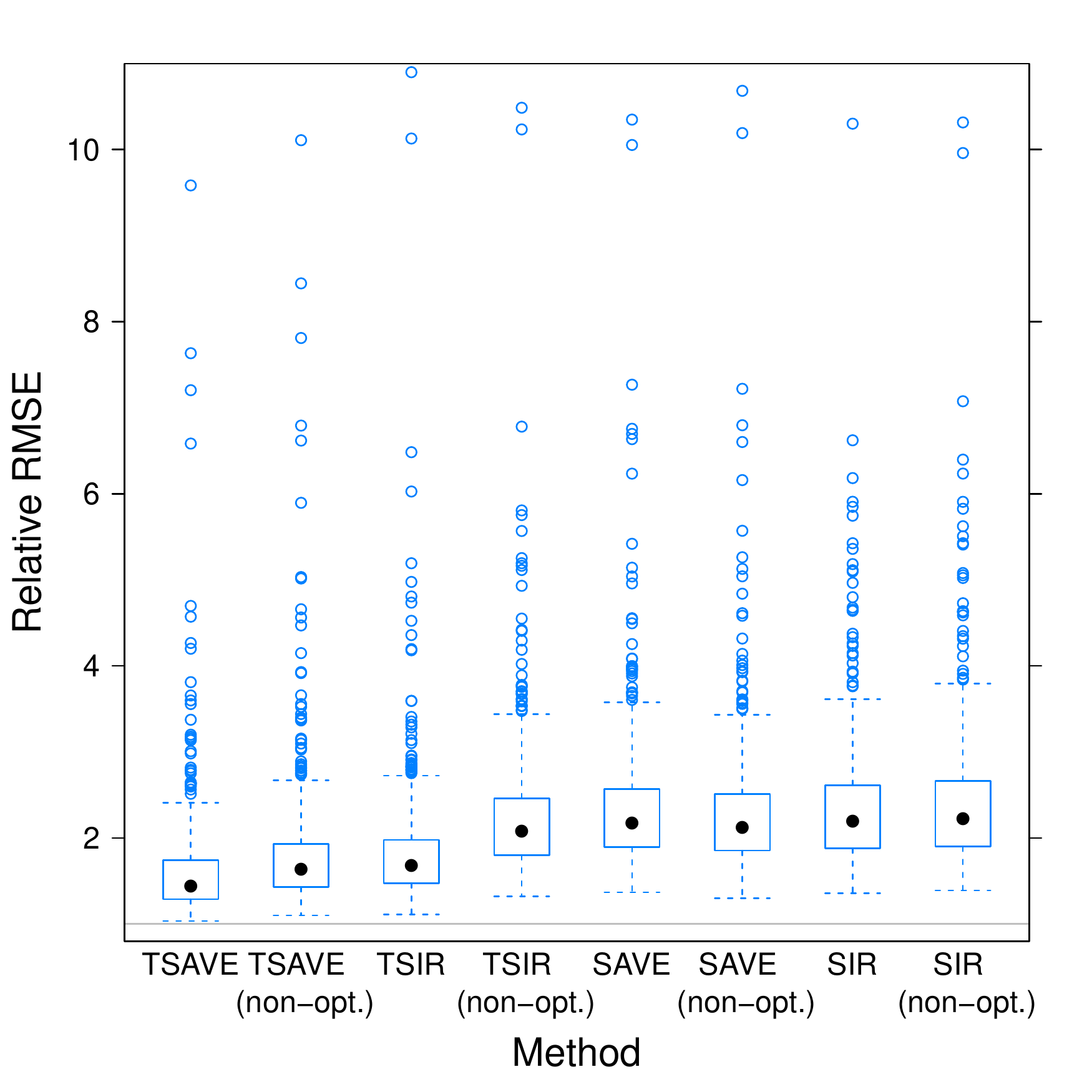}
\caption{$T = 2000$: Big setting with biggest values strategy. Relative RMSE values compared to Oracle estimator.}\label{fig::Sup25}
\end{figure}
\begin{figure}[h]
\centering
\includegraphics[scale=0.44]{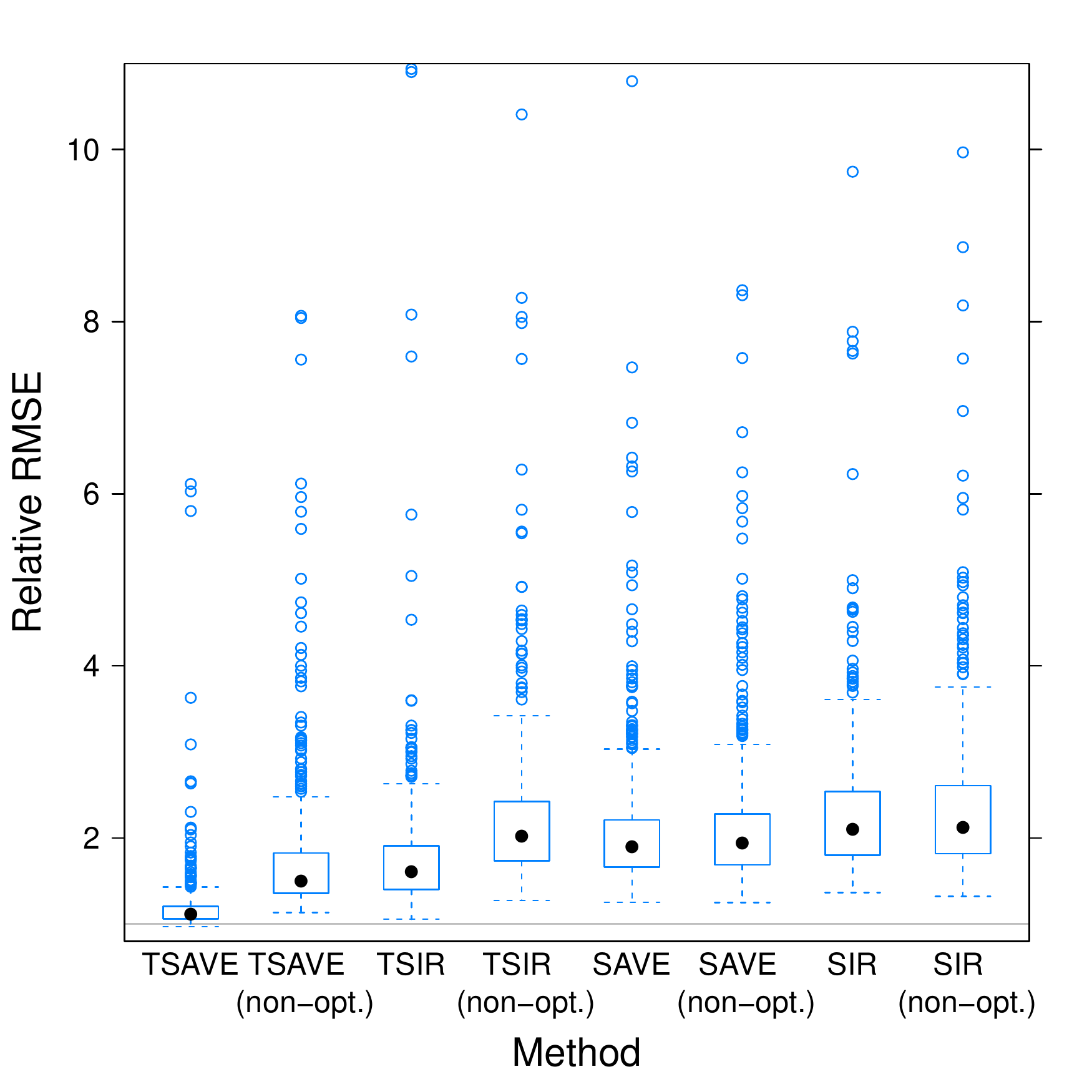}
\caption{$T = 5000$: Big setting with biggest values strategy. Relative RMSE values compared to Oracle estimator.}\label{fig::Sup26}
\end{figure}

\end{document}